\documentclass[aps,prx,reprint,floatfix,superscriptaddress]{revtex4-2}
\usepackage{graphicx}
\usepackage{bm}
\usepackage{dcolumn}
\usepackage{color}
\usepackage{bbm}
\usepackage{amsmath}
\usepackage{amssymb}
\usepackage{braket}
\usepackage[mathlines]{lineno}
\usepackage[colorlinks=true,citecolor=blue,linkcolor=blue,urlcolor=blue]{hyperref}
\begin{document}

\title{Coherent Resonant Coupling between Atoms and a Mechanical Oscillator\\ Mediated by Cavity-Vacuum Fluctuations}

\author{Bo \surname{Wang}}
\affiliation{School of Physics, Sun Yat-sen University, Guangzhou 510275, China}
\author{Jia-Ming \surname{Hu}} 
\affiliation{School of Physics, Sun Yat-sen University, Guangzhou 510275, China}
\author{Vincenzo \surname{Macr\`{\i}}}
\affiliation{Theoretical Quantum Physics Laboratory, RIKEN Cluster for Pioneering Research, Wako-shi, Saitama 351-0198, Japan}
\author{Ze-Liang \surname{Xiang}}
\altaffiliation[xiangzliang@mail.sysu.edu.cn]{}
\affiliation{School of Physics, Sun Yat-sen University, Guangzhou 510275, China}
\author{Franco \surname{Nori}}
\affiliation{Theoretical Quantum Physics Laboratory, RIKEN Cluster for Pioneering Research, Wako-shi, Saitama 351-0198, Japan}
\affiliation{RIKEN Center for Quantum Computing (RQC), Wako-shi, Saitama 351-0198, Japan}
\affiliation{Department of Physics, University of Michigan, Ann Arbor, Michigan 48109-1040, USA}

\begin{abstract}
	
We show that an atom can be coupled to a mechanical oscillator via quantum vacuum fluctuations of a cavity field enabling energy transfer processes between them. In a hybrid quantum system consisting of a cavity resonator with a movable mirror and an atom, these processes are dominated by two pair-creation mechanisms: the counterrotating (atom-cavity system) and dynamical Casimir interaction terms (optomechanical system). Because of these two pair-creation mechanisms, the resonant atom-mirror coupling is the result of high-order virtual processes with different transition paths well described in our theoretical framework. We perform a unitary transformation to the atom-mirror system Hamiltonian, exhibiting two kinds of multiple-order transitions of the pair creation. By tuning the frequency of the atom, we show that photon frequency conversion can be realized within a cavity of multiple modes. Furthermore, when involving two atoms coupled to the same mechanical mode, a single vibrating excitation of the mechanical oscillator can be simultaneously absorbed by the two atoms. Considering recent advances in strong and ultrastrong coupling for cavity optomechanics and other systems, we believe our proposals can be implemented using available technology.


\end{abstract}
	

\maketitle

\section{Introduction}

Hybrid quantum systems consisting of a mechanical oscillator and two-level systems~\cite{2013Xiang,2015kurizki} are increasingly attracting attention because of their growing potential for quantum technologies, including the storage and retrieval of quantum states~\cite{2003Leibfried,2013Pirkkalainen,2020maccabe}. One of the key challenges is how to realize and control the coupling of a mechanical oscillator and an atom at the quantum level, despite the considerable mass imbalance between them~\cite{2014Aspelmeyer}. By reaching the strong-coupling regime, a quantum interface between the mechanical oscillator and atoms can be achieved, allowing coherent energy transfers~\cite{2013Xiang,2015kurizki}. Moreover, measurement and preparation of micromechanical objects via state-of-the-art atomic control can be realized~\cite{2013Xiang,2015kurizki}. So far, the ongoing studies of the interaction mechanisms between them mainly include the electric coupling~\cite{2004Tian}, magnetic coupling~\cite{2007Treutlein,2008Lambert,2009Rabl}, dipole-dipole interaction~\cite{2008Singh}, and the coupling to the potential of optical lattices~\cite{2009Hammerer,2010Hammerer,2010Wallquist,2011Camerer,2015jockel}.

Vacuum fluctuations, which is one of the most striking quantum effects predicted by quantum field theory,  have potential value for exploring quantum technologies~\cite{2021garcia, 2012Nation, 2011Wilson}. Recently, the heat transfer induced by quantum fluctuations between two nanomechanical membranes separated by a vacuum gap has been observed in experiments~\cite{2019fong}. It can be attributed to pairs of virtual photons, acting as messengers to deliver the energy or force between objects~\cite{2019Stefano,2017Zhao,2017Stassi,2000sackett,2003leibfriedExperimental,2009dicarloDemonstration,2017Kockum,2007majer,2019SettineriConversion}. When electromagnetic quantum fluctuations interact with a very fast-oscillating boundary condition, e.g., the harmonically oscillating mirror, pairwise real excitations can be created from the vacuum of the electromagnetic field. Such a purely quantum phenomenon is known as the dynamical Casimir effect (DCE)~\cite{1970Moore,2009Johansson,2010Johansson}, which has been experimentally realized in superconducting circuits~\cite{2011Wilson} and Josephson metamaterials~\cite{2013lahteenmaki}. 

The so-called counterrotating terms (CRT) in the Rabi Hamiltonian allow the simultaneous creation or annihilation of an excitation in both atom and cavity mode. When the system reaches the ultrastrong-coupling regime~\cite{2010Niemczyk,2019Kockum,2019Forn,2019StefanoResolution,2021Settineri,2021salmon,2021Revealing,2021rajabali,2018Settineri,2020Simulating,2021Hughes}, where the coupling rate is comparable to the bare transition frequencies of the system components, the counterrotating terms can be responsible for some interesting physical phenomena~\cite{2012Ridolfo,2013Stassi,2016Garziano,2017Wang,2005Ciuti,2007Liberato,2017Cirio}. Generally, both the DCE and counterrotating terms are responsible for the amplification mechanisms of vacuum fluctuations, respectively, in cavity-optomechanics~\cite{2018Macri} and cavity QED systems~\cite{2005Ciuti}. Thus, one may use a cavity resonator as a bridge to connect these two systems into a hybrid quantum system, where the interaction between atoms and oscillators through two fluctuation mechanisms may be investigated.

In this work, we show how a single or multiple cavity-optomechanics modes can be coupled to an atom by virtual photon pairs, and how mechanical energy can be coherently converted into a two-level excitation by the amplification mechanisms of the cavity-vacuum fluctuations (the DCE and the counterrotating terms). We also develop an analytical method for studying the atom-mirror coupling, which is in good agreement with the numerical results within a simple model consisting of a single atom and a single-cavity mode. When the model involves multiple-cavity modes, it can realize frequency conversion between different cavity modes. In addition, the single-atom model can be extended to the two-atom case. By sending a mechanical drive to the movable mirror, two atoms can be excited even though the mirror and atoms are spatially separated. These results provide an intriguing platform for exploring the manipulation of quantum states by quantum fluctuations.

This paper is organized as follows. In Sec.~\ref{II}, we describe a simple model consisting of a single atom and a single-cavity mode. Then, we derive the effective system Hamiltonian comparing numerical and analytical results. Last, we study the system dynamics in this case. In Sec.~\ref{III}, we extend the simple model in Sec.~\ref{II} from a single-cavity mode to a multiple-cavity mode, to study photon frequency conversion between the two cavity modes. In Sec.~\ref{IV}, we extend the single-atom model to a two-atom case. In this case, we show that a single vibrating excitation of the mechanical oscillator can be absorbed simultaneously by two atoms. In Sec.~\ref{V}, we provide details on potential experimental implementations of proposed effects. Finally, a brief discussion and conclusion are given in Sec.~\ref{VI}. 


\begin{figure}[tpb]
	\centering
	\includegraphics[width = 0.98  \linewidth]{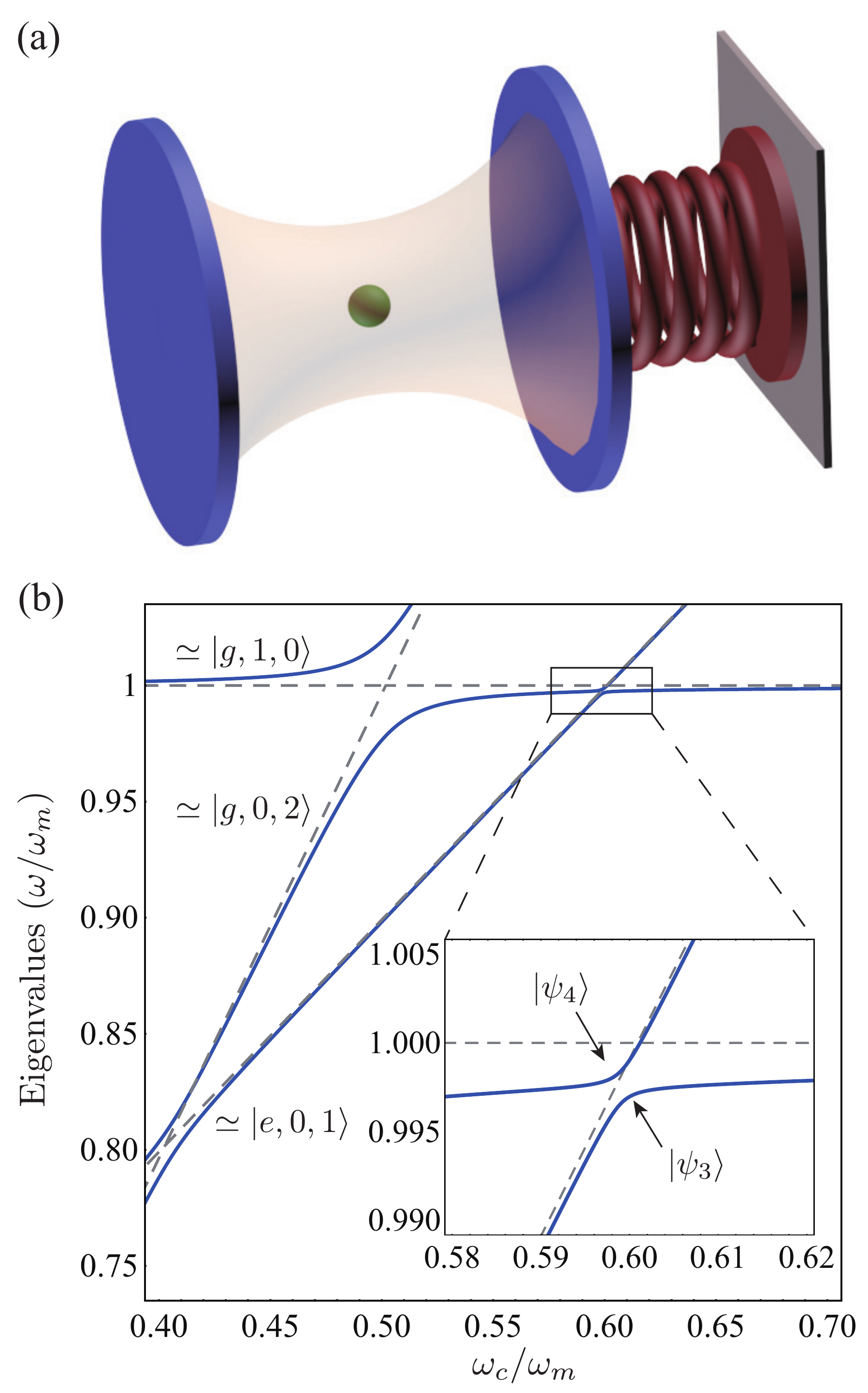}
	\caption{(a) Schematic of an hybrid quantum system consisting of a single atom and a cavity resonator with a movable mirror. (b) Relevant energy levels of the system Hamiltonian versus the ratio between the resonance frequency $\omega_c$ of the cavity mode and the one $\omega_m$ of the mirror; the inset displays the enlarged view of the boxed region showing the avoided-level crossing between $|g,1,0\rangle$ and $|e,0,1\rangle$, where $|\psi_{3,4}\rangle \simeq(1/\sqrt2 )(|g,1,0\rangle \mp |e,0,1\rangle)$ are the symmetric and antisymmetric superposition states at the minimum splitting.}
	\label{figu:1}	 	
\end{figure}

\section{single atom interacting with a cavity-optomechanical mode} \label{II}

In this section we consider a hybrid quantum system consisting of a cavity resonator with a movable mirror and a single two-level atom [see Fig.~\ref{figu:1}(a)]. Here, both the cavity field and the displacement of the mirror are treated as dynamical variables, and a canonical quantization procedure is adopted \cite{2018Macri,1995Law}. 
\subsection{Model}
We first study the case where only one mechanical mode (cavity mode) [with resonance frequency $\omega_m$ ($\omega_c$) and phonon (photon) operators $\hat b$ ($\hat a$) and $\hat b^\dag$ ($\hat a^\dag$)] is considered. The transition frequency of the atom is $\omega_a$, with the raising and lowering operators $\hat \sigma_+$ and $\hat \sigma_-$, respectively. The system Hamiltonian can be written as $\hat H = \hat H_0 + {\hat H}_{\rm I}$, where
\begin{equation}\label{eq1}
\hat H_0 = \hbar \omega_c {\hat a}^\dag {\hat a} + \hbar \omega_m {\hat b}^\dag {\hat b} + \hbar \omega_a{\hat\sigma}_+{\hat\sigma_-}
\end{equation} 
is the unperturbed Hamiltonian and ${\hat H}_{\rm I} = {\hat H}_{\rm I(AF)} +{\hat H}_{\rm I(FM)}$ describes the interaction. Here, 
\begin{equation}\label{eq2}
{\hat H}_{\rm I(AF)} = \hbar \lambda (\hat a + \hat a^\dag)(\hat \sigma_- + \hat \sigma_+)
\end{equation} 
is the atom-field interaction, while the field-mirror interaction can be described as $\hat H_{\rm I(FM)} = \hat V_{\rm om} + \hat V_{\rm DCE}$, where
\begin{equation}\label{eq3}
\hat V_{\rm om} = \hbar g{\hat a}^\dag {\hat a}(\hat b + {\hat b}^\dag)
\end{equation}
is the standard optomechanical interaction and 
\begin{equation}\label{eq4}
\hat V_{\rm DCE} = \frac{\hbar}{2}g({\hat a}^2 + {\hat a}^{\dag 2})({\hat b} + {\hat b}^\dag)
\end{equation}
describes the creation and annihilation of photon pairs~\cite{1995Law}. The parameter $\lambda $ is the coupling rate between the two-level atom and the cavity mode, and $g$ is the optomechanical coupling rate. 

In most cavity-optomechanical experiments, the DCE term is always neglected because the frequency of the oscillating mirror is much smaller than that of the cavity mode~\cite{2009groblacher,2012verhagen,2013bochmann,2014andrews,2014Aspelmeyer}, where the standard Hamiltonian of the field-mirror coupling system~\cite{1995Law} is good enough to describe them. However, when the mechanical frequency is close to the frequency of the cavity mode, the effect of the DCE terms becomes more important. Such conditions can be realized in a superconducting circuit consisting of a coplanar transmission line with a tunable electrical length~\cite{2011Wilson}, or using microwave resonators and ultra-high-frequency mechanical nanoresonators~\cite{2010Connell,2016rouxinol}. 


\subsection{Analytical Method}

From Eq.~\eqref{eq3}, we know that the transitions between different phonon states $|k\rangle$ in each \textit{n}-photon subspace can occur, such as $|k\rangle\xrightarrow{\hat V_{\rm om}} |k+1\rangle\xrightarrow{\hat V_{\rm om}} |k+2\rangle$. Thus, $\hat V_{\rm om}$ involves multiple-order transitions, namely, the transition processes between the different phonon states in a mechanical oscillator. Such transitions can occur in a high-order process with a coupling between the cavity and mechanical modes. If another component, such as an atom, is introduced, then a high-order energy exchange for multiple mixed modes can occur.

Here we can eliminate the standard optomechanical coupling term $\hat V_{\rm om}$ by performing a unitary transformation with the unitary operator $\hat U=\exp[-\beta \hat a^\dag \hat a(\hat b^\dag - \hat b)]$, obtaining the effective system Hamiltonian (details in Appendix \ref{A})
\begin{equation} \label{eq5}
	\hat H^U = \hat U^\dag \hat H \hat U = \hat H_0^U + \hat H_{\rm AFM}^U + \hat H_{\rm FM}^U,
\end{equation}
where 
\begin{equation} \label{eq6}
\begin{split}
&\hat H_0^U = \hbar \omega_c \hat a^\dag \hat a + \hbar \omega _m \hat b^\dag \hat b + \hbar \omega_{a} \hat \sigma_+ \hat \sigma_- - \hbar \frac{g^2}{\omega _m}\hat a^\dag \hat a \hat a^\dag \hat a,\\
&\hat H_{\rm FM}^U \simeq \hbar \frac{g}{2}(\hat a^{\dag 2} + \hat a^2)(\hat b + \hat b^\dag)\\
&\qquad + \hbar \frac{g^2}{\omega_m}\left[ (\hat a^{\dag 2} - \hat a^2)(\hat b^\dag - \hat b)(\hat b + \hat b^\dag) - (\hat a^{\dag 2} + \hat a^2) \hat a^\dag \hat a \right],\\
& \hat H_{\rm AFM}^U \simeq \hbar \lambda (\hat a^\dag + \hat a)(\hat \sigma_-  + \hat \sigma_+) \\
&\qquad\quad + \hbar \left(\frac{g\lambda }{\omega_m}\right)(\hat a^\dag - \hat a)(\hat \sigma_- + \hat \sigma_+)(\hat b^\dag - \hat b).
\end{split} 
\end{equation}
The Hamiltonian $\hat H_0^U$ is now the unperturbed one in the new frame. The last term in $\hat H_0^U$, originating from the optomechanical interaction $\hat V_{\rm om}$, is related to the energy shift ($\sim\hbar g^2n^2/\omega_m$) of the cavity mode.
	

In Eqs.~\eqref{eq6}, the term $\hat H_{\rm FM}^U$, which is related to the DCE, describes the creation and annihilation of photon pairs under the resonance condition $2{\omega _c} = k{\omega _m}$, where $k$ is an integer. Moreover, $\hat H_{\rm FM}^U$ involves multiple-order transitions, which are related to the creation and annihilation of multiple vibrating excitations of the mechanical oscillator. For instance, the term $\hat a^{\dag 2} \hat b^2$ in $\hat H_{\rm FM}^U$ shows the production of a photon pair by the annihilation of two excitations of the mechanical oscillator, indicating a high-order energy exchange between the cavity mode and the mechanical mode.

The last term $\hat H_{\rm AFM}^U$ in Eqs.~\eqref{eq6} describes multiple couplings involving the atom, the cavity mode, and the mechanical mode. Importantly, $\hat H_{\rm AFM}^U$ can be used to investigate the counterrotating terms, which takes into account the coupling between the atom and the mechanical oscillator for the resonance condition 
\begin{equation}
	q\omega_m \simeq \omega_c + \omega_a,
\end{equation}
where $q$ is an integer. Indeed, the term $\hat a^\dag \hat \sigma_+\hat b$ in $\hat H_{\rm AFM}^U$ is responsible for the production of an excitation pair by the annihilation of an excitation of the mechanical oscillator. It is analogous to the DCE, where a photon pair can be created by annihilating an excitation of the mechanical oscillator. Moreover, $\hat H_{\rm AFM}^U$ involves multiple-order transitions, indicating the high-order energy exchange, such as the term $\hat a^\dag \hat \sigma_+\hat b^2$ (see Appendix \ref{A}).

The Hamiltonians of the DCE and the counterrotating terms describe the pair-creation phenomena, involving the annihilation of vibrating excitations of the mechanical oscillator. However, the intrinsic pair-creation mechanisms of these two Hamiltonians are different and involve different transition paths for a coupling between the atom and the mirror.


For the unperturbed Hamiltonian $\hat H_0^{U}$, the eigenstates are described by $|j,k,n \rangle  = |j \rangle  \otimes |k \rangle  \otimes |n \rangle$, where $|k\rangle$ ($|n\rangle$) denotes the Fock state of the mechanical mode (cavity mode) and $\vert j\rangle$ ($j=g, e$) denotes the atom state, with eigenvalues 
\begin{equation}\label{S7}
	E_{j,k,n} = \hbar \omega_cn + \hbar \omega_mk + \hbar \omega_a\langle j|e\rangle-\hbar g^2n^2/\omega_m.
\end{equation}
\begin{figure}[tpb]
	\centering
	\includegraphics[width = 0.98 \linewidth]{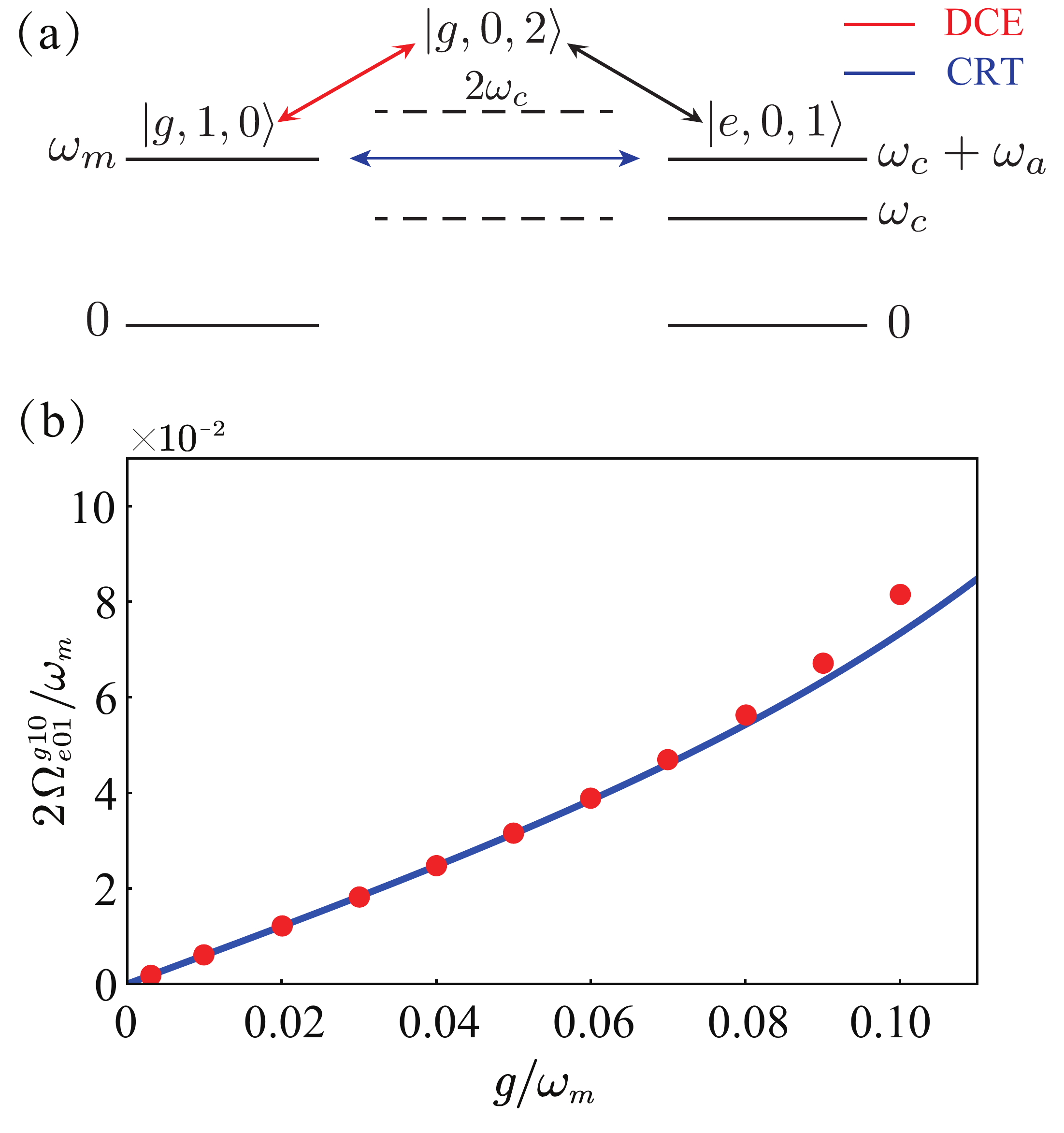}
	\caption{(a) Sketch of the processes giving the main contribution to the  effective coupling between the bare states $|g,1,0\rangle$ and $|e,0,1\rangle$, via the DCE (red arrows) and the counterrotating terms (blue arrows). (b) Comparison between the numerical calculated normalized level splitting (red dots) and the corresponding analytical calculations (blue solid curve), obtained using perturbation theory.}
	\label{figu:2}
\end{figure}


\subsection{Numerics and analysis}

We begin our study by numerically diagonalizing the Hamiltonian $\hat H$ in a truncated finite-dimensional Hilbert space. In Fig.~\ref{figu:1}(b), the blue curves display the energy differences $E_i-E_0$ ($E_0$ is the ground-state energy) of the total system Hamiltonian versus $\omega_c/\omega_m$, with the atom resonance frequency $\omega_a = 0.4\omega_m$, optomechanical coupling rate $g = 0.03\omega_m$, and atom-cavity coupling rate $\lambda=0.005\omega_m$. Compared to the dashed gray curves, which denote the lowest energy-level $E_{j,k,n} $ of the unperturbed $\hat H_0^U$, small energy shifts and an avoided-level crossing can be observed around $\omega_c/\omega_m\sim 0.6 $ (blue curve).

This avoided-level crossing, induced by the coupling between states $|g,1,0\rangle$ and $|e,0,1\rangle$, shows that the vibrating mirror with one phonon can directly excite one atom and radiate one photon, simultaneously, as shown by the inset of Fig.~\ref{figu:1}(b). At the minimum energy-level splitting $2\Omega_{e01}^{g10} = 1.83 \times 10^{-3}\omega_m$, the resulting states are well approximated by
\begin{equation}
	|\psi_{3,4}\rangle \simeq(1/\sqrt2 )(|g,1,0\rangle \mp |e,0,1\rangle).
\end{equation}
In this coupling region, the resonance condition for the DCE is $2\omega_c = \omega_m$, which means the creation of photon pairs. Since one of these photon pairs is converted into one atomic excitation, the resonance condition of the whole system can be written as
\begin{equation}
	\omega_c + \omega_a = \omega_m.
\end{equation}
Note that the result of this process is the same as the resonance condition of the multiple-couplings, $q\omega_m \simeq \omega_c + \omega_a$ with $q=1$.

Using perturbation theory, the coupling between states $|g,1,0\rangle$ and $|e,0,1\rangle$ can be described by the DCE and counterrotating terms, as shown in Fig.~\ref{figu:2}(a). This indicates that the coupling between the atom and the mirror is induced by quantum fluctuations of a cavity field. In particular, one of the processes is dominated by the DCE, as shown in the red inclined double-arrow. If $|g,1,0\rangle$ is the initial state, then it can reach the final state $|e,0,1\rangle $ through the virtual transition involving the out-of-resonance intermediate state $|g,0,2\rangle$. Such transition processes are dominated by the first terms of both $\hat H_{\rm AFM}^U$ and $\hat H_{\rm FM}^U$ in Eqs.~\eqref{eq6}. The other path originates from the counterrotating terms, as shown by the blue horizontal double-arrow. In the interaction picture, when the counterrotating terms (e.g., $\hat a^\dag \hat \sigma_+\exp[i(\omega_c + \omega_a)t]$) are on resonance with the vibrating mirror (e.g., $\hat b\exp[-i\omega_mt]$), the coupling between states $|g,1,0\rangle$ and $|e,0,1\rangle$ can be effectively triggered. Such a phenomenon of multiple-couplings can be justified by the term  $(g\lambda/\omega_m)(\hat a^\dag \hat\sigma_+ \hat b + \hat a \hat \sigma_- \hat b^\dag)$ in $\hat H_{\rm AFM}^U$. Note that the influence from the counterrotating terms grows while the detuning between the frequencies of the atom and the cavity mode increases. This can be indicated by the ratio 
\begin{equation}
	\frac{\omega_a-\omega_c}{\omega_a+\omega_c}=\frac{1}{1+(2\omega_c/\bigtriangleup)}
\end{equation}
 with the detuning $\bigtriangleup=\omega_a-\omega_c$. When the ratio increases, the influence from the counterrotating terms grows.

Therefore, the effective coupling rate between the states $|g,1,0\rangle$ and $|e,0,1\rangle$ becomes, 
\begin{equation}\label{eq8}
	\begin{split}
		\Omega_{e01}^{g10}  = \,&\, \frac{\langle e,0,1|\hat H_{\rm I}^U|g,0,2 \rangle\langle g,0,2|\hat H_{\rm I}^U|g,1,0\rangle}{E_{g,1,0} - E_{g,0,2}}\\  
		\,&\, + \langle e,0,1|\hat H_{\rm AFM}^U|g,1,0\rangle \\
		=\,&\, \frac{\lambda g \left(\sqrt{2}-\frac{\sqrt{2}g^2}{2\omega_m^2}\right) \left(\frac{\sqrt{2}}{2} + \frac{\sqrt{2}g^2}{\omega_m^2}\right)}{\omega_m - 2\omega_c + 4\frac{g^2}{\omega_m}} - \frac{\lambda g}{\omega_m}+\frac{\lambda g^3}{2\omega_m^3},
	\end{split} 
\end{equation}
where $\hat H_{\rm I}^U = \hat H_{\rm AFM}^U + \hat H_{\rm FM}^U$ is the effective perturbative Hamiltonian. In calculating the effective coupling rate, we keep the terms of the expansion of the effective Hamiltonian ($H_{\rm FM}^U$ and $H_{\rm AFM}^U$ in Appendix \ref{A}) up to the third order. The effective coupling rate in Eq.~\ref{eq8} versus $g/\omega _m$ is in good agreement with the numerical results when the normalized optomechanical coupling is not too strong, as shown in Fig.~\ref{figu:2}(b).

Note that, while the optomechanical coupling is increasing to the ultra-strongly coupling regime, the term $\hat V_{\rm DCE}$ can no longer be regarded as a perturbation because this term will cause a nonnegligible energy-shift. For avoided-level crossings at higher energy with the resonance condition $\omega_a + \omega_c = 2\omega_m$, a ladder of increasing level splittings and its corresponding analytical solution are presented in Appendix \ref{A}. When this condition is met, the higher-order DCE and counterrotating terms in the Hamiltonians $\hat H_{\rm FM}^U$ and $\hat H_{\rm AFM}^U$ respectively, are necessary for studying energy splittings.

\subsection{System dynamics}
To investigate the system dynamics, a master equation in the Lindblad form is used (details in Appendix \ref{A}). For temperature $T=0$, by numerically solving the master equation, we can obtain the time evolution of the mean photon number $ \langle {\hat A^\dag } \hat A \rangle $, the mean phonon number  $ \langle {\hat B^\dag }\hat B \rangle $, and the mean atom-excitation number  $ \langle {\hat X^\dag }\hat X \rangle $. The dressed photon, phonon and atom lowering operators $\hat O = \hat A,\hat B,\hat X $ are defined in terms of their bare counterparts $\hat o = \hat a,\hat b,\hat \sigma $ (details in Appendix \ref{A}).

\subsubsection{Weak-coupling regime}
Here, we first study the dynamical evolution when the effective coupling rate (between states $|g,1,0 \rangle $ and $|e,0,1 \rangle $) is smaller than the loss rates of the system, defined as the weak-coupling regime following the terminology of cavity QED. Here we assume the transition frequency of the atom $\omega_a=0.5\omega_m$ and set the atom, photonic, and mechanical loss rate as $\eta = \kappa = \gamma = C$, where $C$ is a constant. To properly describe the system dynamics, a single-tone mechanical drive $F(t) = A\cos (\omega_dt)$, with $\omega_d = \omega_m$, and a weak excitation $A = 0.8\gamma $ is used. Initially, we consider the system in its ground state. 

\begin{figure}[tpb]
	\centering
	\includegraphics[width = 1  \linewidth]{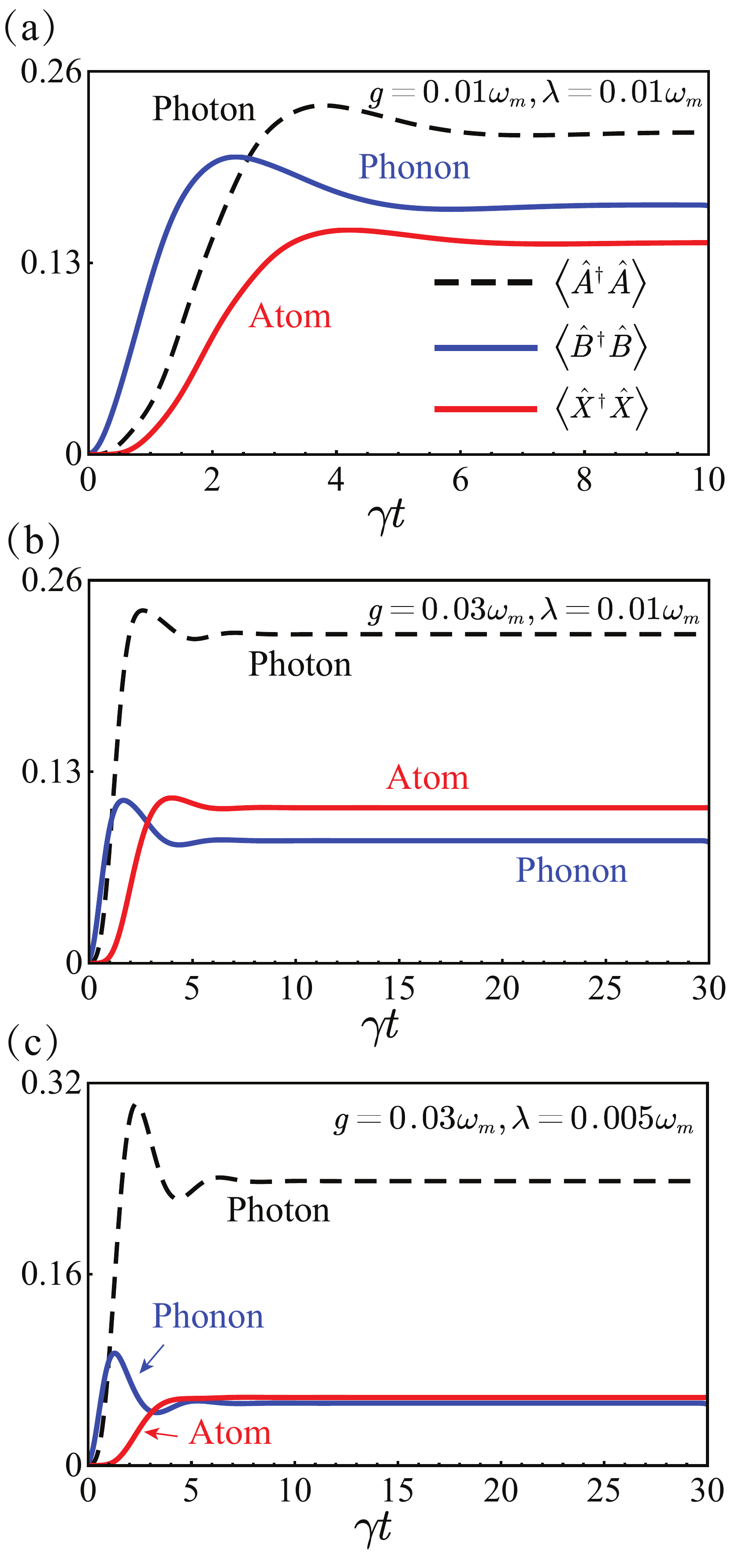}
	\caption{System dynamics for $\omega_c \backsimeq 0.5\omega_m$ with different normalized optomechanical coupling rates and cavity-atom coupling rates, under a sinusoidal drive of the vibrating mirror. The blue and red solid curves denote the mean phonon numbers $ \langle \hat B^\dag \hat B \rangle $ and mean atom-excitation numbers  $ \langle \hat X^\dag \hat X \rangle $, respectively, while the black dashed curve describes the mean photon numbers $ \langle \hat A^\dag \hat A \rangle $}
	\label{figu:3}	 	
\end{figure}

Figure~\ref{figu:3}(a) is obtained using $g{\rm{ = }}0.01{\omega _m}$, $\lambda {\rm{ = }}0.01{\omega _m}$, and $C = {\omega _m}/200$ (with a minimum energy splitting $2\Omega _{e01}^{g10} = 8.37 \times {10^{ - 3}} \omega_m$). Figure~\ref{figu:3}(b) using $g{\rm{ = }}0.03{\omega _m}$, $\lambda {\rm{ = }}0.01{\omega _m}$, and $C = 3{\omega _m}/200$ (with $2\Omega _{e01}^{g10} = 2.46 \times {10^{ - 2}}{\omega _m}$). Figure~\ref{figu:3}(c) using $g{\rm{ = }}0.03{\omega _m}$, $\lambda {\rm{ = }}0.005{\omega _m}$, and $C = 3{\omega _m}/200$ (with $2\Omega _{e01}^{g10} = 2.34 \times {10^{ - 2}}{\omega _m}$). The three plots show that the mechanical and atom excitations almost reach a same intensity, which implies that the excitations transfer between them. In Fig.~\ref{figu:3}, all the results show that the atom and the photon mean excitation number are different. This is because if the state $|g,0,2\rangle$, which is the out-of-resonance intermediate state between states $|g,1,0\rangle$ and $|e,0,1\rangle$ [see Fig.~\ref{figu:2}(a)], decays to the state $|g,0,1\rangle$, the remaining photon will not reach the state $|e,0,1\rangle$ and come back to the state $|g,1,0\rangle$. Thus, the excitation number of photon is larger than that of the atom, as shown in Fig.~\ref{figu:3}(a). 

Furthermore, the greater the coupling rate $g$, the more important is the proportion of the transition path induced by the DCE [Fig.~\ref{figu:2}(a)], resulting in a greater difference of excitation number between the atom and the photon, as shown in Figs.~\ref{figu:3}(a) and (b). 

Moreover, the greater the coupling rate $\lambda$, the less important is the proportion of the transition path induced by the DCE, resulting in a smaller difference of the excitation number between the atom and the photon, as shown in Figs.~\ref{figu:3}(b) and (c).


\begin{figure*}[tpb]
	\begin{center}
	\includegraphics[width = 0.98  \linewidth]{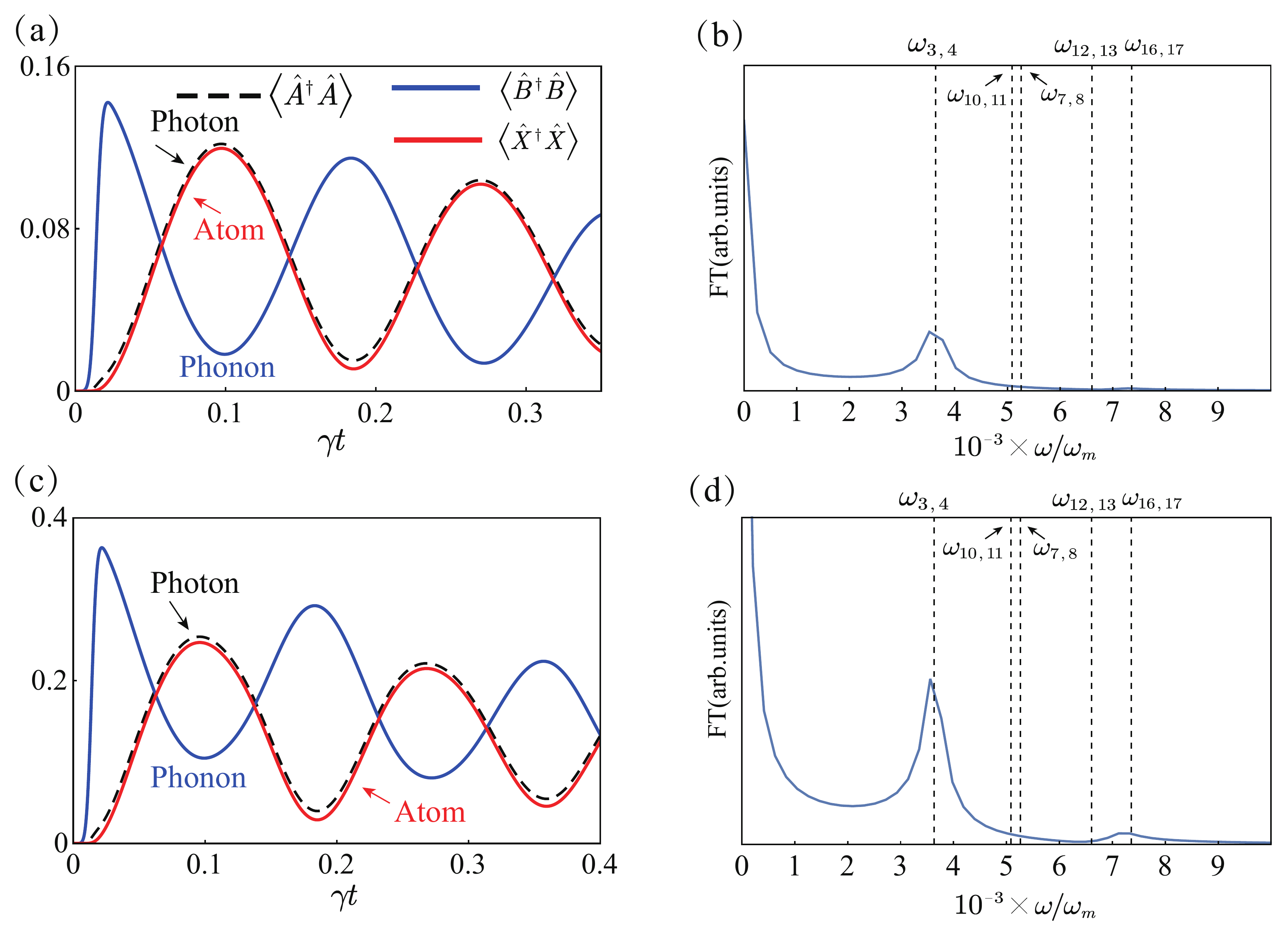}
	\caption{System dynamics for $\omega_c \backsimeq 0.6\omega_m$ after the pulse arrival. The blue and red solid curves denote the mean phonon number $ \langle \hat B^\dag \hat B\rangle$ and the mean atom-excitation number  $\langle \hat X^\dag \hat X \rangle$, respectively, while the black dashed curve describes the mean photon number $\langle \hat A^\dag \hat A\rangle$. (a) The pulse amplitude is $A = 0.25\pi $. (b) The Fourier transform of the mean atom-excitation number of panel (a). (c) The pulse amplitude is $A = 0.40\pi$. (d) The Fourier transform of the mean atom-excitation number of panel (c).}
	\label{figu:4}	 
	\end{center}	
\end{figure*}

\subsubsection{Strong-coupling regime}

Now, we study the system dynamics in the strong-coupling regime, where the effective coupling rate $2\Omega_{e01}^{g10}$ is larger than the loss rate of the system. We consider an ultrafast resonant pulse \cite{2008press} $F(t) = AG(t - t_0)\cos (\omega_dt)$, where $G(t)$ is the Gaussian function and $\omega_d = \omega_m$ is the central driving frequency, directly impinging on the mirror. Here, we set the transition frequency of the atom $\omega_a = 0.4\omega_m$, while the resonance frequency of the cavity mode is fixed at the value providing the minimum level splitting. With the coupling rates $g = 0.03\omega_m$ and $\lambda = 0.01\omega_m$, we obtain a minimum energy splitting $2\Omega_{e01}^{g10} = 3.64 \times 10^{-3}\omega_m$. In this case, we assume the loss rate $\eta = \kappa = \gamma = \omega_m/1000$ and a standard deviation $\sigma = (15\Omega_{e01}^{g10})^{-1}$.

After the pulse arrival, Fig.~\ref{figu:4} shows the system dynamics in the strong-coupling regime and the Fourier transform of the mean atom-excitation number, obtained with pulse amplitudes, $A = 0.25\pi$ for Fig.~\ref{figu:4}(a) and $A=0.40\pi $ for Fig.~\ref{figu:4}(c). In particular, Fig.~\ref{figu:4}(a) shows sinusoidal-like oscillations, exhibiting that the excitations can be reversibly transferred between the mirror and the atom (or photon), at a rate $\omega_{3,4} = E_4 - E_3 = 2\Omega_{e01}^{g10}$.

Note that, at a weak excitation level, the only desired states  $|\psi_{3,4} \rangle=(1/\sqrt{2})(|g,1,0 \rangle \mp |e,0,1 \rangle) $ are excited, as shown in Fig.~\ref{figu:4}(b). When the amplitude of the pulse increases to $A = 0.40\pi$, the system dynamics is no longer a sinusoidal-like oscillation [Fig.~\ref{figu:4}(c)]. This is because the higher-energy state (originating from the coupling between states $|g,2,1 \rangle$ and $ |e,1,2 \rangle $) is excited, indicated by a peak at ${\omega _{16,17}} = {E_{17}} - {E_{16}} $ in Fig.~\ref{figu:4}(d). This process contributes to the system dynamics. The other states indicated by ${\omega _{7,8}}$, ${\omega _{10,11}}$, and ${\omega _{12,13}}$ in Fig.~\ref{figu:4}(d), do not significantly contribute to the dynamics, since the minimum splitting of these states is not provided at the resonance frequency of the cavity mode, such that they have not been effectively excited.

\section{Single atom interacting with multiple cavity-optomechanical modes} \label{III}

In this section, we extend the previous model from a single cavity-optomechanical mode to multiple cavity-optomechanical modes. In this case, we will show how to realize frequency conversion of cavity modes.

\subsection{Model}

In the single-atom model with a single cavity-optomechanical mode, the resonance condition ${\omega _c} + {\omega _{a}} = {\omega _m}$ implies that the photons in the cavity resonator can be affected and even mediated by an atom with adjustable resonance frequency. By introducing more cavity modes into the system, one can transfer the excitations between two or more of them. Here, we study such a case where two cavity modes ($\omega_{c1}$ and $\omega_{c2}$) interact with the atom and mechanical mode, fulfilling the condition of frequency conversion: $2\omega_{c2} = \omega_a + \omega_{c1}$. The acquisition of this condition will be discussed in the next subsection. The atom is assumed to mainly interact with the mode $\omega_{c1}$. Considering the DCE and counterrotating terms, the total Hamiltonian can be described as \cite{1995Law} $\hat H = \hat H_0 + \hat H_{\rm I}$, where                                     
\begin{equation}\label{eq9}
	\begin{split}
		\hat H_0 & = \sum_{i = 1}^2 \hbar \omega_{ci}\hat a_i^\dag \hat a_i  + \hbar \omega_m \hat b^\dag \hat b + \hbar \omega_a\hat \sigma_+\hat \sigma_-, \\
		\hat H_{\rm I} & = \frac{\hbar g}{2}\left[ (\hat a_1^2 + \hat a_1^{\dag2} + 2\hat a_1^\dag \hat a_1) + \frac{\omega_{c2}}{\omega_{c1}}(\hat a_2^2 + \hat a_2^{\dag2} + 2\hat a_2^\dag \hat a_2)\right.\\
		& \left. - 2\sqrt {\frac{\omega_{c2}}{\omega_{c1}}}(\hat a_1\hat a_2 + \hat a_1^\dag \hat a_2^\dag  + \hat a_1^\dag \hat a_2 + \hat a_2^\dag \hat a_1)\right] (\hat b + \hat b^\dag)\\
		& + \hbar \lambda (\hat a_1 + \hat a_1^\dag)(\hat \sigma_- + \hat \sigma_+).
	\end{split}
\end{equation}
The second and third lines of Eqs.~\eqref{eq9} originate from the DCE and imply that the mechanical oscillator can create photon pairs in different cavity modes. By defining the quantized boson operators for hybridized cavity modes, $\hat H$ can be transformed into a similar form as that in the scenario of the single-atom interacting with a single cavity-optomechanical mode (details in Appendix \ref{B}). Consequently, one can analytically study the coupling mechanism of this system (details in Appendix \ref{B}).

\subsection{Frequency conversion}

The mechanism of the frequency conversion in the model of the multiple cavity-optomechanical modes can be introduced by following two processes. In the first process, the energy transfer between the cavity mode $\omega_{c2}$ and the mechanical mode $\omega_m$ can occur with resonance condition $2\omega_{c2} =k\omega_m$. The second process is based on the atom-mirror coupling system. In this process, the energy transfer between the cavity mode $\omega_{c1}$ and the mechanical mode $\omega_m$ can be mediated by an atom at the resonance condition $\omega_a + \omega_{c1} = q\omega_m$. Since these two processes utilize the same mechanical mode, we can obtain the relation $2\omega_{c2} = \omega_a + \omega_{c1}$.
\begin{figure}[tpb]
	\centering
	\includegraphics[width = 0.98  \linewidth]{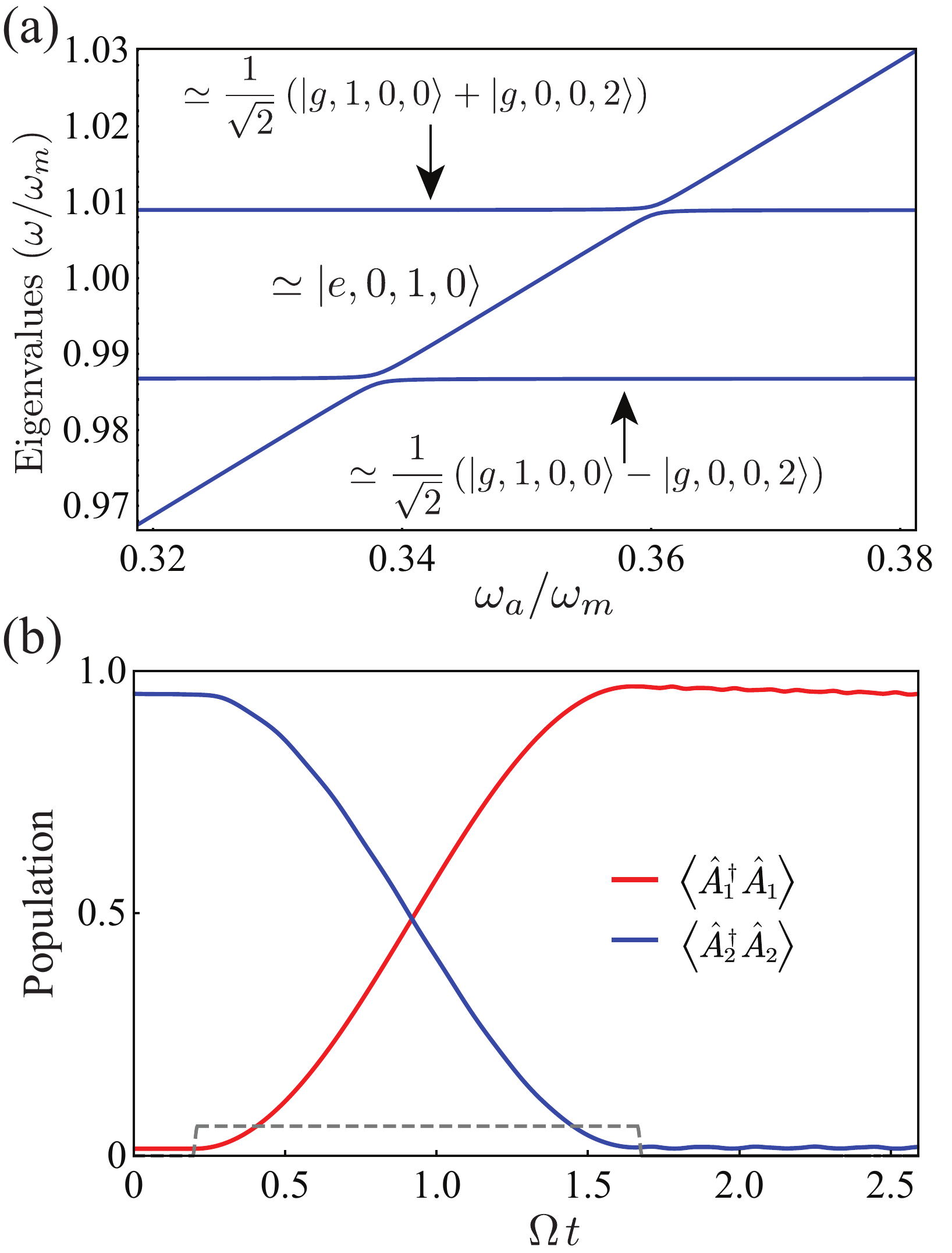}
	\caption{Frequency conversion. (a) Lowest-energy levels of the system Hamiltonian $\hat H$ versus the ratio between the transition frequency $\omega_a$ of the atom and the mechanical frequency $\omega_m$ of the mirror. (b) Time evolution of the mean photon numbers $ \langle {\hat A_1^\dag }\hat A_1 \rangle $ and $ \langle {\hat A_2^\dag }\hat A_2 \rangle $. Here, the dressed photon lowering operators $\hat O = \hat A_1,\hat A_2 $ are defined in terms of their bare counterparts $\hat o = \hat a_1,\hat a_2 $ (details in Appendix \ref{A}) \iffalse The transition frequency of atom is initially detuned from the frequency-conversion resonance condition ($\delta \omega_a=0.18$) and the system is initially prepared in the state $(1/\sqrt{2})(|g,1,0,0 \rangle -|g,0,0,2 \rangle)$. After preparation, the atom would be quickly tuned to the resonance condition ($\delta \omega_a\rightarrow 0$) and then be detuned again.\fi. The detuning $\delta \omega_a=\omega_a-\omega_a^0=0.18 \omega_c^0$, where the $\omega_a^0$ is the value at the minimum splitting, is displayed by a gray dashed curve. Such detuning can be realized by using a superconducting quantum interference device if the two-level system is an $LC$ superconducting qubit~\cite{2007sillanpaa}.}
	\label{figu:5}
\end{figure}

For instance, we consider the case where two cavity modes are $\omega_{c1}=0.65\omega_m$ and $\omega_{c2}=0.5\omega_m$, with coupling rates $g=0.02{\omega _m}$ and $\lambda = 0.01{\omega _m}$. The lowest-energy levels of the system Hamiltonian versus ${\omega _{a}/\omega _{m}}$ are plotted in Fig.~\ref{figu:5}(a). The eigenstates are $|j,k,n_1,n_2 \rangle$,  where the $|n_i\rangle$ is the Fock state of the cavity mode $\omega_{ci}$ ($i=1,2$)(details in Appendix \ref{B}). Note that there are two states $(1/\sqrt{2})(|g,0,0,2 \rangle \pm|g,1,0,0 \rangle)$, which originate from the minimum splitting between $|g,1,0,0\rangle$ and $|g,0,0,2\rangle$. These superposition states indicate the energy exchange between the cavity mode $\omega_{c2}$ and mechanical mode $\omega_m$. If the atom is present and satisfies the relation $2\omega_{c2} = \omega_a + \omega_{c1}$, then the energy of the cavity mode $\omega_{c2}$ will be first transferred to the mechanical oscillator and then to the cavity mode $\omega_{c1}$ and the atom. Specifically, when the avoided-level crossing between states $|e,0,1,0 \rangle $ and $(1/\sqrt{2})(|g,1,0,0 \rangle \pm|g,0,0,2 \rangle)$ appears, the frequency conversion of two cavity modes can occur. The corresponding effective coupling rate $\Omega$ can be calculated by using perturbation theory (details in Appendix \ref{B}).

Now, we use the master equation in the Lindblad form to study the dynamical evolution of this hybrid system in the presence of dissipation (details in Appendix \ref{A}). The initial state is $(1/\sqrt{2})(|g,1,0,0 \rangle -|g,0,0,2 \rangle)$, which can be prepared by utilizing the DCE~\cite{2018Macri}, with $\omega_a$ sufficiently detuned from the frequency-conversion resonance $2\omega_{c2} = \omega_a + \omega_{c1}$. After preparation, the atom is quickly tuned into resonance for half a Rabi oscillation period and then tuned off immediately. This scheme is commonly used for state transfer between resonators (or qubits) in circuit QED~\cite{2007sillanpaa,2008hofheinz,2009hofheinz,2011Wang,2017kockumConversion}. With $\kappa_1=\kappa_2=\gamma=\eta=\Omega/60$ for photonic (cavity modes $\omega_{c1}$ and $\omega_{c2}$), mechanical, and atomic loss rates, Fig.~\ref{figu:5}(b) displays the energy exchange between cavity modes $\omega_{c1}$ and $\omega_{c2}$. 


\section{Two-atom interacting with single cavity-optomechanical mode} \label{IV}

In this section, we extend the single-atom model to the two-atom case. We will show that a single vibrating excitation of the mechanical oscillator can be absorbed simultaneously by two atoms.

\subsection{Model}

A hybrid quantum system consisting of a cavity resonator with a movable mirror and the two atoms is considered. The system Hamiltonian can be written as 
\begin{equation}\label{eq14}
	\begin{split}
		\hat H = \,&\,\hbar {\omega _c}{\hat a^\dag }\hat a + \hbar {\omega _m}{\hat b^\dag }\hat b + \hbar {\omega _{a1}}\hat \sigma_{+}^{(1)} {\hat \sigma_{-} ^{(1)}} + \hbar {\hat \omega _{a2}}\hat \sigma_{+}^{(2)} {\hat \sigma_{-}^{(2)}}\\
		\,&\, +\hbar g{\hat a^\dag }\hat a(\hat b + {\hat b^\dag }) + \frac{\hbar }{2}g({\hat a^2} + {\hat a^{\dag 2}})(\hat b + {\hat b^\dag })\\
		\,&\, + \hbar {\lambda _1}(\hat a + {\hat a^\dag })({\hat \sigma_{+}^{(1)}} + \hat \sigma_{-}^{(1)} ) \\
		\,&\, + \hbar {\lambda _2}(\hat a + {\hat a^\dag })({\hat \sigma_{+}^{(2)}} + \hat \sigma_{-}^{(2)} ).
	\end{split}
\end{equation}
The two-atom model can be analytically solved by performing a unitary transformation (details in Appendix \ref{C}), which is similar to the case of single-atom with single cavity-optomechanical mode. For the unperturbed Hamiltonian $\hat H_0^U$ shown in Appendix \ref{C}, the eigenstate is described as $|j_1,j_2,k,n\rangle$, where $\vert j_1\rangle$ and $\vert j_2\rangle$ ($j=g, e$) denote the states of the atoms $\omega_{a1}$ and $\omega_{a2}$, respectively. For the coupling between the ground state $|g,g,1,0\rangle$ and the excited state $|e,e,0,0\rangle$, the effective coupling rate $\Omega _{ee00}^{gg10}$ can be calculated by using perturbation theory (details in Appendix \ref{C}). By study this coupling, we show that the cavity-vacuum field couples the vibrating mirror via the complete exchange of virtual photons to the two atoms.

\begin{figure}[tpb]
	\centering
	\includegraphics[width = 1  \linewidth]{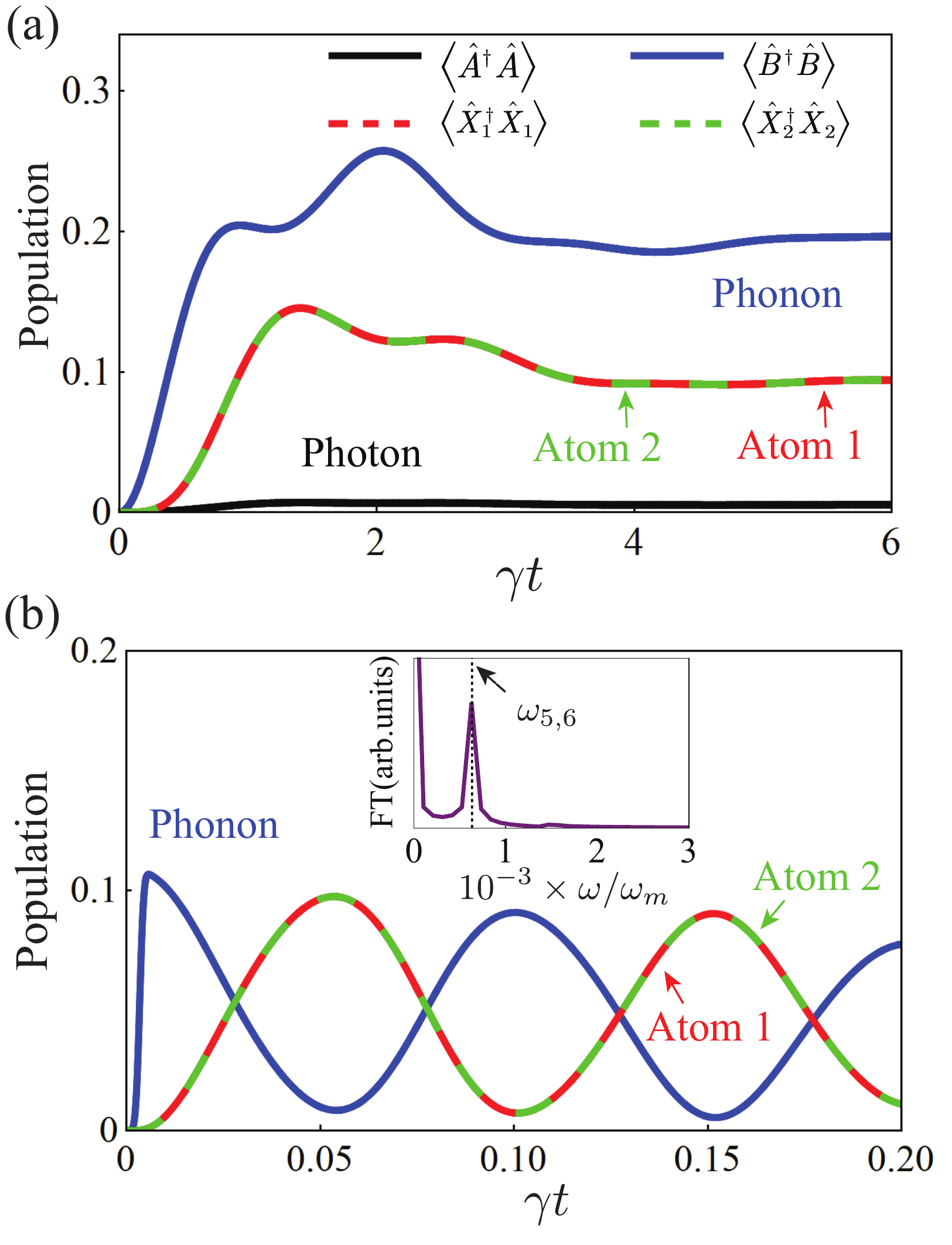}
	\caption{The dynamical evolution of the hybrid quantum system, (a) in the weak-coupling regime with a sinusoidal drive of the vibrating mirror, and (b) in the strong-coupling regime with an ultrafast resonant pulse for the mirror, where the inset displays the Fourier transform (FT) of the mean excitation number of atom 1. The black and blue solid curves describe the mean intracavity photon number $\langle \hat A^\dag \hat A\rangle $ and mean phonon number $\langle \hat B^\dag \hat B\rangle$, respectively, while the red and green dotted curves describe the mean atom-excitation number $\langle \hat X_1^\dag \hat X_1\rangle$ and $\langle \hat X_2^\dag \hat X_2\rangle$, respectively. Here, the dressed photon, phonon and atom lowering operators $\hat O = \hat A,\hat B,\hat X_1,\hat X_2 $ are defined in terms of their bare counterparts $\hat o = \hat a,\hat b,\hat \sigma_{-}^{(1)},\hat \sigma_{-}^{(2)} $ (details in Appendix \ref{A})}
	\label{figu:6}
\end{figure}

\subsection{System dynamics: One phonon being absorbed simultaneously by two atoms}

By using a master equation approach, we study the excitation transfer between the mirror and the two atoms (details in Appendix \ref{A}). For this purpose, a coherent external drive is applied to the movable mirror. The driving Hamiltonian is written as $\hat H_d(t) = F(t)(\hat{b} + \hat{b}^\dag)$, where $F(t)$ is proportional to the external force on the mirror. Moreover, $\eta_1$, $\eta_2$, $\kappa$, and $\gamma$ are the loss rates of the atom $\omega_{a1}$, atom $\omega_{a2}$, cavity mode, and mechanical mode respectively.

Firstly, we study the coherent energy transfer between the two atoms and the mirror in the weak-coupling regime, showing how the excitations of the mirror can be absorbed by the two atoms. A single-tone mechanical drive $F(t) = A\cos (\omega_dt)$ with a weak ($A = 2\gamma $) resonance excitation of the vibrating mirror ($\omega_d = \omega_m$) is adopted. The system is initially in its ground state and the evolution can be obtained, as shown in Fig.~\ref{figu:6}(a), with the transition frequency of two atoms $\omega_{a1}=\omega_{a2}=0.5\omega_m$, the normalized optomechanical coupling rate $g = 0.007\omega_m$, the normalized atom-cavity coupling rate $\lambda_1 = \lambda_2 = 0.007\omega_m$, loss rates $\eta_1 = \eta_2 = \kappa = \gamma = 1 \times 10^{-4}\omega_m$, and $\omega_c = 0.55\omega_m$. 

Figure~\ref{figu:6}(a) shows a remarkable energy transfer from the vibrating mirror to the two atoms via quantum vacuum fluctuations. Note that, at the beginning of the energy transfer, the energy of the two atoms rises simultaneously.  At the steady state, the mean atom-excitation number are detectable, even though the optomechanical coupling is not very strong. Furthermore, the photon population [black curve in Fig.~\ref{figu:6}(a)], is negligible throughout time, implying that the energy transfer is mediated by virtual photons pairs.

Then, we study the \textit{strong-coupling} regime. An ultrafast resonant pulse $F(t) = AG(t - t_0)\cos(\omega_dt)$ is applied on the movable mirror to produce a coherent mechanical state. We consider a pulse with standard deviation $\sigma = (30\Omega_{ee00}^{gg10})^{-1}$, amplitude $A = 0.25\pi$, and central driving frequency $\omega_d = \omega_m$. The system is initially prepared in its ground state with parameters $\kappa = \gamma = \eta_1 = \eta_2 = 1 \times 10^{-5}\omega_m$, $g/\omega_m = 0.03$, $\lambda_1 = \lambda_2 = 0.005\omega_m$, and $\omega_c \simeq 0.55\omega_m$. Figure~\ref{figu:6}(b) displays a coherent energy exchange between the mechanical oscillator and the two atoms after the pulse arrival. The result shows that the energy can be reversibly transferred via quantum fluctuations at a rate $\omega_{5,6} = E_6 - E_5 = 2\Omega_{ee00}^{gg10}$, as indicated by the peak in the inset of Fig.~\ref{figu:6}(b).

\section{Experimental platform for the observation of proposed effects}\label{V}

To demonstrate these results, circuit optomechanics using ultra-high-frequency mechanical resonators~\cite{2010Connell,2016rouxinol} provide a promising platform~\cite{2018Macri,2019Stefano}. An optomechanical system with a radiation-pressure interaction Hamiltonian and a cavity-atom system with Rabi Hamiltonian must be introduced into such a platform. 

The radiation-pressure interaction with a reasonable coupling strength can be obtained by considering a tripartite system consisting of an electromagnetic resonator, an ultra-high-frequency mechanical resonator, and a superconducting charge qubit. In this system, the interaction between the former two parts is mediated by the qubit, which can strongly enhance the optomechanical coupling~\cite{2014Heikkil,2015pirkkalainen}. Specifically, considering one coupling between a generic mechanical oscillator and a charge qubit through a capacitor, the interaction Hamiltonian can be written as~\cite{2019Stefano,2015pirkkalainen} 
\begin{equation}\label{eq15}
	\hat{H}_{\text{qm}}=g_m (\hat{b}+\hat{b^{\dag}})\hat{\sigma}^{(0)}_z,
\end{equation}
where $g_m$ is the coupling strength. Note that this coupling induces a qubit energy shift depending on the mechanical displacement so that $\omega_q \longrightarrow \omega_q +2g_m (\hat{b}+\hat{b}^{\dag}) $, where $\omega_m$ is the mechanical frequency and $\omega_q$ is the transition frequency of the qubit. On the other hand, following Refs.~\cite{2014Heikkil} and ~\cite{2015pirkkalainen}, we consider the additional interaction of the qubit with an electromagnetic resonator, described by  $H_{\text{qc}}=g_c (\hat{a}+\hat{a}^{\dag})\hat{\sigma}^{(0)}_x$. In the dispersive regime, the qubit-cavity interaction can be approximately described by~\cite{2009Zueco}
\begin{equation}\label{eq16}
	\hat{H}_{\text{qm}}=\frac{g^2_c}{2\varDelta} (\hat{a}+\hat{a}^{\dag})^2\hat{\sigma}^{(0)}_z,
\end{equation}
where $\varDelta=\omega_q -\omega_c$ is the detuning frequency. By considering the energy shift of the qubit, the detuning can be rewritten as $\varDelta=\omega_q +2g_m (\hat{b}+\hat{b}^{\dag})-\omega_c$, depending on the mechanical displacement. If we assume a small displacement and consider the qubit in its ground state, then we can obtain the following optomechanical interaction from Eq.~\eqref{eq16},
\begin{equation}\label{eq17}
	\hat{H}_{\text{qm}}=\frac{g}{2} (\hat{a}+\hat{a}^{\dag})^2 (\hat{b}+\hat{b}^{\dag}),
\end{equation}
with
\begin{equation}\label{eq17_1}
	g=\frac{2g_m g^2_c}{(\omega_q -\omega_c)^2}.
\end{equation}

Assuming that the optomechanical system depicted above can be coupled through a capacitor to a superconducting qubit, the effects discussed in this work could be effectively observed. The interaction Hamiltonian can be described by
\begin{equation}\label{eq18}
	\hat{H}_{\text{I}}=\frac{g}{2} (\hat{a}+\hat{a}^{\dag})^2 (\hat{b}+\hat{b}^{\dag})+\lambda (\hat{a}+\hat{a}^{\dag})\hat{\sigma}_x.
\end{equation}
The optomechanical coupling rate $g$ and the cavity-qubit coupling rate $\lambda$ can reach the ultra-strong coupling limit~\cite{2019Kockum,2019Forn}, which should allow us to observe the phenomena predicted in this work.

If the above electromagnetic resonator consists of a coplanar waveguide resonator, which can be designed to trap atomic ensembles~\cite{2017Hattermanna,2006andre}, then it is possible to realize the integration of ultracold atoms (or molecules) with circuit optomechanics. This integration could achieve the atom-mirror coupling  via quantum fluctuations. To obtain a moderate scale of the coupling rate $\lambda$, a rather effective way is to employ an ensemble of $N$ atoms instead of a single atom~\cite{2009Finka,2009Verda,2010Schustera,2010Kuboa,2017Hattermanna,2011zhu} to enhance the atom-cavity coupling by a factor of $\sqrt{N}$. 

In addition, to observe the dynamical evolution of the system, we can apply a coherent driving field to the mechanical resonator through the piezoelectric material, which converts the electrical signal into mechanical vibrations~\cite{2010Connell,2017chu}. In the weak-coupling regime, as in the previous discussion, one can employ the continuous-wave electric field as the coherent drive~\cite{2019Scarlino,2015wollman,2011teufel}. For the system dynamics in the strong-coupling regime, the required ultrafast resonant pulse, like the Gaussian pulse, can be achieved by mixing the fields from the arbitrary waveform generator and a continuous-wave coherent drive~\cite{2019Scarlino}.

%

\begin{table*}[tbp]
	\caption{\label{tab:ex}The coupling between the atoms and the mirror via cavity-vacuum fluctuations with three models described in this work.}
	\begin{ruledtabular}
		\begin{tabular}{llll}
			\rule[-3mm]{0mm}{8mm} Model$^\text{a}$ & $N_a=1$, $N_c=1$, $N_m=1$ & $N_a=1$, $N_c=2$, $N_m=1$ & $N_a=2$, $N_c=1$, $N_m=1$\\
			\hline
			\rule[-1mm]{0mm}{6mm} Main feature & The creation of an excitation   & Frequency conversion & The creation of an excitation  \\
			\rule[-3mm]{0mm}{0mm}  & in both 1 atom and 1 photon  & between two cavity modes & in both atoms \\
			\hline
			\rule[-3mm]{0mm}{8mm} Resonant condition & $N\omega_m=\omega_{c}+\omega_{a}$  & $2\omega_{c2} = \omega_a + \omega_{c1}$ & $N\omega_m=\omega_{a1}+\omega_{a2}$ \\
			\hline
			\rule[-3mm]{0mm}{8mm} Dominant effect$^\text{b}$ & DCE and CRT  & DCE, CRT, and scattering effect & DCE and CRT \\
		\end{tabular}
	\end{ruledtabular}
	\footnotesize $^\text{a}$ \rule[-1.8mm]{0mm}{6mm} $N_a$, $N_c$, and $N_m$ refer to the number of atoms, optomechanical-cavity modes, and mechanical modes, respectively.
	
	$^\text{b}$ \rule[0mm]{0mm}{0mm} DCE and CRT refer to the dynamical Casimir effect and the counterrotating terms, respectively.
	\label{Table1}
\end{table*}

\section{Discussion and Conclusion}\label{VI}

We have demonstrated that the coupling between the atoms and a mechanical oscillator can be induced by vacuum fluctuations of a cavity field, which are amplified by the DCE and counterrotating terms. The novel coupling mechanism has the potential to couple the atom and the mechanical oscillator with a considerable mass imbalance. 

Table~\ref{Table1} lists the three models described in this work. By comparing the zero-atom case~\cite{2018Macri} with the single-atom and the two-atoms ones, we demonstrate that the DCE photons can be transformed to the atomic excitations from one atom to two atoms. It is possible to detect the DCE effect by collecting the readout signal of the quantum state of the atomic system. This approach can be immune to some spurious ambient emission processes for the measured photons and, therefore, can be used as an auxiliary means to estimate the DCE effect. Compared to the two-atom case, the single-atom one may be more suitable for this task because the former (single-atom) requires more high-order processes than the latter.

For the two-atom case, the conditional quantum-state transfer is a possible application. One atom could be encoded with information as a superposition of the ground and excited states. The second atom can be used as a switch to control whether the information is transmitted to the membrane. If the second atom's transition frequency satisfies the resonant condition of the system (e.g., $N\omega_m=\omega_{a1}+\omega_{a2}$), then a single phonon will be created in a membrane that contains the encoded information.

For the two cavity-optomechanical modes model, from the resonant condition of the photon frequency conversion $2\omega_{c2} = \omega_a + \omega_{c1}$, this model can realize frequency up-conversion and down-conversion of optical signals by mediating the atom's transition frequency. In addition, it is also possible to realize wideband wavelength conversion at the expense of requiring a high optomechanical coupling rate and a cavity-qubit coupling rate.

One model with two mechanical modes can also be explored. In this case, we can consider a hybrid quantum system consisting of a cavity resonator with two movable mirrors and two atoms. When one of the two mirrors is driven and encoded with quantum information, the atoms can absorb the excitations. Before reaching the other mirror, we can store these excitations by controlling the atom's transition frequency to get away from the resonant condition of the system. Thus, controlling the arrival time of the excitations between two mirrors is possible. Such a model can also be analytically studied by using the method of Sec.~\ref{II}.

\acknowledgments

We thank Mauro Cirio for a critical reading of the manuscript. This work was supported by the National Natural Science Foundation of China (Grant No. 11874432), the National Key R\&D Program of China (Grant No. 2019YFA0308200), and the China Postdoctoral Science Foundation (Grant No. 2021M693682). F.N. is supported in part by: Nippon Telegraph and Telephone Corporation (NTT) Research, the Japan Science and Technology Agency (JST) [via the Quantum Leap Flagship Program (Q-LEAP) program, and the Moonshot R\&D Grant No. JPMJMS2061], the Japan Society for the Promotion of Science (JSPS) [via the Grants-in-Aid for Scientific Research (KAKENHI) Grant No. JP20H00134], the Asian Office of Aerospace Research and Development (AOARD) (via Grant No. FA2386-20-1-4069), and the Foundational Questions Institute Fund (FQXi) via Grant No. FQXi-IAF19-06.


\medskip
\begin{appendix}
	
\section{Single atom interacting with single cavity-optomechanical mode}\label{A}
\subsection{Effective system Hamiltonian}
By performing a unitary transformation with the unitary operator $\hat U = \exp [ - \beta {\hat a^\dag }\hat a({\hat b^\dag } - \hat b)]$ for $\hat H$ in Sec.~\ref{II} of the main text, we can obtain an effective system Hamiltonian
\begin{equation}\label{S9}
	\hat H^U = \hat U^\dag \hat H\hat U = \hat H_0^U + \hat H_{\rm FM}^U + \hat H_{\rm AFM}^U,
\end{equation}
where
\begin{equation}\label{S12}
	\hat H_0^U = \hat U^\dag(\hat H_0+\hat V_{\rm om})\hat U = \hat H_0 - \hbar \frac{g^2}{\omega _m}\hat a^\dag \hat a\hat a^\dag \hat a,
\end{equation}
\begin{equation}\label{S21}
	\begin{split}
		\hat H_{\rm FM}^U =\,&\, \hat U^\dag \hat V_{\rm DCE}\hat U\\
		= \,&\,\sum_{n = 0}^\infty\frac{1}{n!} \hbar g\left(\frac{g}{\omega _m}\right)^n2^{n-1}\left\{\left[\hat a^{\dag 2} + (-1)^n\hat a^2\right]\right.\\
		\,&\, \left. \times(\hat b^\dag - \hat b)^n (\hat b^\dag + \hat b) - n\left[\hat a^{\dag 2} + (-1)^{n-1}\hat a^2\right] \right.\\
		\,&\, \left. \times\hat a^\dag \hat a(\hat b^\dag - \hat b)^{n-1}\right\},
	\end{split}
\end{equation}
and
\begin{equation}\label{S15}
	\begin{split}
		\hat H_{\rm AFM}^U =\,&\, \hat U^\dag \hat H_{\rm I(AF)}\hat U \\
		=\,&\, \hbar \sum_{n = 0}^\infty \frac{1}{n!} \left(\frac{g}{\omega_m}\right)^n\lambda \left[\hat a^\dag + (-1)^n\hat a\right](\hat \sigma_- + \hat \sigma_+) \\
		\,&\, \times(\hat b^\dag - \hat b)^n.
	\end{split}
\end{equation}

\subsection{Analytical approach to the high-order resonance condition $\omega_a + \omega_c = 2\omega_m$}

\begin{figure}[tpb]
	\centering
	\includegraphics[width = 0.98 \linewidth]{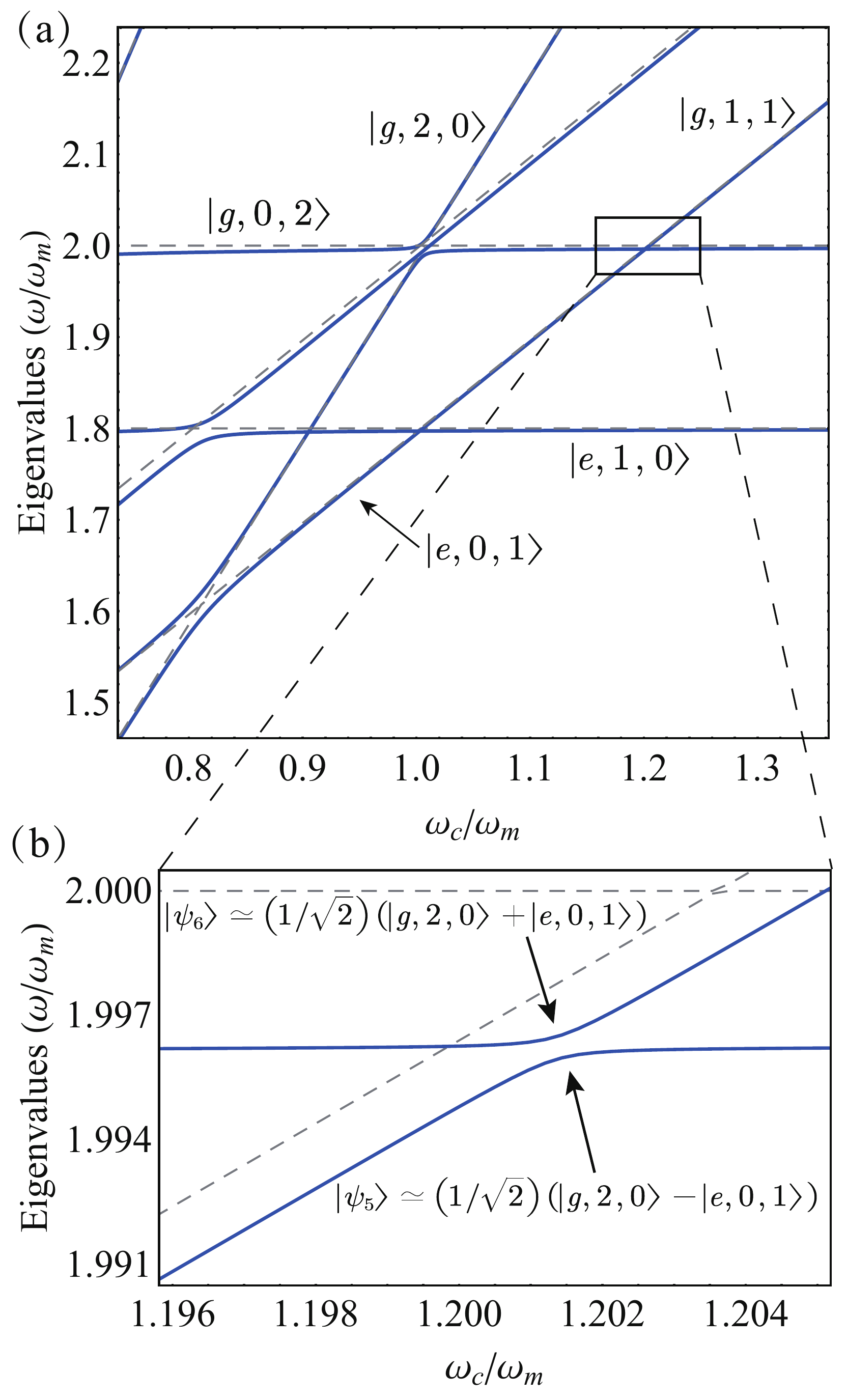}
	\caption{(a) Relevant energy levels of the system Hamiltonian $\hat H$ versus the ratio between the resonance frequencies of the cavity mode $\omega_c$ and the vibrating mirror $\omega_m$. (b) The enlarged view of the boxed region in panel (a) shows the avoided-level crossing between $|g,2,0 \rangle $ and $|e,0,1 \rangle $, which is induced by the DCE and the counterrotating terms. The resulting coupling involves the high-order resonance condition ${\omega _{a}} + {\omega _c} = 2{\omega _m}$.}
	\label{figu:7}	 	
\end{figure}

\begin{figure}[tpb]
	\centering
	\includegraphics[width = 0.9  \linewidth]{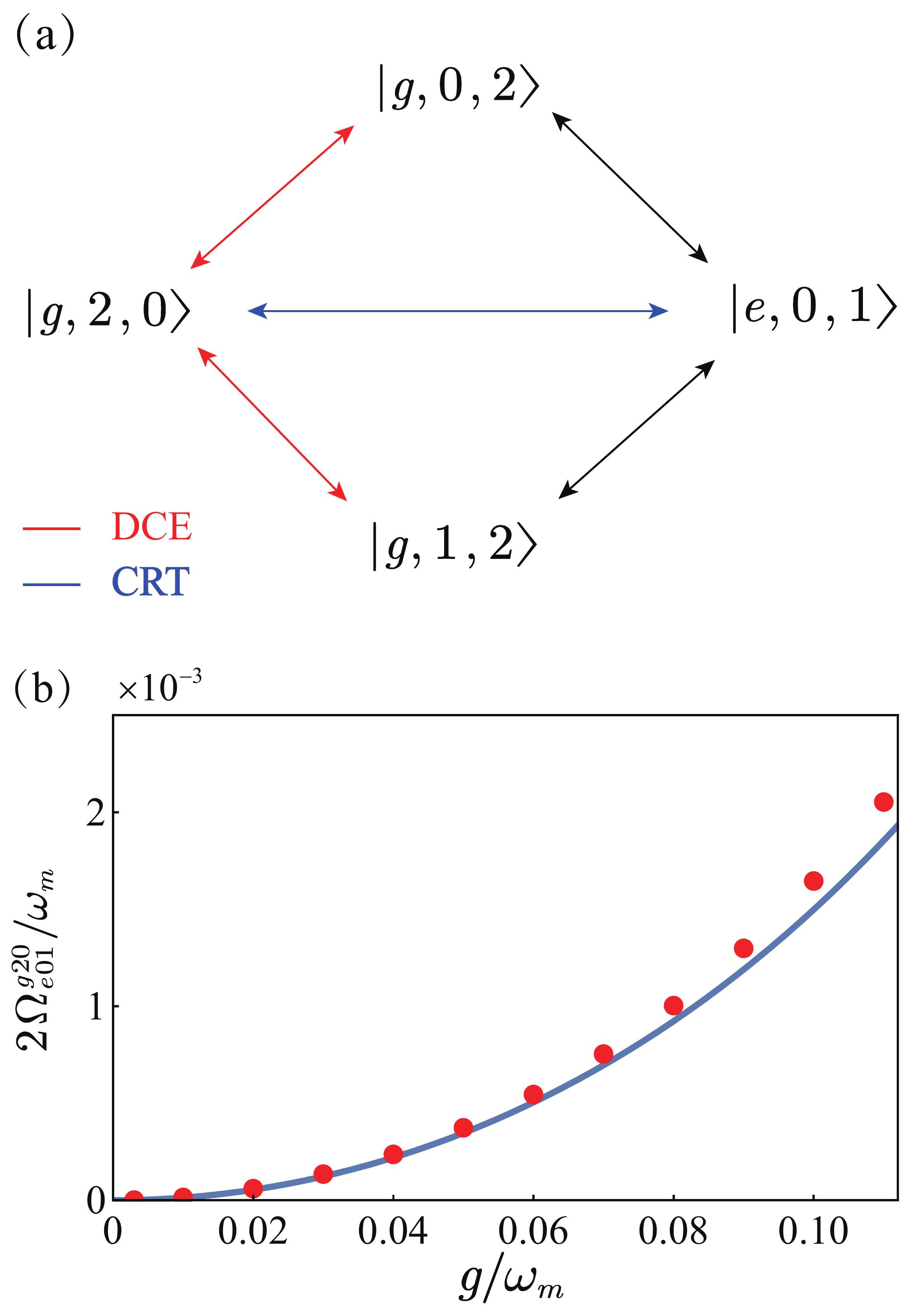}
	\caption{(a) Sketch of the processes giving the main contribution to the effective coupling between the states $|g,2,0 \rangle $ and $|e,0,1 \rangle $, via the DCE  (red arrows) and the counterrotating terms (blue arrows). (b) Comparison between the numerical (red dots) and analytical (blue solid curve) results of the normalized level splitting.}
	\label{figu:8}	 	
\end{figure}

In a ladder of increasing level splitting, the corresponding effective coupling rate can also be calculated by using perturbation theory. For the coupling between states $|g,2,0 \rangle $ and $|e,0,1 \rangle $ in Fig.~\ref{figu:7}, with parameters $g{\rm{ = }}0.06{\omega _m}$, $\lambda {\rm{ = }}0.01{\omega _m}$, and ${\omega _a}{\rm{ = }}0.8{\omega _m}$, the transition paths are shown in Fig.~\ref{figu:8}(a). The top two transition paths involve the creation (or annihilation) of excitation pairs with the high-order processes of the DCE and the counterrotating terms, while the last path shows the standard resonant condition of the DCE. With these transition paths, the effective coupling rate can be obtained
\begin{equation}\label{S24}
	\begin{split}
		\Omega _{e01}^{g20} = \,&\,\frac{\langle e,0,1|\hat H_{\rm I}^U|g,0,2 \rangle \langle g,0,2|\hat H_{\rm I}^U|g,2,0 \rangle }{E_{g,2,0} - E_{g,0,2}}\\
		\,&\, +\frac{ \langle e,0,1|\hat H_{\rm I}^U|g,1,2 \rangle \langle g,1,2|\hat H_{\rm I}^U|g,2,0 \rangle }{E_{g,2,0} - E_{g,1,2}} \\
		\,&\, +  \langle e,0,1|\hat H_{AFM}^U|g,2,0 \rangle \\
		=\,&\, \frac{-\frac{g^2 \lambda}{\omega_m}\left(\sqrt{2}-\frac{\sqrt{2} g^2}{2\omega_m^2}\right)}{\omega_m-\omega_c+\frac{2g^2}{\omega_m}}+\frac{\frac{g^2\lambda}{\omega_m} \left(\sqrt{2}-\frac{\sqrt{2} g^2}{2\omega_m^2}\right)}{\omega_m-2\omega_c+\frac{4g^2}{\omega_m}}\\
		\,&\, + \frac{\sqrt 2 \lambda g^2}{2\omega_m^2}.
	\end{split}
\end{equation}

where $\hat H_{\rm I}^U = \hat H_{\rm AFM}^U + \hat H_{\rm FM}^U$ is the effective perturbative Hamiltonian after the unitary transformation. For calculating the effective coupling rate, we keep the terms of the expansion of the effective Hamiltonian ($\hat H_{\rm FM}^U$ and $\hat H_{\rm AFM}^U$) up to the third order. In Fig.~\ref{figu:8}(b), the numerical and the analytical results of the normalized splitting ($2\Omega _{e01}^{g20}/{\omega_m}$) are in good agreement. This shows that a perturbative approach is still applicable if the normalized optomechanical coupling is not too strong.

\subsection{Master equation approach}

To study the system dynamics, a master equation in the Lindblad form is used. Here, the Born-Markov approximation without the post-trace rotating-wave approximation \cite{2018Macri,2002theory} is considered and the resulting master equation for the density-matrix operator $\hat \rho (t)$ at temperature $T=0$ \cite{2018Settineri} can be described by 
\begin{equation}
	\begin{split}
        \dot {\hat {\rho}} (t) =\,&\, \frac{i}{\hbar}[\hat \rho (t),\hat H + \hat H_d(t)] \\
        \,&\, + \kappa D[\hat A]\hat \rho (t) + \gamma D[\hat B]\hat \rho (t) + {\eta }D[{\hat X}]\hat \rho (t), 
    \end{split}
\end{equation}
where the $\kappa $, $\gamma $, and ${\eta}$ are photonic, mechanical, and atom loss rates, respectively. The driving Hamiltonian is written as $\hat H_{\rm d}(t)=F(t)(\hat b+\hat b^\dag)$, where $F(t)$ is proportional to the external force on the mirror. The dressed photon, phonon and atom lowering operators $\hat O = \hat A,\hat B,\hat X $ are defined in terms of their bare counterparts $\hat o = \hat a,\hat b,\hat \sigma_{-} $ as 
\begin{equation}
     \hat O = \sum\limits_{E_n > E_m} { \langle \psi_m|\hat o + \hat o^\dag|\psi_n \rangle } |\psi_m \rangle  \langle \psi_n|, 
\end{equation}
and the superoperator \textit{D} is defined as
\begin{equation}
    D[\hat O]\hat \rho  = \frac{1}{2}(2\hat O\hat \rho {\hat O^\dag } - {\hat O^\dag }\hat O\hat \rho  - \hat \rho {\hat O^\dag }\hat O). 
\end{equation}
Moreover, we label the $|\psi_n \rangle $ eigenvectors of $\hat H$ and the $E_n$ corresponding eigenvalues, such that $E_n>E_m$ for $n>m$.

\section{Single atom interacting with multiple cavity-optomechanical modes}\label{B}

\subsection{Effective system Hamiltonian}

In this case, the system Hamiltonian is described as \cite{1995Law} $\hat H=\hat H_0+\hat H_{\rm I}$, where
\begin{equation}\label{S25}
	\hat H_0 = \sum_{i = 1}^2 \hbar \omega_{ci}\hat a_i^\dag \hat a_i  + \hbar \omega_m \hat b^\dag \hat b + \hbar \omega_a\hat \sigma_+\hat \sigma_-
\end{equation}
is the unperturbed Hamiltonian and 
\begin{equation}\label{S26}
	\begin{split}
		\hat H_{\rm I} & = \frac{\hbar g}{2}\left[ (\hat a_1^2 + \hat a_1^{\dag2} + 2\hat a_1^\dag \hat a_1) + \frac{\omega_{c2}}{\omega_{c1}}(\hat a_2^2 + \hat a_2^{\dag2} + 2\hat a_2^\dag \hat a_2)\right.\\
		& \left. - 2\sqrt {\frac{\omega_{c2}}{\omega_{c1}}}(\hat a_1\hat a_2 + \hat a_1^\dag \hat a_2^\dag  + \hat a_1^\dag \hat a_2 + \hat a_2^\dag \hat a_1)\right] (\hat b + \hat b^\dag)\\
		& + \hbar \lambda (\hat a_1 + \hat a_1^\dag)(\hat \sigma_- + \hat \sigma_+)
	\end{split}
\end{equation}
describes the interaction. 

To conveniently analyze this quantum system, we need to perform a transformation for the system Hamiltonian. Here, we define the quantized boson operators for the hybridized cavity modes
\begin{equation}\label{S27}
	\begin{split}
		\hat p=\frac{c\hat a_1+\hat a_2}{\sqrt{1+c^2}}\ \ \text{and}\ \ \hat q=\frac{\hat a_1-c\hat a_2}{\sqrt{1+c^2}},
	\end{split}
\end{equation}
where $c=\sqrt{\omega_{c2}/\omega_{c1}}$. They satisfy these relations $[\hat p,\hat q]=0$, $[\hat p,\hat q^\dag]=0$, $[\hat p,\hat p^\dag]=1$, and $[\hat q,\hat q^\dag]=1$. The system Hamiltonian can be rewritten as $\hat {\tilde{H}}=\hat {\tilde{H}}_0+\hat {\tilde{H}}_{\rm I}$, where
\begin{equation}\label{S28}
	\begin{split}
		\hat {\tilde{H}}_0 \,&\, = \hbar \tilde{\omega}_{c1}{\hat p^\dag }\hat p  + \hbar \tilde{\omega}_{c2}{\hat q^\dag }\hat q  +\hbar {\omega _m}{\hat b^\dag }\hat b + \hbar {\omega _{a}}{\hat \sigma_{+} }\hat \sigma_{-} 
	\end{split}
\end{equation}
is the unperturbed Hamiltonian and $\hat {\tilde{H}}_{\rm I}=\hat {\tilde{V}}_{\rm DCE}+\hat {\tilde{V}}_{\rm om}+\hat {\tilde{H}}_{\rm I(AF)}+\hat {\tilde{H}}_{\rm I(FF)}$ describes the interaction. Here,
\begin{equation}\label{S29}
	\begin{split}
		\hat {\tilde{V}}_{\rm DCE} = \frac{{\hbar G}}{2}(\hat q^2 + \hat q^{\dag 2})(\hat b^\dag +\hat b),
	\end{split}
\end{equation}
\begin{equation}\label{S30}
	\begin{split}
		\hat {\tilde{V}}_{\rm om} \,&\, = \hbar G \hat q^\dag \hat q(\hat b^\dag +\hat b),
	\end{split}
\end{equation}
\begin{equation}\label{S31}
	\begin{split}
		\hat {\tilde{H}}_{\rm I(AF)} = \frac{\hbar \lambda}{\sqrt{1+c^2}} (c\hat p^\dag+c\hat p + \hat q^\dag+\hat q )(\hat \sigma_{-}  + {\hat \sigma_{+} }),
	\end{split}
\end{equation}
and
\begin{equation}\label{S32}
	\begin{split}
		\hat {\tilde{H}}_{\rm I(FF)} = \frac{\Delta c \hbar}{1+c^2}(\hat p^\dag \hat q+\hat q^\dag \hat p),
	\end{split}
\end{equation}
where $G=(1+c^2)g$ is the effective optomechanical coupling rate for the hybridized cavity modes and $\Delta=\omega_{c1}-\omega_{c2}$ is the detuning between the cavity mode 1 and mode 2. 
\begin{align*}
	\tilde{\omega}_{c1}=\omega _{c2}+\frac{\Delta c^2}{1+c^2} \ \ \text{and}\ \ \tilde{\omega}_{c2}=\omega _{c2}+\frac{\Delta}{1+c^2}
\end{align*}
are the effective frequency of the hybridized cavity mode 1 and mode 2, respectively. $\hat {\tilde{H}}_{\rm I(FF)}$ describes the scattering interaction between the hybridized cavity modes. Note that the system Hamiltonian $\hat {\tilde{H}}$ is analogous to the scenario of a single-atom with a single cavity-optomechanical mode, so we can utilize the analytical method of Sec.~\ref{II} to this case.

Applying a unitary transformation with the unitary operator $\hat U = \exp [ - (G/\omega_m) {\hat q^\dag }\hat q({\hat b^\dag } - \hat b)]$ for $\hat {\tilde{H}}$, we obtain 
\begin{equation}\label{S33}
	\hat {\tilde{H}}_{\rm 0}^U = {\hat U^\dag }(\hat {\tilde{H_{\rm 0}}}+\hat {\tilde{V}}_{\rm om})\hat U = \hat {\tilde{H_{\rm 0}}} - \hbar \frac{{{G^2}}}{{{\omega _m}}}{\hat q^\dag }\hat q{\hat q^\dag }\hat q,
\end{equation}
\begin{equation}\label{S34}
	\begin{split}
		\hat {\tilde{H}}_{\rm FM}^U =\,&\, {\hat U^\dag }\hat {\tilde{V}}_{\rm DCE}\hat U\\
		= \,&\,\sum\limits_{n = 0}^\infty  {\frac{1}{{n!}}} \hbar G{\left(\frac{G}{{{\omega _m}}}\right)^n}{2^{n - 1}}\left\{ \left[{\hat q^{\dag 2}} + {( - 1)^n}{\hat q^2}\right]\right.\\ 
		\,&\,\left. \times{({\hat b^\dag } - \hat b)^n}({\hat b^\dag } + \hat b)- n\left[{\hat q^{\dag 2}} + {( - 1)^{n - 1}}{\hat q^2}\right]\right.\\
		\,&\,\left. \times{\hat q^\dag }\hat q{({\hat b^\dag } - \hat b)^{n - 1}}\right\},
	\end{split}
\end{equation}
\begin{equation}\label{S35}
	\begin{split}
		 \hat {\tilde{H}}_{\rm AFM}^U =\,&\, {\hat U^\dag }\hat {\tilde{H}}_{\rm AF}\hat U \\
		=\,&\, \hat {\tilde{H}}_{\rm AF} + \frac{\hbar \lambda}{\sqrt{1+c^2}}\sum\limits_{n = 1}^\infty  {\frac{1}{{n!}}} {\left(\frac{G}{{{\omega _m}}}\right)^n}\\
		\,&\, \times\left[{\hat q^\dag } + {( - 1)^n}\hat q\right] (\hat \sigma_{-}  + {\hat \sigma_{+} }){({\hat b^\dag } - \hat b)^n},
	\end{split}
\end{equation}
and
\begin{equation}\label{S36}
	\begin{split}
		 \hat {\tilde{H}}_{\rm FFM}^U =\,&\, {\hat U^\dag }\hat {\tilde{H}}_{\rm FF}\hat U \\
		=\,&\, \frac{\Delta c \hbar}{1+c^2}\sum\limits_{n = 0}^\infty  {\frac{1}{{n!}}} {\left(\frac{G}{{{\omega _m}}}\right)^n} \left[{\hat q^\dag \hat p} + {( - 1)^n}\hat q \hat p^\dag \right]\\
		\,&\,\times{({\hat b^\dag } - \hat b)^n}.
	\end{split}
\end{equation}
Here, $\hat {\tilde{H}}_{\rm FFM}^U$ describes the scattering interaction between the hybridized cavity modes $\tilde{\omega}_{c1}$ and $\tilde{\omega}_{c2}$, involving the multiple-order transition processes induced by $\hat {\tilde{V}}_{\rm om}$.

Now, by substituting the Eqs.~\eqref{S27} to the Eqs.~(\ref{S33})-(\ref{S36}), we can obtain the effective system Hamiltonian $\hat H^U=\hat H_0^U+\hat H_{\rm FM}^U+\hat H_{\rm AFM}^U+\hat H_{\rm FFM}^U$. Here, 
\begin{equation}\label{S37-0}
	\begin{split}
		\hat H_0^U=\,&\,\hat H_0-\frac{\hbar g^2}{\omega_m}\left(\hat a_1^\dag \hat a_1\hat a_1^\dag \hat a_1+2c^2a_1^\dag \hat a_1\hat a_2^\dag \hat a_2\right.\\
		\,&\, \left. +c^4a_2^\dag \hat a_2\hat a_2^\dag \hat a_2+c^2a_1^\dag \hat a_1\hat a_2 \hat a_2^\dag+c^2a_2^\dag \hat a_2\hat a_1 \hat a_1^\dag \right)
	\end{split}
\end{equation}
is the unperturbed Hamiltonian.
\begin{equation}\label{S37}
	\begin{split}
		\hat H_{\rm FM}^U = \,&\,\sum\limits_{n = 0}^\infty  {\frac{1}{{n!}}} \hbar G{\left(\frac{G}{{{\omega _m}}}\right)^n}{2^{n - 1}}\left[\frac{1}{1+c^2}L_1{({\hat b^\dag } - \hat b)^n}\right.\\
		\,&\,\left. \times({\hat b^\dag } + \hat b)- \frac{n}{(1+c^2)^2}L_{2}L_{3}{({\hat b^\dag } - \hat b)^{n - 1}}\right],
	\end{split}
\end{equation}
where 
\begin{equation}\label{S37-1}
	\begin{split}
		L_1=\,&\,\left[\hat a_1^{\dag 2}+(-1)^{n}\hat a_1^{2}\right]+c^{2}\left[\hat a_2^{\dag 2}+(-1)^{n}\hat a_2^{2}\right]\\
		\,&\,-2c\left[\hat a_1^{\dag}\hat a_2^{\dag}+(-1)^{n}\hat a_{1}\hat a_{2}\right],
	\end{split}
\end{equation}
\begin{equation}\label{S37-2}
	\begin{split}
		L_2=\,&\,\left[\hat a_1^{\dag 2}+(-1)^{n-1}\hat a_1^{2}\right]+c^{2}\left[\hat a_2^{\dag 2}+(-1)^{n-1}\hat a_2^{2}\right]\\
		\,&\,-2c\left[\hat a_1^{\dag}\hat a_2^{\dag}+(-1)^{n-1}\hat a_{1}\hat a_{2}\right],
	\end{split}
\end{equation}
and
\begin{equation}\label{S37-3}
	\begin{split}
		L_{3}={\hat a_1^\dag}\hat a_1-c{\hat a_1^\dag}\hat a_2-c{\hat a_2^\dag}\hat a_1+c^2{\hat a_2^\dag}\hat a_2,
	\end{split}
\end{equation}
describes the creation (or annihilation) of photon pairs with different cavity modes, involving the multiple-order transitions. This
\begin{equation}\label{S38}
	\begin{split}
		\hat H_{\rm AFM}^U=\,&\,\hbar \lambda ({\hat a_1} + \hat a_1^\dag )(\hat \sigma_{-}  + {\hat \sigma_{+} })\\
		\,&\,+ \frac{\hbar \lambda}{\sqrt{1+c^2}}\sum\limits_{n = 1}^\infty  {\frac{1}{{n!}}} {\left(\frac{G}{{{\omega _m}}}\right)^n} \left\{ \left[{\hat a_1^\dag } + {( - 1)^n}\hat a_1\right] \right.\\
		\,&\,\left.-c\left[{\hat a_2^\dag } + {( - 1)^n}\hat a_2\right] \right\}(\hat \sigma_{-}  + {\hat \sigma_{+} }){({\hat b^\dag } - \hat b)^n},
	\end{split}
\end{equation}
describes the photon-atom-phonon coupling with different cavity modes, involving multiple-order transitions.
\begin{equation}\label{S39}
	\begin{split}
		\hat H_{\rm FFM}^U \,&\,= \hat H_{\rm FF}+ \frac{\Delta c \hbar}{(1+c^2)^2}\sum\limits_{n = 1}^\infty {\frac{1}{{n!}}} {\left(\frac{G}{{{\omega _m}}}\right)^n}\\ 
		\,&\, \times\left\{\left[1+(-1)^n\right]c({\hat a_1^\dag}\hat a_1-{\hat a_2^\dag}\hat a_2)+\left[1+(-1)^{n-1}c^2\right] \right. \\
		\,&\, \left. \times\left[{\hat a_1^\dag}\hat a_2+(-1)^n{\hat a_1}\hat a_2^\dag \right]\right\}{({\hat b^\dag } - \hat b)^n},
	\end{split}
\end{equation}
where
\begin{equation}\label{S40}
	\begin{split}
		\hat H_{\rm FF}=\,&\,-\frac{g^2 \hbar}{\omega_m}\left[-(c{\hat a_1^\dag}\hat a_1+c^3{\hat a_2^\dag}\hat a_2)({\hat a_1^\dag}\hat a_2+{\hat a_2^\dag}\hat a_1)\right.\\
		\,&\,\left.-({\hat a_1^\dag}\hat a_2+{\hat a_2^\dag}\hat a_1)(c{\hat a_1^\dag}\hat a_1+c^3{\hat a_2^\dag}\hat a_2)\right.\\
		\,&\,\left.+c^2({\hat a_1^{\dag 2}}\hat a_2^2+{\hat a_2^{\dag 2}}\hat a_1^2)\right],
	\end{split}
\end{equation}
describes the scattering interaction between the cavity modes. The scattering effect involves the multiple-order transitions between different phonon states in a mechanical oscillator. 

For the unperturbed Hamiltonian $\hat H_0^U$, the eigenstates can be written as $|j,k,n_1,n_2 \rangle  = |j \rangle  \otimes |k \rangle  \otimes |n_1 \rangle \otimes |n_2 \rangle $ with the eigenvalue 
\begin{equation}\label{S40-1}
	\begin{split}
		{E_{j,k,n}} =\,&\, \hbar {\omega _{c1}}n + \hbar {\omega _{c2}}n + \hbar {\omega _m}k + \hbar {\omega _a}\langle j|e\rangle \\
		\,&\,-\frac{\hbar g^2}{\omega _m}\left[n_1^2+2c^2{n_1 n_2}+c^4{n_2^2} \right.\\
		\,&\, \left. +c^2{n_{1}(n_2+1)}+c^2{n_{2}(n_1+1)}\right].
	\end{split}
\end{equation}

\begin{figure}[tpb]
	\centering
	\includegraphics[width = 1  \linewidth]{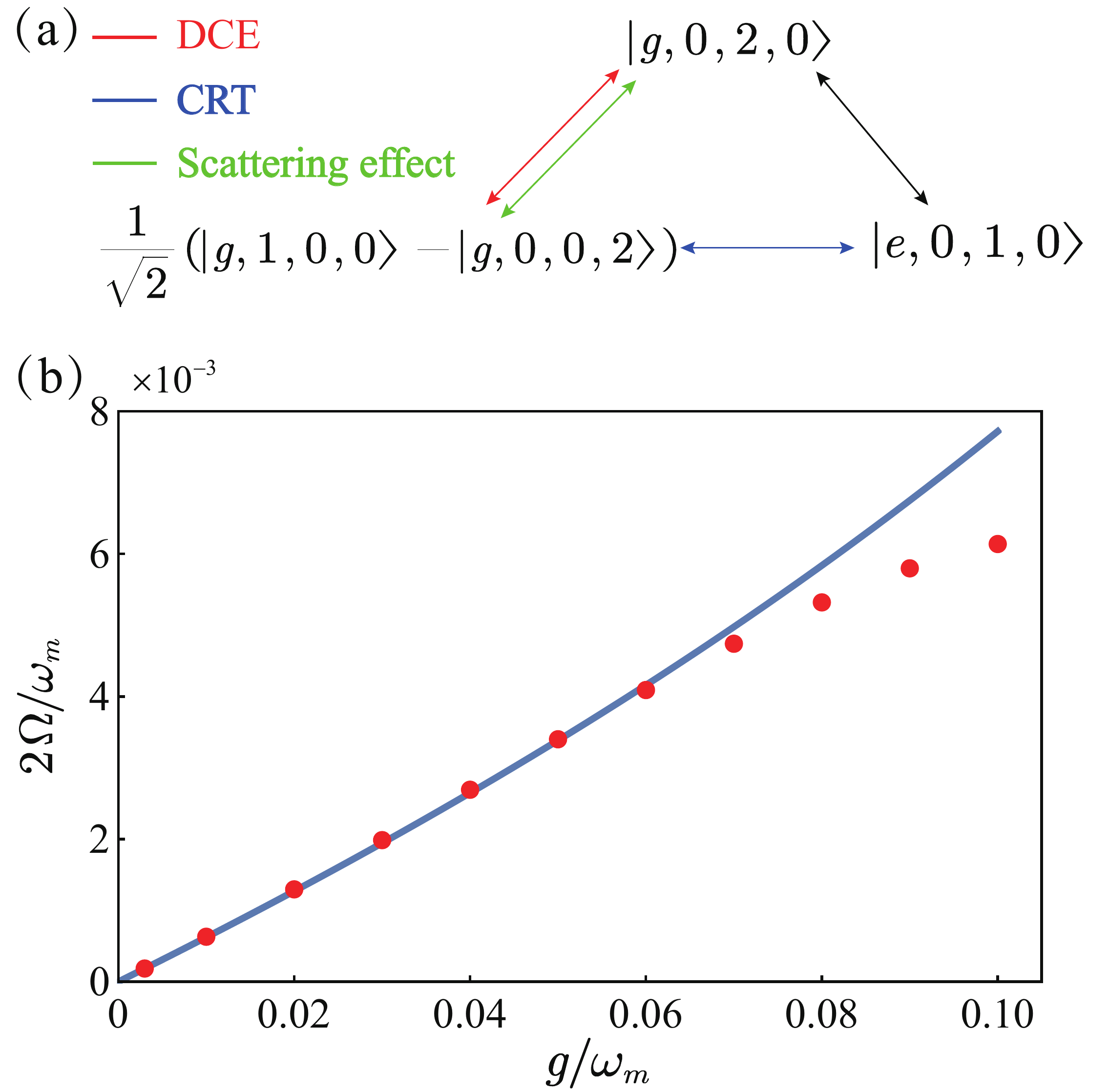}
	\caption{(a) Sketch of the processes giving the main contribution to the effective coupling between the states $(1/\sqrt{2})(|g,1,0,0 \rangle -|g,0,0,2 \rangle)$ and $|e,0,1,0 \rangle$, via DCE (red arrows), counterrotating terms (blue arrows), and scattering effect (green arrows). (b) Comparison between the numerical (red dots) and analytical (blue solid curve) results of the normalized level splitting.}
	\label{figu:9}
\end{figure}

\subsection{Analytical approach to the resonance condition $2\omega_{c2} = \omega_a + \omega_{c1}$} 

For the coupling between states $|e,0,1,0 \rangle $ and $(1/\sqrt{2})(|g,1,0,0 \rangle -|g,0,0,2 \rangle)$ shown in Fig.~\ref{figu:5}(a) in the main text, the transition paths are shown in Fig.~\ref{figu:9}(a). If the state $(1/\sqrt{2})(|g,1,0,0 \rangle -|g,0,0,2 \rangle)$ is the initial state and prepared, then it reaches the final state $|e,0,1,0 \rangle $ through virtual transition involving the intermediate states $|g,0,2,0 \rangle$ by the DCE, or it directly reaches the final state by the counterrotating terms. Besides the two effects, the scattering effect can also contribute to the coupling, as shown in Fig.~\ref{figu:9}(a). Thus, the effective coupling rate can be obtained by using perturbation theory.
\begin{equation}\label{S41}
	\begin{split}
		\Omega  = \,&\,\frac{ \langle e,0,1,0|\hat H_{\rm I}^U|g,0,2,0 \rangle  \langle g,0,2,0|\hat H_{\rm I}^U|\psi_i \rangle }{{E_{\rm i}} - {E_{g,0,2,0}}}\\  
		\,&\, + \langle e,0,1,0|\hat H_{\rm AFM}^U|\psi_i \rangle\\
		=\,&\,\frac{ -\lambda \left(1-\frac{(1+c^2)g^2}{2\omega_m^2}\right)\left(\frac{2g^{2} c^2}{\omega_m}+\frac{\sqrt{2}g}{2}+\frac{(1+c^2)^{2}g^{3}\sqrt{2}}{\omega_m^2}\right)}{2(\omega_{c2}-\omega_{c1})-\frac{4g^{2}(c^{4}-1)}{\omega_m}}\\
		\,&\,-\frac{1}{\sqrt{2}}\left(-\frac{g \lambda}{\omega_m}+\frac{(1+c^2)^{2}\lambda g^3}{2\omega_m^3}\right),
	\end{split} 
\end{equation}
where $\hat H_{\rm I}^U = {\hat H_{\rm AFM}^U + \hat H_{\rm FM}^U+\hat H_{\rm FFM}^U}$ is the effective perturbative Hamiltonian after the unitary transformation and $|\psi_i\rangle=(1/\sqrt{2})(|g,1,0,0 \rangle -|g,0,0,2 \rangle)$ is the initial state. In this case, the transition path of the scattering effect is dominated by the last line of Eq.~\eqref{S40}. Specifically, 
\begin{align*}
	\begin{split}
	\langle g,0,2,0|\hat H_{\rm FFM}^U|g,0,0,2\rangle =\,&\, \langle g,0,2,0|\hat H_{\rm FF}|g,0,0,2\rangle \\
	=\,&\, \frac{-2g^{2} c^2}{\omega_m}.
	\end{split}
\end{align*}
This means that the scattering effect is a second-order process in the coupling. Compared to the DCE and counterrotating terms, the contribution of the scattering effect in the effective coupling rate is relatively small.  Figure~\ref{figu:9}(b) shows a comparison between the numerical and the analytical results of the normalized splitting ($2\Omega/{\omega_m}$). The agreement is good if the normalized optomechanical coupling is not too strong.

\begin{figure}[tpb]
	\centering
	\includegraphics[width = 1  \linewidth]{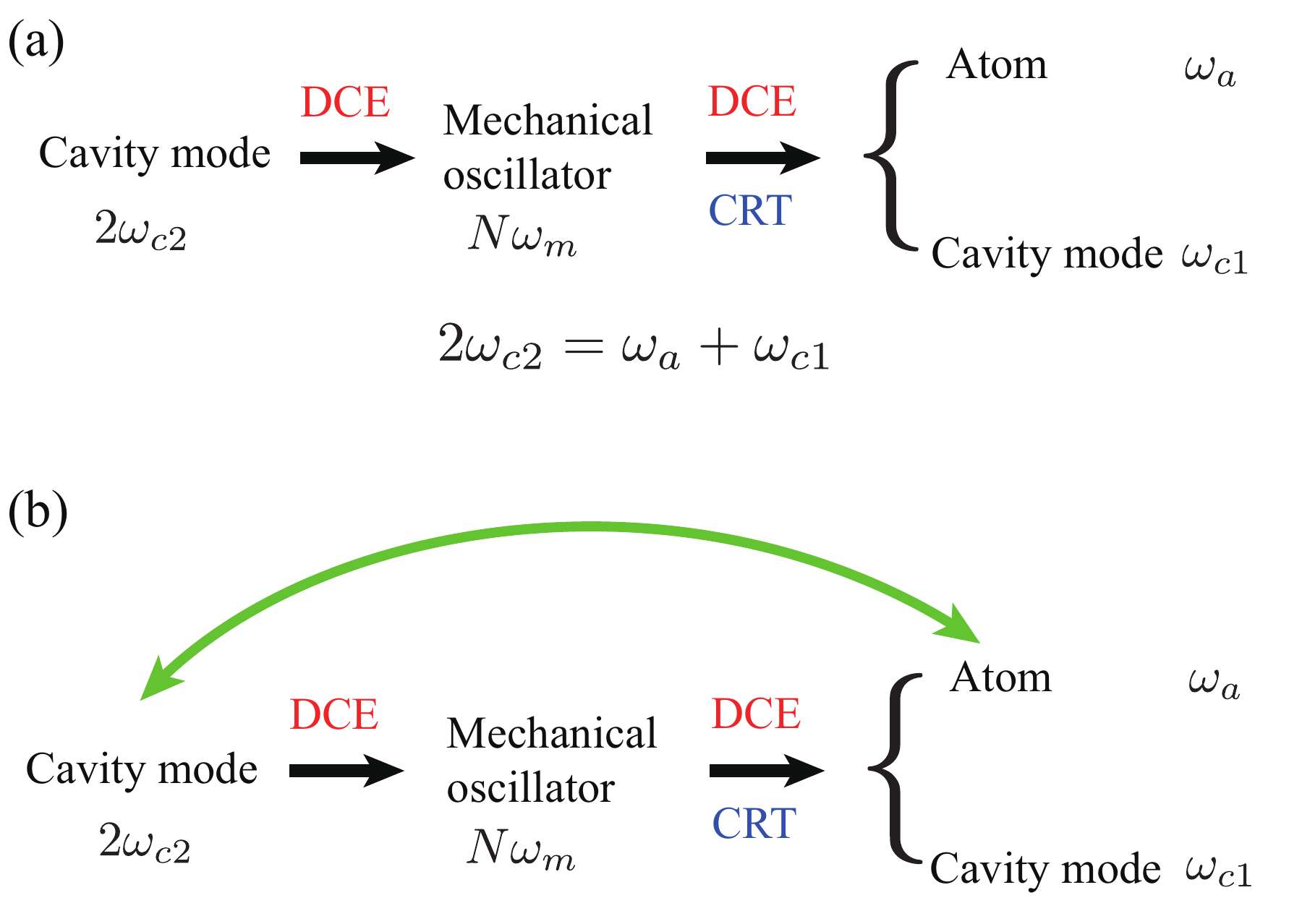}
	\caption{Sketch of the processes of the photon frequency conversion. (a) The path of the energy flow. (b) The paths with the interaction (indicated as the green arrow) between the cavity mode $\omega_{c2}$ and the atom.}
	\label{figu:16}
\end{figure}

Note that, in the above case, the interaction between the cavity mode $\omega_{c2}$ and the atom has not been considered, as shown by Fig.~\ref{figu:16}(a). If this interaction is included [indicated by green arrow in Fig.~\ref{figu:16}(b)], then the unfavorable energy exchange will occur in the photon frequency conversion and then the conversion efficiency will be decreased. To avoid this condition, one can make the dipole moment vector of the atom perpendicular to the field vector of cavity mode $\omega_{c2}$ or keep the atom away from the field distribution of the cavity mode $\omega_{c2}$.

\section{Two-atom interacting with single cavity-optomechanical mode: }\label{C}

\subsection{Analytical approach to the resonance condition $\omega_{a1} + \omega_{a2} = \omega_m$} 

As the description of the two-atom model in Sec.~\ref{IV} of the main text, the system Hamiltonian can be written as
\begin{equation}\label{S42}
	\begin{split}
		\hat H = \,&\,\hbar {\omega _c}{\hat a^\dag }\hat a + \hbar {\omega _m}{\hat b^\dag }\hat b + \hbar {\omega _{a1}}\hat \sigma_{+}^{(1)} {\hat \sigma_{-} ^{(1)}} + \hbar {\hat \omega _{a2}}\hat \sigma_{+}^{(2)} {\hat \sigma_{-}^{(2)}}\\
		\,&\, +\hbar g{\hat a^\dag }\hat a(\hat b + {\hat b^\dag }) + \frac{\hbar }{2}g({\hat a^2} + {\hat a^{\dag 2}})(\hat b + {\hat b^\dag })\\
		\,&\, + \hbar {\lambda _1}(\hat a + {\hat a^\dag })({\hat \sigma_{+}^{(1)}} + \hat \sigma_{-}^{(1)} ) \\
		\,&\, + \hbar {\lambda _2}(\hat a + {\hat a^\dag })({\hat \sigma_{+}^{(2)}} + \hat \sigma_{-}^{(2)} ).
	\end{split}
\end{equation}

\begin{figure}[tpb]
	\centering
	\includegraphics[width = 0.95  \linewidth]{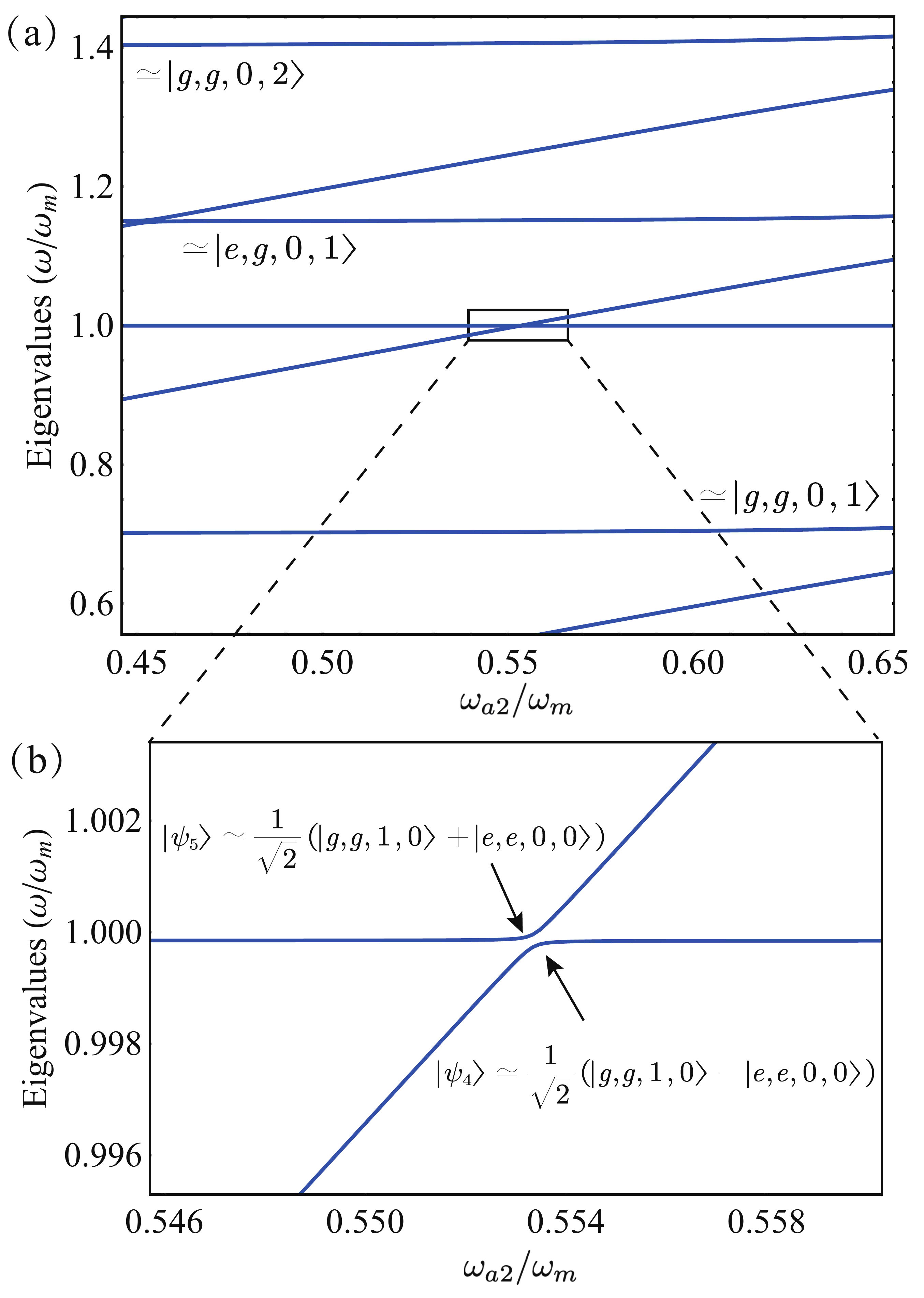}
	\caption{(a) Relevant energy levels of the system Hamiltonian versus ratio between the transition frequency of the atom $\omega_{a2}$ and the frequency of the vibrating mirror $\omega_m$. (b) The enlarged view of the boxed region in panel (a) showing the avoided-level crossing between $|g,g,1,0 \rangle $ and $|e,e,0,0 \rangle $, induced by the DCE and the CRE. The resonance condition in this case is $\omega_{a1} + \omega_{a2} = \omega_m$. Here, the transition frequency of the first atom is $\omega_{a1}=0.45\omega_m$, and the normalized optomechanical coupling rate is $g=0.01\omega_m$}.
	\label{figu:10}	 	
\end{figure}

To derive the effective system Hamiltonian to analyze this model, we can use the method, which is presented in the model of single-atom with single cavity-optomechanical mode. Then, we have 
\begin{equation}\label{S43}
	\begin{split}
		 \hat H_{\rm 0}^U = \,&\,\hbar {\omega _c}{\hat a^\dag }\hat a + \hbar {\omega _m}{\hat b^\dag }\hat b + \hbar {\omega _{a1}}\hat \sigma_{+}^{(1)} {\hat \sigma_{-} ^{(1)}} + \hbar {\omega _{a2}}\hat \sigma_{+}^{(2)} {\hat \sigma_{-}^{(2)}} \\
		\,&\,- \hbar \frac{{{g^2}}}{{{\omega _m}}}{\hat a^\dag }\hat a{\hat a^\dag }\hat a{\rm{ }},
	\end{split}
\end{equation}

\begin{equation}\label{S44}
	\begin{split}
		 \hat H_{\rm {A^{(1)}}FM}^U =\,&\, \hbar \sum\limits_{n = 0}^\infty  {\frac{1}{{n!}}} {\left(\frac{g}{{{\omega _m}}}\right)^n}{\lambda_1} \left[{\hat a^\dag } + {( - 1)^n}\hat a\right]  \\
		\,&\, \times({\hat \sigma_{-}^{(1)}} + \hat \sigma_{+}^{(1)} ){({\hat b^\dag } - \hat b)^n},
	\end{split}
\end{equation}

\begin{equation}\label{S45}
	\begin{split}
		 \hat H_{\rm {A^{(2)}}FM}^U =\,&\, \hbar \sum\limits_{n = 0}^\infty  {\frac{1}{{n!}}} {\left(\frac{g}{{{\omega _m}}}\right)^n}{\lambda_2} \left[{\hat a^\dag } + {( - 1)^n}\hat a\right] \\
		\,&\, \times({\hat \sigma_{-}^{(2)}} + \hat \sigma_{+}^{(2)} ){({\hat b^\dag } - \hat b)^n},
	\end{split}
\end{equation}
and
\begin{equation}\label{S46}
	\begin{split}
		 \hat H_{\rm FM}^U = \,&\,\sum\limits_{n = 0}^\infty  {\frac{1}{{n!}}} \hbar g{\left(\frac{g}{{{\omega _m}}}\right)^n}{2^{n - 1}}\left\{\left[{\hat a^{\dag 2}} + {( - 1)^n}{\hat a^2}\right] \right. \\
		\,&\,\left. \times{({\hat b^\dag } - \hat b)^n}({\hat b^\dag } + \hat b)- n\left[{\hat a^{\dag 2}} + {( - 1)^{n - 1}}{\hat a^2}\right] \right.\\
		\,&\,\left. \times{\hat a^\dag }\hat a{({\hat b^\dag } - \hat b)^{n - 1}}\right\}.
	\end{split}
\end{equation}
In this model, the resonant condition is $\omega_{a1} + \omega_{a2} = k\omega_m$, where $k$ is an integer. This formula implies the exchange of virtual excitations between the atoms and a mirror. For the unperturbed Hamiltonian $\hat H_0^U$, the eigenstate is described as $|j_1,j_2,k,n\rangle$, where $\vert j_1\rangle$ and $\vert j_2\rangle$ ($j=g, e$) denote the states of the atom $\omega_{a1}$ and $\omega_{a2}$, respectively, with eigenvalue 
\begin{equation}\label{S46-1}
	\begin{split}
	E_{j_1.j_2,k,n} =\,&\,  \hbar \omega_cn + \hbar \omega_mk + \hbar \omega_a\langle j_1|e\rangle+ \hbar \omega_a\langle j_2|e\rangle \\
	\,&\, -\hbar g^2n^2/\omega_m.
	\end{split}
\end{equation}

Here, we assume that the two atoms are nonidentical, the mode frequency of the cavity mode is ${\omega _c} = 0.7{\omega _m}$, and that the normalized atom-cavity coupling rates are $\lambda_1 = 0.014\omega_m$ and $\lambda _2 = 0.022\omega _m$. 

\begin{figure}[tpb]
	\centering
	\includegraphics[width = 0.98  \linewidth]{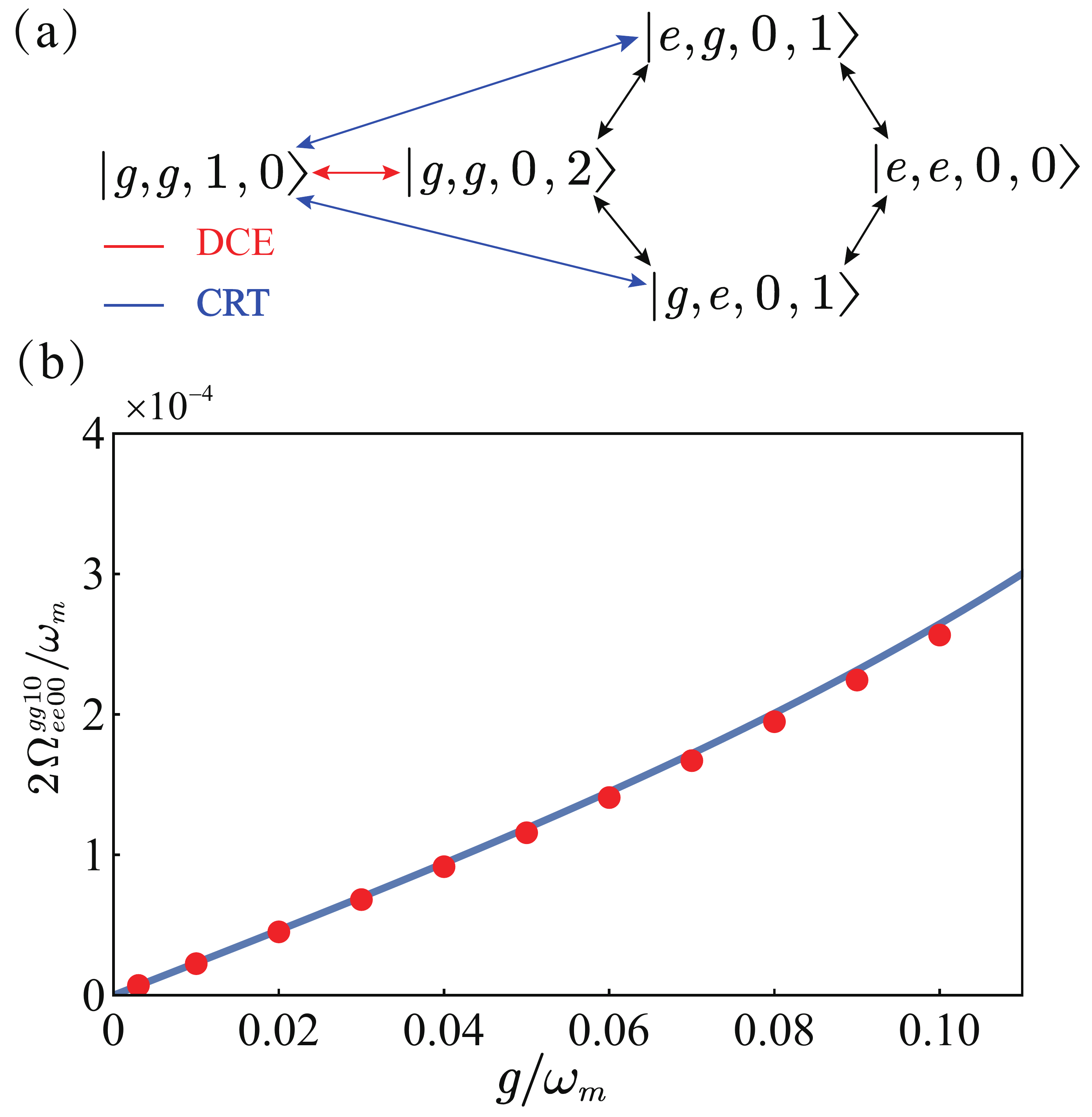}
	\caption{(a) Sketch of the processes giving the main contribution to the effective coupling between the states $|g,g,1,0 \rangle $ and $|e,e,0,0 \rangle $, via the DCE (red arrows) and the counterrotating terms (blue arrows). (b) Comparison between the numerical (red dots) and the analytical (blue solid curve) results of the normalized level splitting.}
	\label{figu:11}	 	
\end{figure}

Figure~\ref{figu:10} displays the lowest energy levels (${E_j} - {E_0}$) of the system Hamiltonian versus the ratio between the transition frequency of the atom $\omega_{a2}$ and the mechanical frequency of the mirror $\omega_m$. The avoided-level crossing in Fig.~\ref{figu:10}(b) originates from the coherent coupling between the states $|g,g,1,0 \rangle $ and $|e,e,0,0 \rangle $. At the minimum energy splitting, the resulting states are well approximated by $|{\varphi _{4,5}} \rangle  \simeq \frac{1}{{\sqrt 2 }}(|g,g,1,0 \rangle  \mp |e,e,0,0 \rangle )$. Obviously, this splitting occurs in a resonant condition $\omega_{a1} + \omega_{a2} = \omega_m$. 

In Fig.~\ref{figu:11}(a), the transition paths for a coupling between states $|g,g,1,0 \rangle $ and $|e,e,0,0 \rangle $  are dominated by the DCE and the counterrotating terms. If $|g,g,1,0 \rangle $ is the initial state of the system, then it reaches the final state $|e,e,0,0 \rangle $ through the virtual transition involving the out-of-resonance intermediate states $|g,g,0,2 \rangle $ or $|e,g,0,1 \rangle $ ($|g,e,0,1 \rangle $). Such a transition process is a detuned resonance transition. This means that when one system satisfies the resonance condition $\omega_{a1} + \omega_{a2} = k\omega_m$ even if $\omega_{a1} \neq \omega_{a2}$, a single vibrating excitation of the mechanical oscillator can be absorbed by the two atoms. Based on the above effective system Hamiltonian, the effective coupling rate can be obtained by using perturbation theory,
\begin{equation}\label{S47}
	\begin{split}
		\Omega _{ee00}^{gg10} =\,&\, \frac{{\frac{g\lambda_1 \lambda_2 }{\omega_m}\left({1 -\frac{ g^2}{2\omega_m^2}}\right)\left(-1 + \frac{{{g^2}}}{{2\omega _m^2}}\right)}}{{\left({\omega _m} - {\omega _c} - {\omega _{a1}} + \frac{{{g^2}}}{{{\omega _m}}}\right)}} \\
		\,&\, +\frac{{\frac{g\lambda_1 \lambda_2 }{\omega_m}\left({1 -\frac{ g^2}{2\omega_m^2}}\right)\left(-1 + \frac{{{g^2}}}{{2\omega _m^2}}\right)}}{{\left({\omega _m} - {\omega _c} - {\omega _{a2}} + \frac{{{g^2}}}{{{\omega _m}}}\right)}} \\
		\,&\, +\frac{{2g\lambda_1 \lambda_2 \left({1 -\frac{g^2}{2\omega_m^2}}\right)^{2}\left(\frac{1}{2}+\frac{g^2}{\omega_m^2}\right)}}{{\left({\omega _m} - {\omega _c} - {\omega _{a1}} + \frac{{{g^2}}}{{{\omega _m}}}\right)\left({\omega _m} - {2\omega _c}+ \frac{{{4g^2}}}{{{\omega _m}}}\right)}} \\
		\,&\, +\frac{{2g\lambda_1 \lambda_2 \left({1 -\frac{g^2}{2\omega_m^2}}\right)^{2}\left(\frac{1}{2}+\frac{g^2}{\omega_m^2}\right)}}{{\left({\omega _m} - {\omega _c} - {\omega _{a2}} + \frac{{{g^2}}}{{{\omega _m}}}\right)\left({\omega _m} - {2\omega _c}+ \frac{{{4g^2}}}{{{\omega _m}}}\right)}}. \\
	\end{split}
\end{equation}

In Fig.~\ref{figu:11}(b), it is in good agreement between the numerical and the analytical results of the normalized splitting ($2\Omega _{ee01}^{gg10}\omega_m$) versus the normalized optomechanical coupling ($g/{\omega _m}$). This demonstrates that two atoms can couple to the mirror via a complete exchange of virtual excitations.

\subsection{System dynamics for two nonidentical atoms with different coupling rates}

\begin{figure}[tpb]
	\centering
	\includegraphics[width = 1  \linewidth]{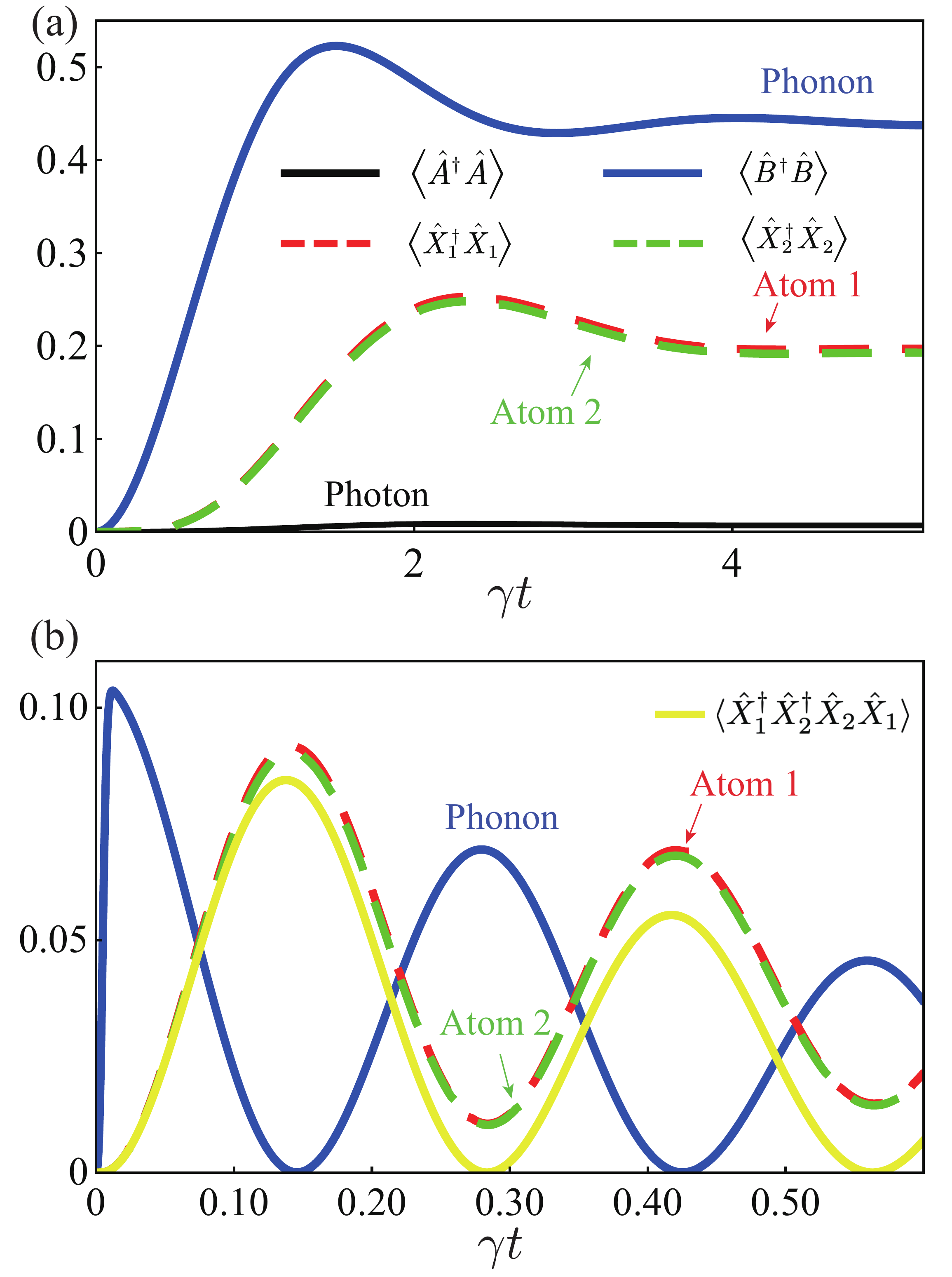}
	\caption{The dynamical evolution of the hybrid quantum system, (a) in the weak-coupling regime (loss rate $\eta_1 = \eta_2 = \kappa = \gamma = 1.3 \times 10^{-4}\omega_m$) with a single-tone mechanical drive $F(t)=A\cos(\omega_d t)$ (a weak resonance excitation $A = 2\gamma $ and $\omega_d = \omega_m$), and (b) in the strong-coupling regime ( $\kappa = \gamma = \eta_1 = \eta_2 = 1 \times 10^{-5}\omega_m$) with an ultrafast resonant pulse $F(t)=A G(t-t_0)\cos(\omega_d t)$ applied on the movable mirror (standard deviation $\sigma = (30\Omega_{ee00}^{gg10})^{-1}$, amplitude $A = 0.25\pi$, and central driving frequency $\omega_d = \omega_m$).  The system is initially prepared at the ground state.  }
	\label{figu:14}
\end{figure}

As in the previous discussion, from the resonance condition $\omega_{a1} + \omega_{a2} = k\omega_m$, we know that the two atoms do not need to be identical. In this subsection, we study the system dynamics in a more general case, where the two atoms are nonidentical with different coupling rates $\lambda_i$.

Figure~\ref{figu:14} displays the system dynamics [corresponding to the minimum level splitting shown in Fig.~\ref{figu:10}(b)] of the mean photon number $\langle \hat A^\dag \hat A\rangle $, the mean phonon number $\langle \hat B^\dag \hat B\rangle$, the mean atom-excitation number $\langle \hat X_1^\dag \hat X_1\rangle$, and $\langle \hat X_2^\dag \hat X_2\rangle$. These are done in the weak-coupling regime with the continuous-wave mechanical drive and the strong-coupling regime with the ultrafast resonant mechanical pulse. Figure~\ref{figu:14} shows that, at the beginning of the dynamics, the energy of two atoms rises simultaneously after the mechanical oscillator is driven. This phenomenon is similar to the case shown in Fig.~\ref{figu:6}, where the two atoms and corresponding coupling rates $\lambda$ are all identical. The difference is that, after a short time, the excitations of the two atoms have not coincided with each other in Fig.~\ref{figu:14}. This can be attributed to the differences of proportion of the transition paths [Fig.~\ref{figu:11}(a)], which are induced by the differences between the coupling rates of the cavity-atom.

In addition, we observe that the single-atom excitation $\langle \hat X_i^\dag \hat X_i\rangle$ is close to the two-atom correlation $G_a^{(2)}=\langle X_1^{\dag} X_2^{\dag} X_2 X_1 \rangle$. By excluding the effect of system decay, we have also calculated $G_a^{(2)}$ and found that  $\langle \hat X_i^\dag \hat X_i\rangle$ and $G_a^{(2)}=\langle X_1^{\dag} X_2^{\dag} X_2 X_1 \rangle$ almost coincide over time. This result is a clear signature of joint excitation: if one atom becomes excited, the probability of the other atom being excited is very close to one.

\section{Additional discussions} \label{D}

Very recently, we noticed a work considering the case when the mechanical oscillator can excite an atom through a cavity-vacuum field~\cite{2021yin}. The cavity-atom part of this work is based on the single-photon Rabi model and the two-photon Rabi model. Reference \cite{2021yin} gives the analytical solution of the two-photon Rabi model by using the method of Ref~\cite{2018Macri}. However it neglects the transition processes between the different phonon states of one mechanical oscillator, which is essential in this method. In the single-photon Rabi model, Ref.~\cite{2021yin} only presents the numerical results. Actually, the method of Ref.~\cite{2018Macri} is not suitable for the single-photon Rabi model because the energy transfer between an atom and a mechanical oscillator is at the single-photon level. In this section, we will compare the analytical methods used in our works and in Ref.~\cite{2018Macri} for studying the coupling between an atom and a mechanical oscillator, with the single-photon Rabi model or two-photon Rabi model.

\subsection{Single-photon Rabi model} 

\begin{figure}[tpb]
	\centering
	\includegraphics[width = 0.98  \linewidth]{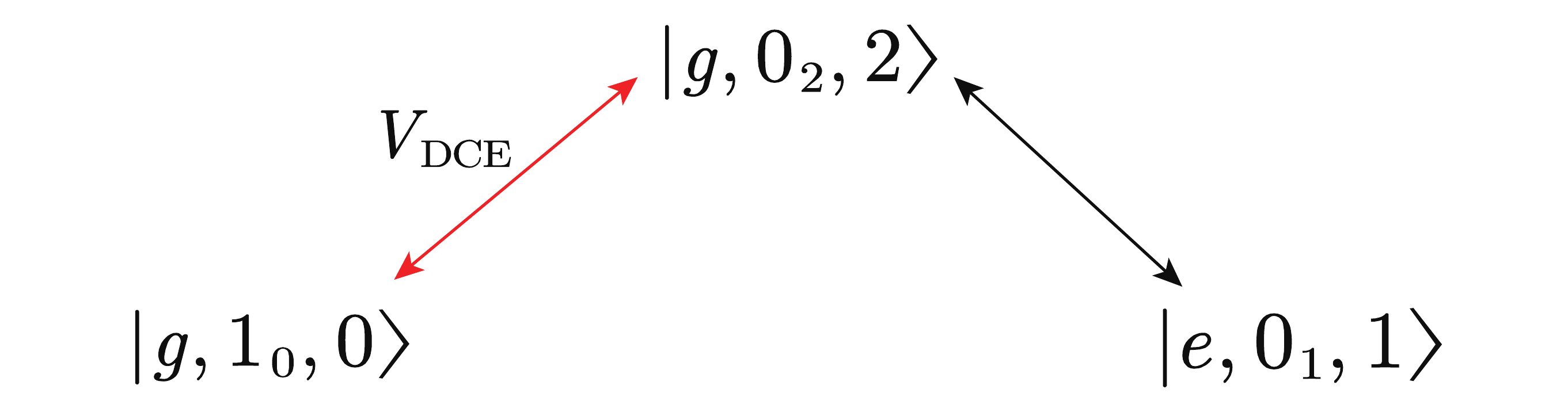}
	\caption{Sketch of the processes giving the main contribution to the  effective coupling between the states $|g,1_0,0 \rangle $ and $|e,0_1,1 \rangle $, via the DCE (red arrows), using the method of Ref.~\cite{2018Macri}.}
	\label{figu:12}	 	
\end{figure}

Using the method presented in Ref.~\cite{2018Macri} for the case of single atom interacting with a single cavity-optomechanical mode, $\hat{H}_0+\hat{V}_{\rm om}$ (details in Sec.~\ref{II}) is considered as an unperturbed Hamiltonian and can be analytically diagonalized by defining the displacement operator  $D(n\beta ) = \exp [n\beta (b - {b^\dag })]$ with $\beta  = g/{\omega _m}$. Therefore, for the unperturbed Hamiltonian $\hat{H}_0+\hat{V}_{\rm om}$, the system eigenstates can be written as 
\begin{equation}\label{S48}
	\begin{split}
		|j,k_n,n \rangle  = |j \rangle  \otimes D(n\beta) |k \rangle  \otimes |n \rangle,
	\end{split}
\end{equation}
where $|k\rangle$ ($|n\rangle$) denotes the Fock state of the mechanical (cavity) mode and $\vert j\rangle$ ($j=g, e$) denotes the atom states, with the eigenvalue 
\begin{equation}\label{S49}
	{E_{j,k,n}} = \hbar {\omega _c}n + \hbar {\omega _m}k + \hbar {\omega _a}\langle j|e\rangle-\hbar {{\rm{g}}^2}{n^2}/{\omega _m}.
\end{equation}
The perturbed Hamiltonian is described by
\begin{equation}\label{S50}
	\begin{split}
		\hat{H}_{\rm p}=\dfrac{\hbar g}{2}(\hat{a}^2+\hat{a}^{\dag 2})(\hat{b}+\hat{b}^{\dag})+\hbar \lambda(\hat{a}+\hat{a}^{\dag})(\hat{\sigma}_{-}+\hat{\sigma}_{+}).
	\end{split}
\end{equation}

According to this analytical model, the coupling between states $|g,1_0,0\rangle$ and $|e,0_1,1\rangle$ can only be analyzed by one transition path, as shown in Fig.~\ref{figu:12}, which is dominated by the DCE. The effective coupling rate can be written as,
\begin{equation}\label{S50_2}
	\begin{split}
		\Omega_p  = \,&\, \frac{\langle e,0_1,1|\hat H_{\rm p}|g,0_2,2 \rangle\langle g,0_2,2|\hat H_{\rm p}|g,1_0,0\rangle}{E_{g,1,0} - E_{g,0,2}} \\
		=\,&\, \frac{\sqrt{2}\lambda D_{0,0}(\beta)[\frac{\sqrt{2}}{2}g D_{0,0}(-2\beta)+g D_{0,2}(-2\beta)]}{\omega_m-2\omega_c+\frac{4g^2}{\omega_m}},
	\end{split} 
\end{equation}
where the matrix elements of the displacement operators can be expressed in terms of associated Laguerre polynomials:
\begin{equation}\label{S50_3}
	D_{k',k}(\alpha)=\sqrt{k!/k'!}\alpha^{k'-k}e^{-|\alpha|^2}L^{k'-k}_k (|\alpha|^2).
\end{equation}

Actually, the counterrotating terms in the cavity-atom system Hamiltonian can also contribute to the coupling if it can combine with the annihilation of one vibrating excitation of the mechanical oscillator, which is involved in $V_{\rm om}$. However, $V_{\rm om}$ is not in the perturbed Hamiltonian. 

Performing a unitary transformation to eliminate $V_{\rm om}$ can better deal with this problem, because the influence of this term $V_{\rm om}$ on both $H_0$ and $H_{\rm p}$ can then be revealed (details in Appendix~\ref{A}). By defining a unitary operator $U = \exp [ - \beta {a^\dag }a({b^\dag } - b)]$, the eigenvalue equation can be described as
\begin{equation}\label{S50_1}
	\begin{split}
		E_{j,k,n}\,&\,=\langle j,k_n,n|UU^\dag(H_0+V_{\rm om})UU^\dag|j,k_n,n \rangle \\
		\,&\,=\langle j,k,n|U^\dag(H_0+V_{\rm om})U|j,k,n \rangle,
	\end{split}
\end{equation}
Then, for the unperturbed Hamiltonian $U^\dag(H_0+V_{\rm om})U$, the eigenstate is described by $|j,k,n \rangle$. In this method, the coupling between states $|g,1,0\rangle$ and $|e,0,1\rangle$ can be analyzed by two transition paths, which is dominated by the DCE and the counterrotating terms independently, as shown in Fig.~\ref{figu:2}(a).

\begin{figure}[tpb]
	\centering
	\includegraphics[width = 0.98 \linewidth]{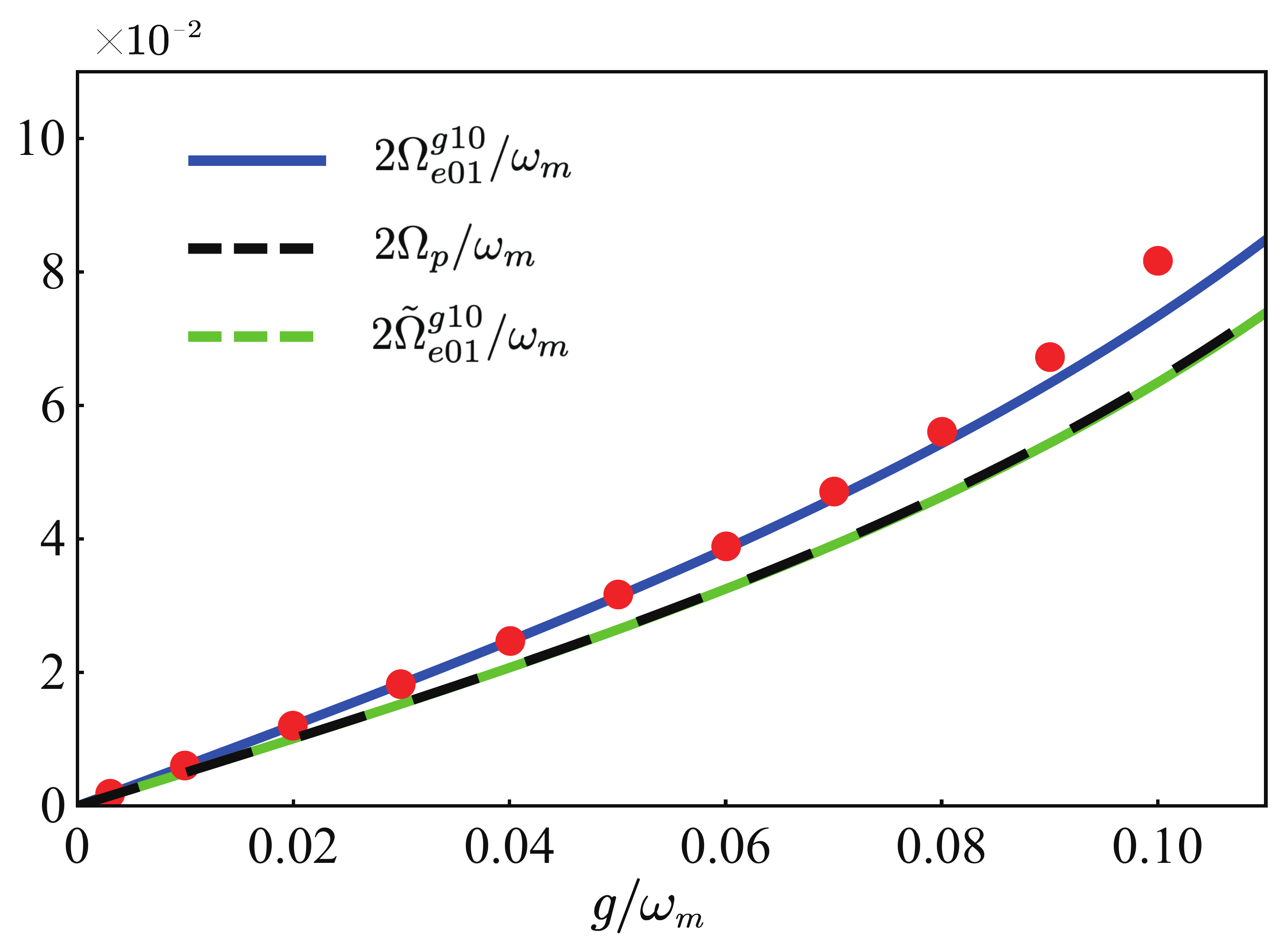}
	\caption{Comparison between the numerical calculated normalized level splitting (red dots) and the corresponding analytical calculations [2$\Omega_{e01}^{g10}/\omega_m$ (blue curve), 2$\Omega_p/\omega_m$ (black dotted curve), 2$\tilde{\Omega}_{e01}^{g10}/\omega_m$ (green dotted curve)], obtained using perturbation theory. These results are obtained by using the parameters of Fig.~\ref{figu:2}(b) }
	\label{figu:15}
\end{figure}

Figure~\ref{figu:15} shows the comparison between the numerical calculated normalized level splitting and the three kinds of analytical calculations, where $\Omega_{e01}^{g10}$ and $\Omega_p$ are respectively described by Eq.~\eqref{eq8} and Eq.~\eqref{S50_2}. We have also presented the effective coupling rate, 
\begin{equation}\label{S50_3}
	\begin{split}
		\tilde{\Omega}_{e01}^{g10}  = \,&\, \frac{\langle e,0,1|\hat H_{\rm I}^U|g,0,2 \rangle\langle g,0,2|\hat H_{\rm I}^U|g,1,0\rangle}{E_{g,1,0} - E_{g,0,2}}\\  
		=\,&\, \frac{\lambda g \left(\sqrt{2}-\frac{\sqrt{2}g^2}{2\omega_m^2}\right) \left(\frac{\sqrt{2}}{2} + \frac{\sqrt{2}g^2}{\omega_m^2}\right)}{\omega_m - 2\omega_c + 4\frac{g^2}{\omega_m}},
	\end{split} 
\end{equation}
using the method of Sec.~\ref{II}, without calculating the transition path of the counterrotating term [Fig.~\ref{figu:2}(a)]. The results show that the effect of the counterrotating term has an important contribution to the coupling between the atom and the vibrating mirror by comparing $\Omega_{e01}^{g10}$ with $\Omega_p$ ($\tilde{\Omega}_{e01}^{g10}$). In addition, the two curves $\Omega_p$ and $\tilde{\Omega}_{e01}^{g10}$ almost coincide at the range $g/\omega_m \leq 0.1$, illustrating that the method used in Sec.~\ref{II} can also exactly describe the DCE effect.

\subsection{Two-photon Rabi model} 

Reference~\cite{2021yin} utilizes the method of Ref.~\cite{2018Macri} to analyze its model. Specifically, the unperturbed Hamiltonian is $H_{0}+V_{\rm om}$, while the perturbed one is written as
\begin{equation}\label{S51}
	\begin{split}
		\hat{H}_{\rm p} =\dfrac{\hbar g}{2}(\hat{a}^2+\hat{a}^{\dag 2})(\hat{b}+\hat{b}^{\dag})+\hbar \lambda(\hat{a}^2+\hat{a}^{\dag 2})(\hat{\sigma}_{-}+\hat{\sigma}_{+}).
	\end{split}
\end{equation}

\begin{figure}[tpb]
	\centering
	\includegraphics[width = 0.98  \linewidth]{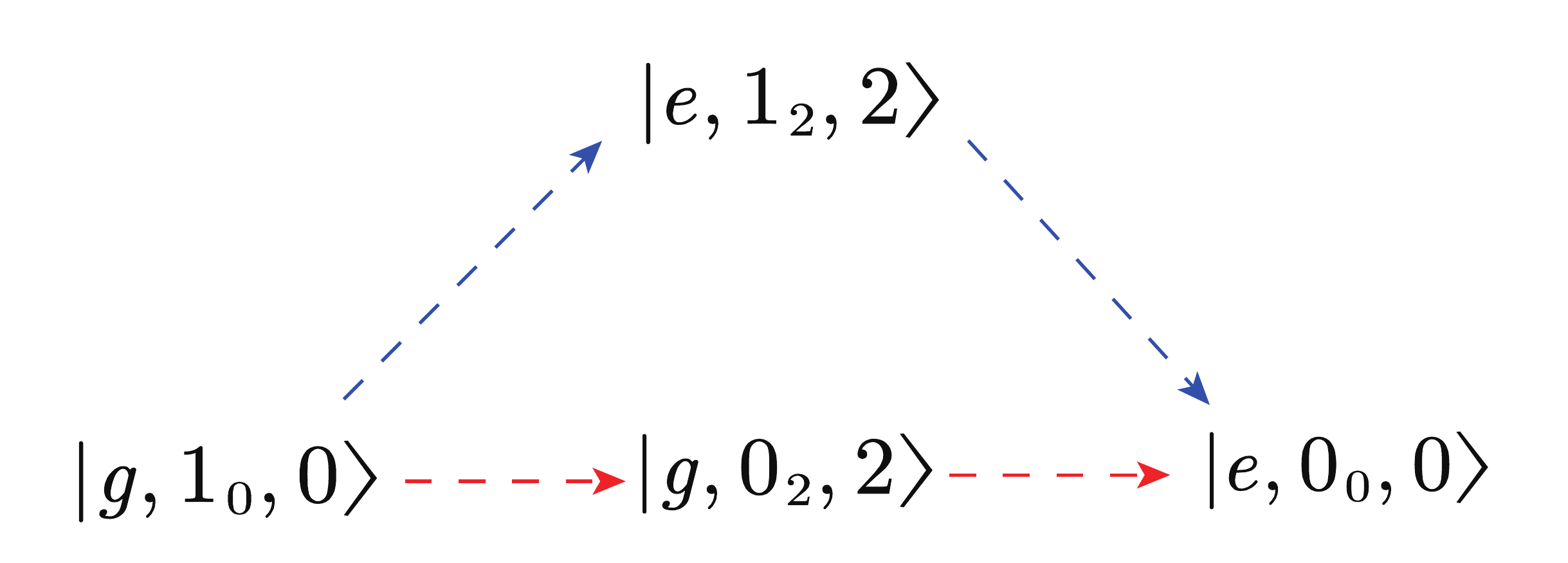}
	\caption{Sketch of the processes giving the main contribution to the  effective coupling between the states $|g,1_0,0 \rangle $ and $|e,0_0,0 \rangle $~\cite{2021yin}, using the method of Ref.~\cite{2018Macri}. }
	\label{figu:13}	 	
\end{figure}

The system eigenstates are described by $|j,k_n n\rangle  = |j \rangle  \otimes D(n\beta) |k \rangle  \otimes |n \rangle$, where $|k\rangle$ ($|n\rangle$) denotes the Fock state of the mechanical (cavity) mode and $\vert j\rangle$ ($j=g, e$) denotes the atom states. With the two-photon Rabi model, the energy transfer between an atom and a mechanical oscillator occurs in the form of a two-photon process, as shown in Fig.~\ref{figu:13}. This is a special case where the form of the energy transfer matches the creation of photon pairs induced by the DCE [or the effect of counterrotating terms of cavity-atom system Hamiltonian in Eq.~\eqref{S51}]. Thus, the method of Ref.~\cite{2018Macri} can analyze this case. However, in Ref.~\cite{2021yin}, all these states shown in Fig.~\ref{figu:13} are treated as the bare states $|j,k,n\rangle$. It means that Ref.~\cite{2021yin} neglects the transition processes between the phonon states within a $n$-photon subspace, which are indicated by the matrix elements of the displacement operator, $\langle k_{0}^{'}|D(n\beta)|k_0\rangle=D_{k^{'},k}(n\beta)$. Thus, the analytical solution for the effective coupling rate in Ref.~\cite{2021yin} is obtained by incorrectly using the method in Ref.~\cite{2018Macri}. Actually, the effective coupling rate should be written as 
\begin{equation}\label{S52}
	\begin{split}
		{\Omega_{\rm eff}} =\,&\,\dfrac{\biggl[g\lambda D_{0,0}(2\beta)-2g\lambda \beta D_{0,1}(2\beta)\biggr]D_{1,1}(-2\beta)}{-\omega_a-2\omega_c+4g^2/\omega_m} \\
		\,&\, +\dfrac{\lambda g D_{0,0}(2\beta)D_{0,0}(-2\beta)}{\omega_m-2\omega_c+4g^2/\omega_m}.
	\end{split}
\end{equation}

Note that Ref.~\cite{2021yin} utilizes the two photons produced by the counterrotating term of the two-photon Rabi Hamiltonian in Eq.~\eqref{S51} to transfer energy between the mirror and the atom, as indicated by the transition path (blue arrows) in Fig.~\ref{figu:13}. In our works, the atom and photon are treated as a whole, which interacts with the mirror. This is a significant difference between this work and Ref.~\cite{2021yin}. Moreover, because the transition path (blue arrows) in Fig.~\ref{figu:13} also utilizes such a counterrotating term $a^2 b$ in Eq.~\eqref{S51}, the contribution of this path to the effective coupling rate $\Omega_{\rm eff}$ is too small.

Importantly, the two-photon Rabi model can also be analyzed in our method. We can eliminate the term $\hat V_{\rm om}$ by performing a unitary transformation with the unitary operator $\hat U =\exp[-\beta \hat a^\dag \hat a(\hat b^\dag - \hat b)]$. Then, we can obtain the effective system Hamiltonian:
\begin{equation}\label{S53}
	\hat H_{0}^U =  {\hat H_0} - \hbar \frac{{{g^2}}}{{{\omega _m}}}{\hat a^\dag }\hat a{\hat a^\dag }\hat a{\rm{  }},
\end{equation}
\begin{equation}\label{S54}
	\begin{split}
		 \hat H_{\rm AF_{(2)}M}^U =\,&\, {\hat U^\dag }{\left[\hbar \lambda(\hat a^2+\hat a^{\dag 2})(\hat \sigma_{-}+\hat \sigma_{+})\right]}\hat U \\
		=\,&\, \hbar \sum\limits_{n = 0}^\infty  {\frac{1}{{n!}}} \lambda {\left(\frac{g}{{\omega _m}}\right)^n} 2^{n} \left[{\hat a^{\dag 2}} + {( - 1)^n}\hat a^2\right]\\
		\,&\, \times(\hat \sigma_{-}  + {\hat \sigma_{+} }){({\hat b^\dag } - \hat b)^n},
	\end{split}
\end{equation}
and
\begin{equation}\label{S55}
	\begin{split}
		 \hat H_{F_{(2)}M}^U = \,&\,\sum\limits_{n = 0}^\infty  {\frac{1}{{n!}}} \hbar g{\left(\frac{g}{{{\omega _m}}}\right)^n}{2^{n - 1}}\left\{\left[{\hat a^{\dag 2}} + {( - 1)^n}{\hat a^2}\right] \right.\\
		\,&\, \left. \times{({\hat b^\dag } - \hat b)^n}({\hat b^\dag } + \hat b)- n\left[{\hat a^{\dag 2}} + {( - 1)^{n - 1}}{\hat a^2}\right]\right.\\
		\,&\, \left.\times{\hat a^\dag }\hat a{({\hat b^\dag } - \hat b)^{n - 1}}\right\}.
	\end{split}
\end{equation}
For the unperturbed Hamiltonian $\hat H_0^U$, the eigenstates are described by $|j,k,n \rangle  = |j \rangle  \otimes |k \rangle  \otimes |n \rangle $, with the eigenvalue 
\begin{equation}\label{S56}
	{E_{j,k,n}} = \hbar {\omega _c}n + \hbar {\omega _m}k + \hbar {\omega _a}\langle j|e\rangle-\hbar {{\rm{g}}^2}{n^2}/{\omega _m}.
\end{equation}
In this analytical method, the transition processes indicated by $D_{k^{'},k}(n\beta)$ can be directly exhibited in $\hat H_{\rm AF_{(2)}M}^U$ and $\hat H_{F_{(2)}M}^U$. Then, the effective coupling rate in our method can be written as 
\begin{equation}\label{S57}
	\Omega_{\rm eff}=\dfrac{\lambda g-6\frac{\lambda g^3}{\omega_m^2}}{-\omega_a-2\omega_c+\frac{4g^2}{\omega_m}}+\frac{\lambda g+\frac{2\lambda g^3}{\omega_m^2}}{\omega_m-2\omega_c+\frac{4g^2}{\omega_m}}.
\end{equation}

In summary, the analytical method used in this work is more suitable for studying the coupling between the atoms and a mechanical oscillator induced by quantum vacuum fluctuations. 

\end{appendix}
\bibliography{FIPAref.bib}

\begin{thebibliography}{81}%
\makeatletter
\providecommand \@ifxundefined [1]{%
 \@ifx{#1\undefined}
}%
\providecommand \@ifnum [1]{%
 \ifnum #1\expandafter \@firstoftwo
 \else \expandafter \@secondoftwo
 \fi
}%
\providecommand \@ifx [1]{%
 \ifx #1\expandafter \@firstoftwo
 \else \expandafter \@secondoftwo
 \fi
}%
\providecommand \natexlab [1]{#1}%
\providecommand \enquote  [1]{``#1''}%
\providecommand \bibnamefont  [1]{#1}%
\providecommand \bibfnamefont [1]{#1}%
\providecommand \citenamefont [1]{#1}%
\providecommand \href@noop [0]{\@secondoftwo}%
\providecommand \href [0]{\begingroup \@sanitize@url \@href}%
\providecommand \@href[1]{\@@startlink{#1}\@@href}%
\providecommand \@@href[1]{\endgroup#1\@@endlink}%
\providecommand \@sanitize@url [0]{\catcode `\\12\catcode `\$12\catcode
  `\&12\catcode `\#12\catcode `\^12\catcode `\_12\catcode `\%12\relax}%
\providecommand \@@startlink[1]{}%
\providecommand \@@endlink[0]{}%
\providecommand \url  [0]{\begingroup\@sanitize@url \@url }%
\providecommand \@url [1]{\endgroup\@href {#1}{\urlprefix }}%
\providecommand \urlprefix  [0]{URL }%
\providecommand \Eprint [0]{\href }%
\providecommand \doibase [0]{https://doi.org/}%
\providecommand \selectlanguage [0]{\@gobble}%
\providecommand \bibinfo  [0]{\@secondoftwo}%
\providecommand \bibfield  [0]{\@secondoftwo}%
\providecommand \translation [1]{[#1]}%
\providecommand \BibitemOpen [0]{}%
\providecommand \bibitemStop [0]{}%
\providecommand \bibitemNoStop [0]{.\EOS\space}%
\providecommand \EOS [0]{\spacefactor3000\relax}%
\providecommand \BibitemShut  [1]{\csname bibitem#1\endcsname}%
\let\auto@bib@innerbib\@empty
\bibitem [{\citenamefont {Xiang}\ \emph {et~al.}(2013)\citenamefont {Xiang},
  \citenamefont {Ashhab}, \citenamefont {You},\ and\ \citenamefont
  {Nori}}]{2013Xiang}%
  \BibitemOpen
  \bibfield  {author} {\bibinfo {author} {\bibfnamefont {Z.-L.}\ \bibnamefont
  {Xiang}}, \bibinfo {author} {\bibfnamefont {S.}~\bibnamefont {Ashhab}},
  \bibinfo {author} {\bibfnamefont {J.Q.}\ \bibnamefont {You}},\ and\ \bibinfo
  {author} {\bibfnamefont {F.}~\bibnamefont {Nori}},\ }\bibfield  {title}
  {\bibinfo {title} {Hybrid quantum circuits: Superconducting circuits
  interacting with other quantum systems},\ }\href
  {https://doi.org/10.1103/RevModPhys.85.623} {\bibfield  {journal} {\bibinfo
  {journal} {Rev. Mod. Phys.}\ }\textbf {\bibinfo {volume} {85}},\ \bibinfo
  {pages} {623} (\bibinfo {year} {2013})}\BibitemShut {NoStop}%
\bibitem [{\citenamefont {Kurizki}\ \emph {et~al.}(2015)\citenamefont
  {Kurizki}, \citenamefont {Bertet}, \citenamefont {Kubo}, \citenamefont
  {M{\o}lmer}, \citenamefont {Petrosyan}, \citenamefont {Rabl},\ and\
  \citenamefont {Schmiedmayer}}]{2015kurizki}%
  \BibitemOpen
  \bibfield  {author} {\bibinfo {author} {\bibfnamefont {G.}~\bibnamefont
  {Kurizki}}, \bibinfo {author} {\bibfnamefont {P.}~\bibnamefont {Bertet}},
  \bibinfo {author} {\bibfnamefont {Y.}~\bibnamefont {Kubo}}, \bibinfo {author}
  {\bibfnamefont {K.}~\bibnamefont {M{\o}lmer}}, \bibinfo {author}
  {\bibfnamefont {D.}~\bibnamefont {Petrosyan}}, \bibinfo {author}
  {\bibfnamefont {P.}~\bibnamefont {Rabl}},\ and\ \bibinfo {author}
  {\bibfnamefont {J.}~\bibnamefont {Schmiedmayer}},\ }\bibfield  {title}
  {\bibinfo {title} {Quantum technologies with hybrid systems},\ }\href
  {https://doi.org/10.1073/pnas.1419326112} {\bibfield  {journal} {\bibinfo
  {journal} {Proceedings of the National Academy of Sciences}\ }\textbf
  {\bibinfo {volume} {112}},\ \bibinfo {pages} {3866} (\bibinfo {year}
  {2015})}\BibitemShut {NoStop}%
\bibitem [{\citenamefont {Leibfried}\ \emph
  {et~al.}(2003{\natexlab{a}})\citenamefont {Leibfried}, \citenamefont {Blatt},
  \citenamefont {Monroe},\ and\ \citenamefont {Wineland}}]{2003Leibfried}%
  \BibitemOpen
  \bibfield  {author} {\bibinfo {author} {\bibfnamefont {D.}~\bibnamefont
  {Leibfried}}, \bibinfo {author} {\bibfnamefont {R.}~\bibnamefont {Blatt}},
  \bibinfo {author} {\bibfnamefont {C.}~\bibnamefont {Monroe}},\ and\ \bibinfo
  {author} {\bibfnamefont {D.}~\bibnamefont {Wineland}},\ }\bibfield  {title}
  {\bibinfo {title} {Quantum dynamics of single trapped ions},\ }\href
  {https://doi.org/10.1103/RevModPhys.75.281} {\bibfield  {journal} {\bibinfo
  {journal} {Rev. Mod. Phys.}\ }\textbf {\bibinfo {volume} {75}},\ \bibinfo
  {pages} {281} (\bibinfo {year} {2003}{\natexlab{a}})}\BibitemShut {NoStop}%
\bibitem [{\citenamefont {Pirkkalainen}\ \emph {et~al.}(2013)\citenamefont
  {Pirkkalainen}, \citenamefont {Cho}, \citenamefont {Li}, \citenamefont
  {Paraoanu}, \citenamefont {Hakonen},\ and\ \citenamefont
  {Sillanp{\"a}{\"a}}}]{2013Pirkkalainen}%
  \BibitemOpen
  \bibfield  {author} {\bibinfo {author} {\bibfnamefont {J.-M.}\ \bibnamefont
  {Pirkkalainen}}, \bibinfo {author} {\bibfnamefont {S.}~\bibnamefont {Cho}},
  \bibinfo {author} {\bibfnamefont {J.}~\bibnamefont {Li}}, \bibinfo {author}
  {\bibfnamefont {G.}~\bibnamefont {Paraoanu}}, \bibinfo {author}
  {\bibfnamefont {P.}~\bibnamefont {Hakonen}},\ and\ \bibinfo {author}
  {\bibfnamefont {M.}~\bibnamefont {Sillanp{\"a}{\"a}}},\ }\bibfield  {title}
  {\bibinfo {title} {Hybrid circuit cavity quantum electrodynamics with a
  micromechanical resonator},\ }\href {https://doi.org/10.1038/nature11821}
  {\bibfield  {journal} {\bibinfo  {journal} {Nature}\ }\textbf {\bibinfo
  {volume} {494}},\ \bibinfo {pages} {211} (\bibinfo {year}
  {2013})}\BibitemShut {NoStop}%
\bibitem [{\citenamefont {MacCabe}\ \emph {et~al.}(2020)\citenamefont
  {MacCabe}, \citenamefont {Ren}, \citenamefont {Luo}, \citenamefont {Cohen},
  \citenamefont {Zhou}, \citenamefont {Sipahigil}, \citenamefont
  {Mirhosseini},\ and\ \citenamefont {Painter}}]{2020maccabe}%
  \BibitemOpen
  \bibfield  {author} {\bibinfo {author} {\bibfnamefont {G.S.}\ \bibnamefont
  {MacCabe}}, \bibinfo {author} {\bibfnamefont {H.}~\bibnamefont {Ren}},
  \bibinfo {author} {\bibfnamefont {J.}~\bibnamefont {Luo}}, \bibinfo {author}
  {\bibfnamefont {J.D.}\ \bibnamefont {Cohen}}, \bibinfo {author}
  {\bibfnamefont {H.}~\bibnamefont {Zhou}}, \bibinfo {author} {\bibfnamefont
  {A.}~\bibnamefont {Sipahigil}}, \bibinfo {author} {\bibfnamefont
  {M.}~\bibnamefont {Mirhosseini}},\ and\ \bibinfo {author} {\bibfnamefont
  {O.}~\bibnamefont {Painter}},\ }\bibfield  {title} {\bibinfo {title}
  {Nano-acoustic resonator with ultralong phonon lifetime},\ }\href
  {https://www.science.org/doi/full/10.1126/science.abc7312} {\bibfield
  {journal} {\bibinfo  {journal} {Science}\ }\textbf {\bibinfo {volume}
  {370}},\ \bibinfo {pages} {840} (\bibinfo {year} {2020})}\BibitemShut
  {NoStop}%
\bibitem [{\citenamefont {Aspelmeyer}\ \emph {et~al.}(2014)\citenamefont
  {Aspelmeyer}, \citenamefont {Kippenberg},\ and\ \citenamefont
  {Marquardt}}]{2014Aspelmeyer}%
  \BibitemOpen
  \bibfield  {author} {\bibinfo {author} {\bibfnamefont {M.}~\bibnamefont
  {Aspelmeyer}}, \bibinfo {author} {\bibfnamefont {T.J.}\ \bibnamefont
  {Kippenberg}},\ and\ \bibinfo {author} {\bibfnamefont {F.}~\bibnamefont
  {Marquardt}},\ }\bibfield  {title} {\bibinfo {title} {Cavity optomechanics},\
  }\href {https://doi.org/10.1103/RevModPhys.86.1391} {\bibfield  {journal}
  {\bibinfo  {journal} {Rev. Mod. Phys.}\ }\textbf {\bibinfo {volume} {86}},\
  \bibinfo {pages} {1391} (\bibinfo {year} {2014})}\BibitemShut {NoStop}%
\bibitem [{\citenamefont {Tian}\ and\ \citenamefont {Zoller}(2004)}]{2004Tian}%
  \BibitemOpen
  \bibfield  {author} {\bibinfo {author} {\bibfnamefont {L.}~\bibnamefont
  {Tian}}\ and\ \bibinfo {author} {\bibfnamefont {P.}~\bibnamefont {Zoller}},\
  }\bibfield  {title} {\bibinfo {title} {Coupled ion-nanomechanical systems},\
  }\href {https://doi.org/10.1103/PhysRevLett.93.266403} {\bibfield  {journal}
  {\bibinfo  {journal} {Phys. Rev. Lett.}\ }\textbf {\bibinfo {volume} {93}},\
  \bibinfo {pages} {266403} (\bibinfo {year} {2004})}\BibitemShut {NoStop}%
\bibitem [{\citenamefont {Treutlein}\ \emph {et~al.}(2007)\citenamefont
  {Treutlein}, \citenamefont {Hunger}, \citenamefont {Camerer}, \citenamefont
  {H\"ansch},\ and\ \citenamefont {Reichel}}]{2007Treutlein}%
  \BibitemOpen
  \bibfield  {author} {\bibinfo {author} {\bibfnamefont {P.}~\bibnamefont
  {Treutlein}}, \bibinfo {author} {\bibfnamefont {D.}~\bibnamefont {Hunger}},
  \bibinfo {author} {\bibfnamefont {S.}~\bibnamefont {Camerer}}, \bibinfo
  {author} {\bibfnamefont {T.W.}\ \bibnamefont {H\"ansch}},\ and\ \bibinfo
  {author} {\bibfnamefont {J.}~\bibnamefont {Reichel}},\ }\bibfield  {title}
  {\bibinfo {title} {Bose-{E}instein condensate coupled to a nanomechanical
  resonator on an atom chip},\ }\href
  {https://doi.org/10.1103/PhysRevLett.99.140403} {\bibfield  {journal}
  {\bibinfo  {journal} {Phys. Rev. Lett.}\ }\textbf {\bibinfo {volume} {99}},\
  \bibinfo {pages} {140403} (\bibinfo {year} {2007})}\BibitemShut {NoStop}%
\bibitem [{\citenamefont {Lambert}\ \emph {et~al.}(2008)\citenamefont
  {Lambert}, \citenamefont {Mahboob}, \citenamefont {Pioro-Ladri\`ere},
  \citenamefont {Tokura}, \citenamefont {Tarucha},\ and\ \citenamefont
  {Yamaguchi}}]{2008Lambert}%
  \BibitemOpen
  \bibfield  {author} {\bibinfo {author} {\bibfnamefont {N.}~\bibnamefont
  {Lambert}}, \bibinfo {author} {\bibfnamefont {I.}~\bibnamefont {Mahboob}},
  \bibinfo {author} {\bibfnamefont {M.}~\bibnamefont {Pioro-Ladri\`ere}},
  \bibinfo {author} {\bibfnamefont {Y.}~\bibnamefont {Tokura}}, \bibinfo
  {author} {\bibfnamefont {S.}~\bibnamefont {Tarucha}},\ and\ \bibinfo {author}
  {\bibfnamefont {H.}~\bibnamefont {Yamaguchi}},\ }\bibfield  {title} {\bibinfo
  {title} {Electron-{S}pin manipulation and resonator readout in a
  double-quantum-dot nanoelectromechanical system},\ }\href
  {https://doi.org/10.1103/PhysRevLett.100.136802} {\bibfield  {journal}
  {\bibinfo  {journal} {Phys. Rev. Lett.}\ }\textbf {\bibinfo {volume} {100}},\
  \bibinfo {pages} {136802} (\bibinfo {year} {2008})}\BibitemShut {NoStop}%
\bibitem [{\citenamefont {Rabl}\ \emph {et~al.}(2009)\citenamefont {Rabl},
  \citenamefont {Cappellaro}, \citenamefont {Dutt}, \citenamefont {Jiang},
  \citenamefont {Maze},\ and\ \citenamefont {Lukin}}]{2009Rabl}%
  \BibitemOpen
  \bibfield  {author} {\bibinfo {author} {\bibfnamefont {P.}~\bibnamefont
  {Rabl}}, \bibinfo {author} {\bibfnamefont {P.}~\bibnamefont {Cappellaro}},
  \bibinfo {author} {\bibfnamefont {M.V.G.}\ \bibnamefont {Dutt}}, \bibinfo
  {author} {\bibfnamefont {L.}~\bibnamefont {Jiang}}, \bibinfo {author}
  {\bibfnamefont {J.R.}\ \bibnamefont {Maze}},\ and\ \bibinfo {author}
  {\bibfnamefont {M.D.}\ \bibnamefont {Lukin}},\ }\bibfield  {title} {\bibinfo
  {title} {Strong magnetic coupling between an electronic spin qubit and a
  mechanical resonator},\ }\href {https://doi.org/10.1103/PhysRevB.79.041302}
  {\bibfield  {journal} {\bibinfo  {journal} {Phys. Rev. B}\ }\textbf {\bibinfo
  {volume} {79}},\ \bibinfo {pages} {041302} (\bibinfo {year}
  {2009})}\BibitemShut {NoStop}%
\bibitem [{\citenamefont {Singh}\ \emph {et~al.}(2008)\citenamefont {Singh},
  \citenamefont {Bhattacharya}, \citenamefont {Dutta},\ and\ \citenamefont
  {Meystre}}]{2008Singh}%
  \BibitemOpen
  \bibfield  {author} {\bibinfo {author} {\bibfnamefont {S.}~\bibnamefont
  {Singh}}, \bibinfo {author} {\bibfnamefont {M.}~\bibnamefont {Bhattacharya}},
  \bibinfo {author} {\bibfnamefont {O.}~\bibnamefont {Dutta}},\ and\ \bibinfo
  {author} {\bibfnamefont {P.}~\bibnamefont {Meystre}},\ }\bibfield  {title}
  {\bibinfo {title} {Coupling nanomechanical cantilevers to dipolar
  molecules},\ }\href {https://doi.org/10.1103/PhysRevLett.101.263603}
  {\bibfield  {journal} {\bibinfo  {journal} {Phys. Rev. Lett.}\ }\textbf
  {\bibinfo {volume} {101}},\ \bibinfo {pages} {263603} (\bibinfo {year}
  {2008})}\BibitemShut {NoStop}%
\bibitem [{\citenamefont {Hammerer}\ \emph {et~al.}(2009)\citenamefont
  {Hammerer}, \citenamefont {Wallquist}, \citenamefont {Genes}, \citenamefont
  {Ludwig}, \citenamefont {Marquardt}, \citenamefont {Treutlein}, \citenamefont
  {Zoller}, \citenamefont {Ye},\ and\ \citenamefont {Kimble}}]{2009Hammerer}%
  \BibitemOpen
  \bibfield  {author} {\bibinfo {author} {\bibfnamefont {K.}~\bibnamefont
  {Hammerer}}, \bibinfo {author} {\bibfnamefont {M.}~\bibnamefont {Wallquist}},
  \bibinfo {author} {\bibfnamefont {C.}~\bibnamefont {Genes}}, \bibinfo
  {author} {\bibfnamefont {M.}~\bibnamefont {Ludwig}}, \bibinfo {author}
  {\bibfnamefont {F.}~\bibnamefont {Marquardt}}, \bibinfo {author}
  {\bibfnamefont {P.}~\bibnamefont {Treutlein}}, \bibinfo {author}
  {\bibfnamefont {P.}~\bibnamefont {Zoller}}, \bibinfo {author} {\bibfnamefont
  {J.}~\bibnamefont {Ye}},\ and\ \bibinfo {author} {\bibfnamefont {H.J.}\
  \bibnamefont {Kimble}},\ }\bibfield  {title} {\bibinfo {title} {Strong
  coupling of a mechanical oscillator and a single atom},\ }\href
  {https://doi.org/10.1103/PhysRevLett.103.063005} {\bibfield  {journal}
  {\bibinfo  {journal} {Phys. Rev. Lett.}\ }\textbf {\bibinfo {volume} {103}},\
  \bibinfo {pages} {063005} (\bibinfo {year} {2009})}\BibitemShut {NoStop}%
\bibitem [{\citenamefont {Hammerer}\ \emph {et~al.}(2010)\citenamefont
  {Hammerer}, \citenamefont {Stannigel}, \citenamefont {Genes}, \citenamefont
  {Zoller}, \citenamefont {Treutlein}, \citenamefont {Camerer}, \citenamefont
  {Hunger},\ and\ \citenamefont {H\"ansch}}]{2010Hammerer}%
  \BibitemOpen
  \bibfield  {author} {\bibinfo {author} {\bibfnamefont {K.}~\bibnamefont
  {Hammerer}}, \bibinfo {author} {\bibfnamefont {K.}~\bibnamefont {Stannigel}},
  \bibinfo {author} {\bibfnamefont {C.}~\bibnamefont {Genes}}, \bibinfo
  {author} {\bibfnamefont {P.}~\bibnamefont {Zoller}}, \bibinfo {author}
  {\bibfnamefont {P.}~\bibnamefont {Treutlein}}, \bibinfo {author}
  {\bibfnamefont {S.}~\bibnamefont {Camerer}}, \bibinfo {author} {\bibfnamefont
  {D.}~\bibnamefont {Hunger}},\ and\ \bibinfo {author} {\bibfnamefont {T.W.}\
  \bibnamefont {H\"ansch}},\ }\bibfield  {title} {\bibinfo {title} {Optical
  lattices with micromechanical mirrors},\ }\href
  {https://doi.org/10.1103/PhysRevA.82.021803} {\bibfield  {journal} {\bibinfo
  {journal} {Phys. Rev. A}\ }\textbf {\bibinfo {volume} {82}},\ \bibinfo
  {pages} {021803} (\bibinfo {year} {2010})}\BibitemShut {NoStop}%
\bibitem [{\citenamefont {Wallquist}\ \emph {et~al.}(2010)\citenamefont
  {Wallquist}, \citenamefont {Hammerer}, \citenamefont {Zoller}, \citenamefont
  {Genes}, \citenamefont {Ludwig}, \citenamefont {Marquardt}, \citenamefont
  {Treutlein}, \citenamefont {Ye},\ and\ \citenamefont
  {Kimble}}]{2010Wallquist}%
  \BibitemOpen
  \bibfield  {author} {\bibinfo {author} {\bibfnamefont {M.}~\bibnamefont
  {Wallquist}}, \bibinfo {author} {\bibfnamefont {K.}~\bibnamefont {Hammerer}},
  \bibinfo {author} {\bibfnamefont {P.}~\bibnamefont {Zoller}}, \bibinfo
  {author} {\bibfnamefont {C.}~\bibnamefont {Genes}}, \bibinfo {author}
  {\bibfnamefont {M.}~\bibnamefont {Ludwig}}, \bibinfo {author} {\bibfnamefont
  {F.}~\bibnamefont {Marquardt}}, \bibinfo {author} {\bibfnamefont
  {P.}~\bibnamefont {Treutlein}}, \bibinfo {author} {\bibfnamefont
  {J.}~\bibnamefont {Ye}},\ and\ \bibinfo {author} {\bibfnamefont {H.J.}\
  \bibnamefont {Kimble}},\ }\bibfield  {title} {\bibinfo {title} {Single-atom
  cavity {QED} and optomicromechanics},\ }\href
  {https://doi.org/10.1103/PhysRevA.81.023816} {\bibfield  {journal} {\bibinfo
  {journal} {Phys. Rev. A}\ }\textbf {\bibinfo {volume} {81}},\ \bibinfo
  {pages} {023816} (\bibinfo {year} {2010})}\BibitemShut {NoStop}%
\bibitem [{\citenamefont {Camerer}\ \emph {et~al.}(2011)\citenamefont
  {Camerer}, \citenamefont {Korppi}, \citenamefont {J\"ockel}, \citenamefont
  {Hunger}, \citenamefont {H\"ansch},\ and\ \citenamefont
  {Treutlein}}]{2011Camerer}%
  \BibitemOpen
  \bibfield  {author} {\bibinfo {author} {\bibfnamefont {S.}~\bibnamefont
  {Camerer}}, \bibinfo {author} {\bibfnamefont {M.}~\bibnamefont {Korppi}},
  \bibinfo {author} {\bibfnamefont {A.}~\bibnamefont {J\"ockel}}, \bibinfo
  {author} {\bibfnamefont {D.}~\bibnamefont {Hunger}}, \bibinfo {author}
  {\bibfnamefont {T.W.}\ \bibnamefont {H\"ansch}},\ and\ \bibinfo {author}
  {\bibfnamefont {P.}~\bibnamefont {Treutlein}},\ }\bibfield  {title} {\bibinfo
  {title} {Realization of an optomechanical interface between ultracold atoms
  and a membrane},\ }\href {https://doi.org/10.1103/PhysRevLett.107.223001}
  {\bibfield  {journal} {\bibinfo  {journal} {Phys. Rev. Lett.}\ }\textbf
  {\bibinfo {volume} {107}},\ \bibinfo {pages} {223001} (\bibinfo {year}
  {2011})}\BibitemShut {NoStop}%
\bibitem [{\citenamefont {J{\"o}ckel}\ \emph {et~al.}(2015)\citenamefont
  {J{\"o}ckel}, \citenamefont {Faber}, \citenamefont {Kampschulte},
  \citenamefont {Korppi}, \citenamefont {Rakher},\ and\ \citenamefont
  {Treutlein}}]{2015jockel}%
  \BibitemOpen
  \bibfield  {author} {\bibinfo {author} {\bibfnamefont {A.}~\bibnamefont
  {J{\"o}ckel}}, \bibinfo {author} {\bibfnamefont {A.}~\bibnamefont {Faber}},
  \bibinfo {author} {\bibfnamefont {T.}~\bibnamefont {Kampschulte}}, \bibinfo
  {author} {\bibfnamefont {M.}~\bibnamefont {Korppi}}, \bibinfo {author}
  {\bibfnamefont {M.T.}\ \bibnamefont {Rakher}},\ and\ \bibinfo {author}
  {\bibfnamefont {P.}~\bibnamefont {Treutlein}},\ }\bibfield  {title} {\bibinfo
  {title} {Sympathetic cooling of a membrane oscillator in a hybrid
  mechanical--atomic system},\ }\href {https://doi.org/10.1038/nnano.2014.278}
  {\bibfield  {journal} {\bibinfo  {journal} {Nature nanotechnology}\ }\textbf
  {\bibinfo {volume} {10}},\ \bibinfo {pages} {55} (\bibinfo {year}
  {2015})}\BibitemShut {NoStop}%
\bibitem [{\citenamefont {Garcia-Vidal}\ \emph {et~al.}(2021)\citenamefont
  {Garcia-Vidal}, \citenamefont {Ciuti},\ and\ \citenamefont
  {Ebbesen}}]{2021garcia}%
  \BibitemOpen
  \bibfield  {author} {\bibinfo {author} {\bibfnamefont {F.J.}\ \bibnamefont
  {Garcia-Vidal}}, \bibinfo {author} {\bibfnamefont {C.}~\bibnamefont
  {Ciuti}},\ and\ \bibinfo {author} {\bibfnamefont {T.W.}\ \bibnamefont
  {Ebbesen}},\ }\bibfield  {title} {\bibinfo {title} {Manipulating matter by
  strong coupling to vacuum fields},\ }\href
  {https://www.science.org/doi/full/10.1126/science.abd0336} {\bibfield
  {journal} {\bibinfo  {journal} {Science}\ }\textbf {\bibinfo {volume}
  {373}},\ \bibinfo {pages} {1} (\bibinfo {year} {2021})}\BibitemShut {NoStop}%
\bibitem [{\citenamefont {Nation}\ \emph {et~al.}(2012)\citenamefont {Nation},
  \citenamefont {Johansson}, \citenamefont {Blencowe},\ and\ \citenamefont
  {Nori}}]{2012Nation}%
  \BibitemOpen
  \bibfield  {author} {\bibinfo {author} {\bibfnamefont {P.D.}\ \bibnamefont
  {Nation}}, \bibinfo {author} {\bibfnamefont {J.R.}\ \bibnamefont
  {Johansson}}, \bibinfo {author} {\bibfnamefont {M.P.}\ \bibnamefont
  {Blencowe}},\ and\ \bibinfo {author} {\bibfnamefont {F.}~\bibnamefont
  {Nori}},\ }\bibfield  {title} {\bibinfo {title} {Colloquium: Stimulating
  uncertainty: Amplifying the quantum vacuum with superconducting circuits},\
  }\href {https://doi.org/10.1103/RevModPhys.84.1} {\bibfield  {journal}
  {\bibinfo  {journal} {Rev. Mod. Phys.}\ }\textbf {\bibinfo {volume} {84}},\
  \bibinfo {pages} {1} (\bibinfo {year} {2012})}\BibitemShut {NoStop}%
\bibitem [{\citenamefont {Wilson}\ \emph {et~al.}(2011)\citenamefont {Wilson},
  \citenamefont {Johansson}, \citenamefont {Pourkabirian}, \citenamefont
  {Simoen}, \citenamefont {Johansson}, \citenamefont {Duty}, \citenamefont
  {Nori},\ and\ \citenamefont {Delsing}}]{2011Wilson}%
  \BibitemOpen
  \bibfield  {author} {\bibinfo {author} {\bibfnamefont {C.M.}\ \bibnamefont
  {Wilson}}, \bibinfo {author} {\bibfnamefont {G.}~\bibnamefont {Johansson}},
  \bibinfo {author} {\bibfnamefont {A.}~\bibnamefont {Pourkabirian}}, \bibinfo
  {author} {\bibfnamefont {M.}~\bibnamefont {Simoen}}, \bibinfo {author}
  {\bibfnamefont {J.R.}\ \bibnamefont {Johansson}}, \bibinfo {author}
  {\bibfnamefont {T.}~\bibnamefont {Duty}}, \bibinfo {author} {\bibfnamefont
  {F.}~\bibnamefont {Nori}},\ and\ \bibinfo {author} {\bibfnamefont
  {P.}~\bibnamefont {Delsing}},\ }\bibfield  {title} {\bibinfo {title}
  {Observation of the dynamical {C}asimir effect in a superconducting
  circuit},\ }\href {https://doi.org/10.1038/nature10561} {\bibfield  {journal}
  {\bibinfo  {journal} {Nature}\ }\textbf {\bibinfo {volume} {479}},\ \bibinfo
  {pages} {376} (\bibinfo {year} {2011})}\BibitemShut {NoStop}%
\bibitem [{\citenamefont {Fong}\ \emph {et~al.}(2019)\citenamefont {Fong},
  \citenamefont {Li}, \citenamefont {Zhao}, \citenamefont {Yang}, \citenamefont
  {Wang},\ and\ \citenamefont {Zhang}}]{2019fong}%
  \BibitemOpen
  \bibfield  {author} {\bibinfo {author} {\bibfnamefont {K.Y.}\ \bibnamefont
  {Fong}}, \bibinfo {author} {\bibfnamefont {H.-K.}\ \bibnamefont {Li}},
  \bibinfo {author} {\bibfnamefont {R.}~\bibnamefont {Zhao}}, \bibinfo {author}
  {\bibfnamefont {S.}~\bibnamefont {Yang}}, \bibinfo {author} {\bibfnamefont
  {Y.}~\bibnamefont {Wang}},\ and\ \bibinfo {author} {\bibfnamefont
  {X.}~\bibnamefont {Zhang}},\ }\bibfield  {title} {\bibinfo {title} {Phonon
  heat transfer across a vacuum through quantum fluctuations},\ }\href
  {https://doi.org/10.1038/s41586-019-1800-4} {\bibfield  {journal} {\bibinfo
  {journal} {Nature}\ }\textbf {\bibinfo {volume} {576}},\ \bibinfo {pages}
  {243} (\bibinfo {year} {2019})}\BibitemShut {NoStop}%
\bibitem [{\citenamefont {Di~Stefano{,
  }et~al.}(2019{\natexlab{a}})}]{2019Stefano}%
  \BibitemOpen
  \bibfield  {author} {\bibinfo {author} {\bibfnamefont {O.}~\bibnamefont
  {Di~Stefano{, }et~al.}},\ }\bibfield  {title} {\bibinfo {title} {Interaction
  of mechanical oscillators mediated by the exchange of virtual photon pairs},\
  }\href {https://doi.org/10.1103/PhysRevLett.122.030402} {\bibfield  {journal}
  {\bibinfo  {journal} {Phys. Rev. Lett.}\ }\textbf {\bibinfo {volume} {122}},\
  \bibinfo {pages} {030402} (\bibinfo {year} {2019}{\natexlab{a}})}\BibitemShut
  {NoStop}%
\bibitem [{\citenamefont {Zhao}\ \emph {et~al.}(2017)\citenamefont {Zhao},
  \citenamefont {Tan}, \citenamefont {Yu}, \citenamefont {Zhu},\ and\
  \citenamefont {Yu}}]{2017Zhao}%
  \BibitemOpen
  \bibfield  {author} {\bibinfo {author} {\bibfnamefont {P.}~\bibnamefont
  {Zhao}}, \bibinfo {author} {\bibfnamefont {X.}~\bibnamefont {Tan}}, \bibinfo
  {author} {\bibfnamefont {H.}~\bibnamefont {Yu}}, \bibinfo {author}
  {\bibfnamefont {S.-L.}\ \bibnamefont {Zhu}},\ and\ \bibinfo {author}
  {\bibfnamefont {Y.}~\bibnamefont {Yu}},\ }\bibfield  {title} {\bibinfo
  {title} {Circuit {QED} with qutrits: Coupling three or more atoms via
  virtual-photon exchange},\ }\href
  {https://doi.org/10.1103/PhysRevA.96.043833} {\bibfield  {journal} {\bibinfo
  {journal} {Phys. Rev. A}\ }\textbf {\bibinfo {volume} {96}},\ \bibinfo
  {pages} {043833} (\bibinfo {year} {2017})}\BibitemShut {NoStop}%
\bibitem [{\citenamefont {Stassi{, }et~al.}(2017)}]{2017Stassi}%
  \BibitemOpen
  \bibfield  {author} {\bibinfo {author} {\bibfnamefont {R.}~\bibnamefont
  {Stassi{, }et~al.}},\ }\bibfield  {title} {\bibinfo {title} {Quantum
  nonlinear optics without photons},\ }\href
  {https://doi.org/10.1103/PhysRevA.96.023818} {\bibfield  {journal} {\bibinfo
  {journal} {Phys. Rev. A}\ }\textbf {\bibinfo {volume} {96}},\ \bibinfo
  {pages} {023818} (\bibinfo {year} {2017})}\BibitemShut {NoStop}%
\bibitem [{\citenamefont {Sackett}\ \emph {et~al.}(2000)\citenamefont
  {Sackett}, \citenamefont {Kielpinski}, \citenamefont {King}, \citenamefont
  {Langer}, \citenamefont {Meyer}, \citenamefont {Myatt}, \citenamefont {Rowe},
  \citenamefont {Turchette}, \citenamefont {Itano}, \citenamefont {Wineland},\
  and\ \citenamefont {Monroe}}]{2000sackett}%
  \BibitemOpen
  \bibfield  {author} {\bibinfo {author} {\bibfnamefont {C.A.}\ \bibnamefont
  {Sackett}}, \bibinfo {author} {\bibfnamefont {D.}~\bibnamefont {Kielpinski}},
  \bibinfo {author} {\bibfnamefont {B.E.}\ \bibnamefont {King}}, \bibinfo
  {author} {\bibfnamefont {C.}~\bibnamefont {Langer}}, \bibinfo {author}
  {\bibfnamefont {V.}~\bibnamefont {Meyer}}, \bibinfo {author} {\bibfnamefont
  {C.J.}\ \bibnamefont {Myatt}}, \bibinfo {author} {\bibfnamefont
  {M.}~\bibnamefont {Rowe}}, \bibinfo {author} {\bibfnamefont {Q.}~\bibnamefont
  {Turchette}}, \bibinfo {author} {\bibfnamefont {W.M.}\ \bibnamefont
  {Itano}}, \bibinfo {author} {\bibfnamefont {D.J.}\ \bibnamefont
  {Wineland}},\ and\ \bibinfo {author} {\bibfnamefont {C.}~\bibnamefont
  {Monroe}},\ }\bibfield  {title} {\bibinfo {title} {Experimental entanglement
  of four particles},\ }\href {https://doi.org/10.1038/35005011} {\bibfield
  {journal} {\bibinfo  {journal} {Nature}\ }\textbf {\bibinfo {volume} {404}},\
  \bibinfo {pages} {256} (\bibinfo {year} {2000})}\BibitemShut {NoStop}%
\bibitem [{\citenamefont {Leibfried}\ \emph
  {et~al.}(2003{\natexlab{b}})\citenamefont {Leibfried}, \citenamefont
  {DeMarco}, \citenamefont {Meyer}, \citenamefont {Lucas}, \citenamefont
  {Barrett}, \citenamefont {Britton}, \citenamefont {Itano}, \citenamefont
  {Jelenkovi{\'c}}, \citenamefont {Langer}, \citenamefont {Rosenband},\ and\
  \citenamefont {Wineland}}]{2003leibfriedExperimental}%
  \BibitemOpen
  \bibfield  {author} {\bibinfo {author} {\bibfnamefont {D.}~\bibnamefont
  {Leibfried}}, \bibinfo {author} {\bibfnamefont {B.}~\bibnamefont {DeMarco}},
  \bibinfo {author} {\bibfnamefont {V.}~\bibnamefont {Meyer}}, \bibinfo
  {author} {\bibfnamefont {D.}~\bibnamefont {Lucas}}, \bibinfo {author}
  {\bibfnamefont {M.}~\bibnamefont {Barrett}}, \bibinfo {author} {\bibfnamefont
  {J.}~\bibnamefont {Britton}}, \bibinfo {author} {\bibfnamefont {W.M.}\
  \bibnamefont {Itano}}, \bibinfo {author} {\bibfnamefont {B.}~\bibnamefont
  {Jelenkovi{\'c}}}, \bibinfo {author} {\bibfnamefont {C.}~\bibnamefont
  {Langer}}, \bibinfo {author} {\bibfnamefont {T.}~\bibnamefont {Rosenband}},\
  and\ \bibinfo {author} {\bibfnamefont {D.J.}\ \bibnamefont {Wineland}},\
  }\bibfield  {title} {\bibinfo {title} {Experimental demonstration of a
  robust, high-fidelity geometric two ion-qubit phase gate},\ }\href
  {https://doi.org/10.1038/nature01492} {\bibfield  {journal} {\bibinfo
  {journal} {Nature}\ }\textbf {\bibinfo {volume} {422}},\ \bibinfo {pages}
  {412} (\bibinfo {year} {2003}{\natexlab{b}})}\BibitemShut {NoStop}%
\bibitem [{\citenamefont {DiCarlo}\ \emph {et~al.}(2009)\citenamefont
  {DiCarlo}, \citenamefont {Chow}, \citenamefont {Gambetta}, \citenamefont
  {Bishop}, \citenamefont {Johnson}, \citenamefont {Schuster}, \citenamefont
  {Majer}, \citenamefont {Blais}, \citenamefont {Frunzio}, \citenamefont
  {Girvin},\ and\ \citenamefont {Schoelkopf}}]{2009dicarloDemonstration}%
  \BibitemOpen
  \bibfield  {author} {\bibinfo {author} {\bibfnamefont {L.}~\bibnamefont
  {DiCarlo}}, \bibinfo {author} {\bibfnamefont {J.M.}\ \bibnamefont {Chow}},
  \bibinfo {author} {\bibfnamefont {J.M.}\ \bibnamefont {Gambetta}}, \bibinfo
  {author} {\bibfnamefont {L.S.}\ \bibnamefont {Bishop}}, \bibinfo {author}
  {\bibfnamefont {B.R.}\ \bibnamefont {Johnson}}, \bibinfo {author}
  {\bibfnamefont {D.}~\bibnamefont {Schuster}}, \bibinfo {author}
  {\bibfnamefont {J.}~\bibnamefont {Majer}}, \bibinfo {author} {\bibfnamefont
  {A.}~\bibnamefont {Blais}}, \bibinfo {author} {\bibfnamefont
  {L.}~\bibnamefont {Frunzio}}, \bibinfo {author} {\bibfnamefont
  {S.}~\bibnamefont {Girvin}},\ and\ \bibinfo {author} {\bibfnamefont {R.J.}\
  \bibnamefont {Schoelkopf}},\ }\bibfield  {title} {\bibinfo {title}
  {Demonstration of two-qubit algorithms with a superconducting quantum
  processor},\ }\href {https://doi.org/10.1038/nature08121} {\bibfield
  {journal} {\bibinfo  {journal} {Nature}\ }\textbf {\bibinfo {volume} {460}},\
  \bibinfo {pages} {240} (\bibinfo {year} {2009})}\BibitemShut {NoStop}%
\bibitem [{\citenamefont {Kockum{, }et~al.}(2017{\natexlab{a}})}]{2017Kockum}%
  \BibitemOpen
  \bibfield  {author} {\bibinfo {author} {\bibfnamefont {A.F.}\ \bibnamefont
  {Kockum{, }et~al.}},\ }\bibfield  {title} {\bibinfo {title} {Deterministic
  quantum nonlinear optics with single atoms and virtual photons},\ }\href
  {https://doi.org/10.1103/PhysRevA.95.063849} {\bibfield  {journal} {\bibinfo
  {journal} {Phys. Rev. A}\ }\textbf {\bibinfo {volume} {95}},\ \bibinfo
  {pages} {063849} (\bibinfo {year} {2017}{\natexlab{a}})}\BibitemShut
  {NoStop}%
\bibitem [{\citenamefont {Majer}\ \emph {et~al.}(2007)\citenamefont {Majer},
  \citenamefont {Chow}, \citenamefont {Gambetta}, \citenamefont {Koch},
  \citenamefont {Johnson}, \citenamefont {Schreier}, \citenamefont {Frunzio},
  \citenamefont {Schuster}, \citenamefont {Houck}, \citenamefont {Wallraff},\
  and\ \citenamefont {Schoelkopf}}]{2007majer}%
  \BibitemOpen
  \bibfield  {author} {\bibinfo {author} {\bibfnamefont {J.}~\bibnamefont
  {Majer}}, \bibinfo {author} {\bibfnamefont {J.M.}\ \bibnamefont {Chow}},
  \bibinfo {author} {\bibfnamefont {J.M.}\ \bibnamefont {Gambetta}}, \bibinfo
  {author} {\bibfnamefont {J.}~\bibnamefont {Koch}}, \bibinfo {author}
  {\bibfnamefont {B.R.}\ \bibnamefont {Johnson}}, \bibinfo {author}
  {\bibfnamefont {J.A.}\ \bibnamefont {Schreier}}, \bibinfo {author}
  {\bibfnamefont {L.}~\bibnamefont {Frunzio}}, \bibinfo {author} {\bibfnamefont
  {D.I.}\ \bibnamefont {Schuster}}, \bibinfo {author} {\bibfnamefont {A.A.}\
  \bibnamefont {Houck}}, \bibinfo {author} {\bibfnamefont {A.}~\bibnamefont
  {Wallraff}},\ and\ \bibinfo {author} {\bibfnamefont {R.J.}\ \bibnamefont
  {Schoelkopf}},\ }\bibfield  {title} {\bibinfo {title} {Coupling
  superconducting qubits via a cavity bus},\ }\href
  {https://doi.org/10.1038/nature06184} {\bibfield  {journal} {\bibinfo
  {journal} {Nature}\ }\textbf {\bibinfo {volume} {449}},\ \bibinfo {pages}
  {443} (\bibinfo {year} {2007})}\BibitemShut {NoStop}%
\bibitem [{\citenamefont {Settineri{,
  }et~al.}(2019)}]{2019SettineriConversion}%
  \BibitemOpen
  \bibfield  {author} {\bibinfo {author} {\bibfnamefont {A.}~\bibnamefont
  {Settineri{, }et~al.}},\ }\bibfield  {title} {\bibinfo {title} {Conversion of
  mechanical noise into correlated photon pairs: Dynamical {C}asimir effect
  from an incoherent mechanical drive},\ }\href
  {https://doi.org/10.1103/PhysRevA.100.022501} {\bibfield  {journal} {\bibinfo
   {journal} {Phys. Rev. A}\ }\textbf {\bibinfo {volume} {100}},\ \bibinfo
  {pages} {022501} (\bibinfo {year} {2019})}\BibitemShut {NoStop}%
\bibitem [{\citenamefont {Moore}(1970)}]{1970Moore}%
  \BibitemOpen
  \bibfield  {author} {\bibinfo {author} {\bibfnamefont {G.T.}\ \bibnamefont
  {Moore}},\ }\bibfield  {title} {\bibinfo {title} {Quantum theory of the
  electromagnetic field in a variable-length one-dimensional cavity},\ }\href
  {https://doi.org/10.1063/1.1665432} {\bibfield  {journal} {\bibinfo
  {journal} {Journal of Mathematical Physics}\ }\textbf {\bibinfo {volume}
  {11}},\ \bibinfo {pages} {2679} (\bibinfo {year} {1970})}\BibitemShut
  {NoStop}%
\bibitem [{\citenamefont {Johansson{, }et~al.}(2009)}]{2009Johansson}%
  \BibitemOpen
  \bibfield  {author} {\bibinfo {author} {\bibfnamefont {J.R.}\ \bibnamefont
  {Johansson{, }et~al.}},\ }\bibfield  {title} {\bibinfo {title} {Dynamical
  {C}asimir effect in a superconducting coplanar waveguide},\ }\href
  {https://doi.org/10.1103/PhysRevLett.103.147003} {\bibfield  {journal}
  {\bibinfo  {journal} {Phys. Rev. Lett.}\ }\textbf {\bibinfo {volume} {103}},\
  \bibinfo {pages} {147003} (\bibinfo {year} {2009})}\BibitemShut {NoStop}%
\bibitem [{\citenamefont {Johansson{, }et~al.}(2010)}]{2010Johansson}%
  \BibitemOpen
  \bibfield  {author} {\bibinfo {author} {\bibfnamefont {J.R.}\ \bibnamefont
  {Johansson{, }et~al.}},\ }\bibfield  {title} {\bibinfo {title} {Dynamical
  {C}asimir effect in superconducting microwave circuits},\ }\href
  {https://doi.org/10.1103/PhysRevA.82.052509} {\bibfield  {journal} {\bibinfo
  {journal} {Phys. Rev. A}\ }\textbf {\bibinfo {volume} {82}},\ \bibinfo
  {pages} {052509} (\bibinfo {year} {2010})}\BibitemShut {NoStop}%
\bibitem [{\citenamefont {L{\"a}hteenm{\"a}ki}\ \emph
  {et~al.}(2013)\citenamefont {L{\"a}hteenm{\"a}ki}, \citenamefont {Paraoanu},
  \citenamefont {Hassel},\ and\ \citenamefont {Hakonen}}]{2013lahteenmaki}%
  \BibitemOpen
  \bibfield  {author} {\bibinfo {author} {\bibfnamefont {P.}~\bibnamefont
  {L{\"a}hteenm{\"a}ki}}, \bibinfo {author} {\bibfnamefont {G.}~\bibnamefont
  {Paraoanu}}, \bibinfo {author} {\bibfnamefont {J.}~\bibnamefont {Hassel}},\
  and\ \bibinfo {author} {\bibfnamefont {P.J.}\ \bibnamefont {Hakonen}},\
  }\bibfield  {title} {\bibinfo {title} {Dynamical {C}asimir effect in a
  {J}osephson metamaterial},\ }\href {https://doi.org/10.1073/pnas.1212705110}
  {\bibfield  {journal} {\bibinfo  {journal} {Proceedings of the National
  Academy of Sciences}\ }\textbf {\bibinfo {volume} {110}},\ \bibinfo {pages}
  {4234} (\bibinfo {year} {2013})}\BibitemShut {NoStop}%
\bibitem [{\citenamefont {Niemczyk}\ \emph {et~al.}(2010)\citenamefont
  {Niemczyk}, \citenamefont {Deppe}, \citenamefont {Huebl}, \citenamefont
  {Menzel}, \citenamefont {Hocke}, \citenamefont {Schwarz}, \citenamefont
  {Garcia-Ripoll}, \citenamefont {Zueco}, \citenamefont {H{\"u}mmer},
  \citenamefont {Solano}, \citenamefont {Marx},\ and\ \citenamefont
  {Gross}}]{2010Niemczyk}%
  \BibitemOpen
  \bibfield  {author} {\bibinfo {author} {\bibfnamefont {T.}~\bibnamefont
  {Niemczyk}}, \bibinfo {author} {\bibfnamefont {F.}~\bibnamefont {Deppe}},
  \bibinfo {author} {\bibfnamefont {H.}~\bibnamefont {Huebl}}, \bibinfo
  {author} {\bibfnamefont {E.}~\bibnamefont {Menzel}}, \bibinfo {author}
  {\bibfnamefont {F.}~\bibnamefont {Hocke}}, \bibinfo {author} {\bibfnamefont
  {M.}~\bibnamefont {Schwarz}}, \bibinfo {author} {\bibfnamefont
  {J.}~\bibnamefont {Garcia-Ripoll}}, \bibinfo {author} {\bibfnamefont
  {D.}~\bibnamefont {Zueco}}, \bibinfo {author} {\bibfnamefont
  {T.}~\bibnamefont {H{\"u}mmer}}, \bibinfo {author} {\bibfnamefont
  {E.}~\bibnamefont {Solano}}, \bibinfo {author} {\bibfnamefont
  {A.}~\bibnamefont {Marx}},\ and\ \bibinfo {author} {\bibfnamefont
  {R.}~\bibnamefont {Gross}},\ }\bibfield  {title} {\bibinfo {title} {Circuit
  quantum electrodynamics in the ultrastrong-coupling regime},\ }\href
  {https://doi.org/10.1038/nphys1730} {\bibfield  {journal} {\bibinfo
  {journal} {Nature Physics}\ }\textbf {\bibinfo {volume} {6}},\ \bibinfo
  {pages} {772} (\bibinfo {year} {2010})}\BibitemShut {NoStop}%
\bibitem [{\citenamefont {Kockum}\ \emph {et~al.}(2019)\citenamefont {Kockum},
  \citenamefont {Miranowicz}, \citenamefont {De~Liberato}, \citenamefont
  {Savasta},\ and\ \citenamefont {Nori}}]{2019Kockum}%
  \BibitemOpen
  \bibfield  {author} {\bibinfo {author} {\bibfnamefont {A.F.}\ \bibnamefont
  {Kockum}}, \bibinfo {author} {\bibfnamefont {A.}~\bibnamefont {Miranowicz}},
  \bibinfo {author} {\bibfnamefont {S.}~\bibnamefont {De~Liberato}}, \bibinfo
  {author} {\bibfnamefont {S.}~\bibnamefont {Savasta}},\ and\ \bibinfo {author}
  {\bibfnamefont {F.}~\bibnamefont {Nori}},\ }\bibfield  {title} {\bibinfo
  {title} {Ultrastrong coupling between light and matter},\ }\href
  {https://doi.org/10.1038/s42254-018-0006-2} {\bibfield  {journal} {\bibinfo
  {journal} {Nature Reviews Physics}\ }\textbf {\bibinfo {volume} {1}},\
  \bibinfo {pages} {19} (\bibinfo {year} {2019})}\BibitemShut {NoStop}%
\bibitem [{\citenamefont {Forn-D\'{\i}az}\ \emph {et~al.}(2019)\citenamefont
  {Forn-D\'{\i}az}, \citenamefont {Lamata}, \citenamefont {Rico}, \citenamefont
  {Kono},\ and\ \citenamefont {Solano}}]{2019Forn}%
  \BibitemOpen
  \bibfield  {author} {\bibinfo {author} {\bibfnamefont {P.}~\bibnamefont
  {Forn-D\'{\i}az}}, \bibinfo {author} {\bibfnamefont {L.}~\bibnamefont
  {Lamata}}, \bibinfo {author} {\bibfnamefont {E.}~\bibnamefont {Rico}},
  \bibinfo {author} {\bibfnamefont {J.}~\bibnamefont {Kono}},\ and\ \bibinfo
  {author} {\bibfnamefont {E.}~\bibnamefont {Solano}},\ }\bibfield  {title}
  {\bibinfo {title} {Ultrastrong coupling regimes of light-matter
  interaction},\ }\href {https://doi.org/10.1103/RevModPhys.91.025005}
  {\bibfield  {journal} {\bibinfo  {journal} {Rev. Mod. Phys.}\ }\textbf
  {\bibinfo {volume} {91}},\ \bibinfo {pages} {025005} (\bibinfo {year}
  {2019})}\BibitemShut {NoStop}%
\bibitem [{\citenamefont {Di~Stefano{,
  }et~al.}(2019{\natexlab{b}})}]{2019StefanoResolution}%
  \BibitemOpen
  \bibfield  {author} {\bibinfo {author} {\bibfnamefont {O.}~\bibnamefont
  {Di~Stefano{, }et~al.}},\ }\bibfield  {title} {\bibinfo {title} {Resolution
  of gauge ambiguities in ultrastrong-coupling cavity quantum
  electrodynamics},\ }\href {https://doi.org/10.1038/s41567-019-0534-4}
  {\bibfield  {journal} {\bibinfo  {journal} {Nature Physics}\ }\textbf
  {\bibinfo {volume} {15}},\ \bibinfo {pages} {803} (\bibinfo {year}
  {2019}{\natexlab{b}})}\BibitemShut {NoStop}%
\bibitem [{\citenamefont {Settineri{, }et~al.}(2021)}]{2021Settineri}%
  \BibitemOpen
  \bibfield  {author} {\bibinfo {author} {\bibfnamefont {A.}~\bibnamefont
  {Settineri{, }et~al.}},\ }\bibfield  {title} {\bibinfo {title} {Gauge
  freedom, quantum measurements, and time-dependent interactions in cavity
  {QED}},\ }\href {https://doi.org/10.1103/PhysRevResearch.3.023079} {\bibfield
   {journal} {\bibinfo  {journal} {Phys. Rev. Research}\ }\textbf {\bibinfo
  {volume} {3}},\ \bibinfo {pages} {023079} (\bibinfo {year}
  {2021})}\BibitemShut {NoStop}%
\bibitem [{\citenamefont {Salmon{, }et~al.}(2022)}]{2021salmon}%
  \BibitemOpen
  \bibfield  {author} {\bibinfo {author} {\bibfnamefont {W.}~\bibnamefont
  {Salmon{, }et~al.}},\ }\bibfield  {title} {\bibinfo {title}
  {Gauge-independent emission spectra and quantum correlations in the
  ultrastrong coupling regime of open system cavity{-QED}},\ }\href
  {https://doi.org/10.1515/nanoph-2021-0718} {\bibfield  {journal} {\bibinfo
  {journal} {Nanophotonics}\ } (\bibinfo {year} {2022})}\BibitemShut {NoStop}%
\bibitem [{\citenamefont {Macrì{, }et~al.}(2022)}]{2021Revealing}%
  \BibitemOpen
  \bibfield  {author} {\bibinfo {author} {\bibfnamefont {V.}~\bibnamefont
  {Macrì{, }et~al.}},\ }\bibfield  {title} {\bibinfo {title} {Revealing
  higher-order light and matter energy exchanges using quantum trajectories in
  ultrastrong coupling},\ }\href {https://doi.org/10.1103/PhysRevA.105.023720}
  {\bibfield  {journal} {\bibinfo  {journal} {Phys. Rev. A}\ }\textbf {\bibinfo
  {volume} {105}},\ \bibinfo {pages} {023720} (\bibinfo {year}
  {2022})}\BibitemShut {NoStop}%
\bibitem [{\citenamefont {Rajabali}\ \emph {et~al.}(2021)\citenamefont
  {Rajabali}, \citenamefont {Cortese}, \citenamefont {Beck}, \citenamefont
  {De~Liberato}, \citenamefont {Faist},\ and\ \citenamefont
  {Scalari}}]{2021rajabali}%
  \BibitemOpen
  \bibfield  {author} {\bibinfo {author} {\bibfnamefont {S.}~\bibnamefont
  {Rajabali}}, \bibinfo {author} {\bibfnamefont {E.}~\bibnamefont {Cortese}},
  \bibinfo {author} {\bibfnamefont {M.}~\bibnamefont {Beck}}, \bibinfo {author}
  {\bibfnamefont {S.}~\bibnamefont {De~Liberato}}, \bibinfo {author}
  {\bibfnamefont {J.}~\bibnamefont {Faist}},\ and\ \bibinfo {author}
  {\bibfnamefont {G.}~\bibnamefont {Scalari}},\ }\bibfield  {title} {\bibinfo
  {title} {Polaritonic nonlocality in light-matter interaction},\ }\href
  {https://doi.org/10.1038/s41566-021-00854-3} {\bibfield  {journal} {\bibinfo
  {journal} {Nature Photonics}\ }\textbf {\bibinfo {volume} {15}},\ \bibinfo
  {pages} {690} (\bibinfo {year} {2021})}\BibitemShut {NoStop}%
\bibitem [{\citenamefont {Settineri{, }et~al.}(2018)}]{2018Settineri}%
  \BibitemOpen
  \bibfield  {author} {\bibinfo {author} {\bibfnamefont {A.}~\bibnamefont
  {Settineri{, }et~al.}},\ }\bibfield  {title} {\bibinfo {title} {Dissipation
  and thermal noise in hybrid quantum systems in the ultrastrong-coupling
  regime},\ }\href {https://doi.org/10.1103/PhysRevA.98.053834} {\bibfield
  {journal} {\bibinfo  {journal} {Phys. Rev. A}\ }\textbf {\bibinfo {volume}
  {98}},\ \bibinfo {pages} {053834} (\bibinfo {year} {2018})}\BibitemShut
  {NoStop}%
\bibitem [{\citenamefont {S\'anchez Mu\~noz{, }et al.}(2020)}]{2020Simulating}%
  \BibitemOpen
  \bibfield  {author} {\bibinfo {author} {\bibfnamefont {C.}~\bibnamefont
  {S\'anchez Mu\~noz{, }et al.}},\ }\bibfield  {title} {\bibinfo {title}
  {Simulating ultrastrong-coupling processes breaking parity conservation in
  {J}aynes-{C}ummings systems},\ }\href
  {https://doi.org/10.1103/PhysRevA.102.033716} {\bibfield  {journal} {\bibinfo
   {journal} {Phys. Rev. A}\ }\textbf {\bibinfo {volume} {102}},\ \bibinfo
  {pages} {033716} (\bibinfo {year} {2020})}\BibitemShut {NoStop}%
\bibitem [{\citenamefont {Hughes}\ \emph {et~al.}(2021)\citenamefont {Hughes},
  \citenamefont {Settineri}, \citenamefont {Savasta},\ and\ \citenamefont
  {Nori}}]{2021Hughes}%
  \BibitemOpen
  \bibfield  {author} {\bibinfo {author} {\bibfnamefont {S.}~\bibnamefont
  {Hughes}}, \bibinfo {author} {\bibfnamefont {A.}~\bibnamefont {Settineri}},
  \bibinfo {author} {\bibfnamefont {S.}~\bibnamefont {Savasta}},\ and\ \bibinfo
  {author} {\bibfnamefont {F.}~\bibnamefont {Nori}},\ }\bibfield  {title}
  {\bibinfo {title} {Resonant raman scattering of single molecules under
  simultaneous strong cavity coupling and ultrastrong optomechanical coupling
  in plasmonic resonators: Phonon-dressed polaritons},\ }\href
  {https://doi.org/10.1103/PhysRevB.104.045431} {\bibfield  {journal} {\bibinfo
   {journal} {Phys. Rev. B}\ }\textbf {\bibinfo {volume} {104}},\ \bibinfo
  {pages} {045431} (\bibinfo {year} {2021})}\BibitemShut {NoStop}%
\bibitem [{\citenamefont {Ridolfo}\ \emph {et~al.}(2012)\citenamefont
  {Ridolfo}, \citenamefont {Leib}, \citenamefont {Savasta},\ and\ \citenamefont
  {Hartmann}}]{2012Ridolfo}%
  \BibitemOpen
  \bibfield  {author} {\bibinfo {author} {\bibfnamefont {A.}~\bibnamefont
  {Ridolfo}}, \bibinfo {author} {\bibfnamefont {M.}~\bibnamefont {Leib}},
  \bibinfo {author} {\bibfnamefont {S.}~\bibnamefont {Savasta}},\ and\ \bibinfo
  {author} {\bibfnamefont {M.J.}\ \bibnamefont {Hartmann}},\ }\bibfield
  {title} {\bibinfo {title} {Photon blockade in the ultrastrong coupling
  regime},\ }\href {https://doi.org/10.1103/PhysRevLett.109.193602} {\bibfield
  {journal} {\bibinfo  {journal} {Phys. Rev. Lett.}\ }\textbf {\bibinfo
  {volume} {109}},\ \bibinfo {pages} {193602} (\bibinfo {year}
  {2012})}\BibitemShut {NoStop}%
\bibitem [{\citenamefont {Stassi{, }et~al.}(2013)}]{2013Stassi}%
  \BibitemOpen
  \bibfield  {author} {\bibinfo {author} {\bibfnamefont {R.}~\bibnamefont
  {Stassi{, }et~al.}},\ }\bibfield  {title} {\bibinfo {title} {Spontaneous
  conversion from virtual to real photons in the ultrastrong-coupling regime},\
  }\href {https://doi.org/10.1103/PhysRevLett.110.243601} {\bibfield  {journal}
  {\bibinfo  {journal} {Phys. Rev. Lett.}\ }\textbf {\bibinfo {volume} {110}},\
  \bibinfo {pages} {243601} (\bibinfo {year} {2013})}\BibitemShut {NoStop}%
\bibitem [{\citenamefont {Garziano{, }et~al.}(2016)}]{2016Garziano}%
  \BibitemOpen
  \bibfield  {author} {\bibinfo {author} {\bibfnamefont {L.}~\bibnamefont
  {Garziano{, }et~al.}},\ }\bibfield  {title} {\bibinfo {title} {One photon can
  simultaneously excite two or more atoms},\ }\href
  {https://doi.org/10.1103/PhysRevLett.117.043601} {\bibfield  {journal}
  {\bibinfo  {journal} {Phys. Rev. Lett.}\ }\textbf {\bibinfo {volume} {117}},\
  \bibinfo {pages} {043601} (\bibinfo {year} {2016})}\BibitemShut {NoStop}%
\bibitem [{\citenamefont {Wang{, }et~al.}(2017)}]{2017Wang}%
  \BibitemOpen
  \bibfield  {author} {\bibinfo {author} {\bibfnamefont {X.}~\bibnamefont
  {Wang{, }et~al.}},\ }\bibfield  {title} {\bibinfo {title} {Observing pure
  effects of counter-rotating terms without ultrastrong coupling: A single
  photon can simultaneously excite two qubits},\ }\href
  {https://doi.org/10.1103/PhysRevA.96.063820} {\bibfield  {journal} {\bibinfo
  {journal} {Phys. Rev. A}\ }\textbf {\bibinfo {volume} {96}},\ \bibinfo
  {pages} {063820} (\bibinfo {year} {2017})}\BibitemShut {NoStop}%
\bibitem [{\citenamefont {Ciuti}\ \emph {et~al.}(2005)\citenamefont {Ciuti},
  \citenamefont {Bastard},\ and\ \citenamefont {Carusotto}}]{2005Ciuti}%
  \BibitemOpen
  \bibfield  {author} {\bibinfo {author} {\bibfnamefont {C.}~\bibnamefont
  {Ciuti}}, \bibinfo {author} {\bibfnamefont {G.}~\bibnamefont {Bastard}},\
  and\ \bibinfo {author} {\bibfnamefont {I.}~\bibnamefont {Carusotto}},\
  }\bibfield  {title} {\bibinfo {title} {Quantum vacuum properties of the
  intersubband cavity polariton field},\ }\href
  {https://doi.org/10.1103/PhysRevB.72.115303} {\bibfield  {journal} {\bibinfo
  {journal} {Phys. Rev. B}\ }\textbf {\bibinfo {volume} {72}},\ \bibinfo
  {pages} {115303} (\bibinfo {year} {2005})}\BibitemShut {NoStop}%
\bibitem [{\citenamefont {Liberato}\ \emph {et~al.}(2007)\citenamefont
  {Liberato}, \citenamefont {Ciuti},\ and\ \citenamefont
  {Carusotto}}]{2007Liberato}%
  \BibitemOpen
  \bibfield  {author} {\bibinfo {author} {\bibfnamefont {S.D.}\ \bibnamefont
  {Liberato}}, \bibinfo {author} {\bibfnamefont {C.}~\bibnamefont {Ciuti}},\
  and\ \bibinfo {author} {\bibfnamefont {I.}~\bibnamefont {Carusotto}},\
  }\bibfield  {title} {\bibinfo {title} {Quantum vacuum radiation spectra from
  a semiconductor microcavity with a time-modulated vacuum rabi frequency},\
  }\href {https://doi.org/10.1103/PhysRevLett.98.103602} {\bibfield  {journal}
  {\bibinfo  {journal} {Phys. Rev. Lett.}\ }\textbf {\bibinfo {volume} {98}},\
  \bibinfo {pages} {103602} (\bibinfo {year} {2007})}\BibitemShut {NoStop}%
\bibitem [{\citenamefont {Cirio{, }et~al.}(2017)}]{2017Cirio}%
  \BibitemOpen
  \bibfield  {author} {\bibinfo {author} {\bibfnamefont {M.}~\bibnamefont
  {Cirio{, }et~al.}},\ }\bibfield  {title} {\bibinfo {title} {Amplified
  optomechanical transduction of virtual radiation pressure},\ }\href
  {https://doi.org/10.1103/PhysRevLett.119.053601} {\bibfield  {journal}
  {\bibinfo  {journal} {Phys. Rev. Lett.}\ }\textbf {\bibinfo {volume} {119}},\
  \bibinfo {pages} {053601} (\bibinfo {year} {2017})}\BibitemShut {NoStop}%
\bibitem [{\citenamefont {Macr\`{\i}{, }et~al.}(2018)}]{2018Macri}%
  \BibitemOpen
  \bibfield  {author} {\bibinfo {author} {\bibfnamefont {V.}~\bibnamefont
  {Macr\`{\i}{, }et~al.}},\ }\bibfield  {title} {\bibinfo {title}
  {Nonperturbative dynamical {C}asimir effect in optomechanical systems: Vacuum
  {C}asimir-{R}abi splittings},\ }\href
  {https://doi.org/10.1103/PhysRevX.8.011031} {\bibfield  {journal} {\bibinfo
  {journal} {Phys. Rev. X}\ }\textbf {\bibinfo {volume} {8}},\ \bibinfo {pages}
  {011031} (\bibinfo {year} {2018})}\BibitemShut {NoStop}%
\bibitem [{\citenamefont {Law}(1995)}]{1995Law}%
  \BibitemOpen
  \bibfield  {author} {\bibinfo {author} {\bibfnamefont {C.K.}\ \bibnamefont
  {Law}},\ }\bibfield  {title} {\bibinfo {title} {Interaction between a moving
  mirror and radiation pressure: A hamiltonian formulation},\ }\href
  {https://doi.org/10.1103/PhysRevA.51.2537} {\bibfield  {journal} {\bibinfo
  {journal} {Phys. Rev. A}\ }\textbf {\bibinfo {volume} {51}},\ \bibinfo
  {pages} {2537} (\bibinfo {year} {1995})}\BibitemShut {NoStop}%
\bibitem [{\citenamefont {Gr{\"o}blacher}\ \emph {et~al.}(2009)\citenamefont
  {Gr{\"o}blacher}, \citenamefont {Hammerer}, \citenamefont {Vanner},\ and\
  \citenamefont {Aspelmeyer}}]{2009groblacher}%
  \BibitemOpen
  \bibfield  {author} {\bibinfo {author} {\bibfnamefont {S.}~\bibnamefont
  {Gr{\"o}blacher}}, \bibinfo {author} {\bibfnamefont {K.}~\bibnamefont
  {Hammerer}}, \bibinfo {author} {\bibfnamefont {M.R.}\ \bibnamefont
  {Vanner}},\ and\ \bibinfo {author} {\bibfnamefont {M.}~\bibnamefont
  {Aspelmeyer}},\ }\bibfield  {title} {\bibinfo {title} {Observation of strong
  coupling between a micromechanical resonator and an optical cavity field},\
  }\href {https://doi.org/10.1038/nature08171} {\bibfield  {journal} {\bibinfo
  {journal} {Nature}\ }\textbf {\bibinfo {volume} {460}},\ \bibinfo {pages}
  {724} (\bibinfo {year} {2009})}\BibitemShut {NoStop}%
\bibitem [{\citenamefont {Verhagen}\ \emph {et~al.}(2012)\citenamefont
  {Verhagen}, \citenamefont {Del{\'e}glise}, \citenamefont {Weis},
  \citenamefont {Schliesser},\ and\ \citenamefont {Kippenberg}}]{2012verhagen}%
  \BibitemOpen
  \bibfield  {author} {\bibinfo {author} {\bibfnamefont {E.}~\bibnamefont
  {Verhagen}}, \bibinfo {author} {\bibfnamefont {S.}~\bibnamefont
  {Del{\'e}glise}}, \bibinfo {author} {\bibfnamefont {S.}~\bibnamefont {Weis}},
  \bibinfo {author} {\bibfnamefont {A.}~\bibnamefont {Schliesser}},\ and\
  \bibinfo {author} {\bibfnamefont {T.J.}\ \bibnamefont {Kippenberg}},\
  }\bibfield  {title} {\bibinfo {title} {Quantum-coherent coupling of a
  mechanical oscillator to an optical cavity mode},\ }\href
  {https://doi.org/10.1038/nature10787} {\bibfield  {journal} {\bibinfo
  {journal} {Nature}\ }\textbf {\bibinfo {volume} {482}},\ \bibinfo {pages}
  {63} (\bibinfo {year} {2012})}\BibitemShut {NoStop}%
\bibitem [{\citenamefont {Bochmann}\ \emph {et~al.}(2013)\citenamefont
  {Bochmann}, \citenamefont {Vainsencher}, \citenamefont {Awschalom},\ and\
  \citenamefont {Cleland}}]{2013bochmann}%
  \BibitemOpen
  \bibfield  {author} {\bibinfo {author} {\bibfnamefont {J.}~\bibnamefont
  {Bochmann}}, \bibinfo {author} {\bibfnamefont {A.}~\bibnamefont
  {Vainsencher}}, \bibinfo {author} {\bibfnamefont {D.D.}\ \bibnamefont
  {Awschalom}},\ and\ \bibinfo {author} {\bibfnamefont {A.N.}\ \bibnamefont
  {Cleland}},\ }\bibfield  {title} {\bibinfo {title} {Nanomechanical coupling
  between microwave and optical photons},\ }\href
  {https://doi.org/10.1038/nphys2748} {\bibfield  {journal} {\bibinfo
  {journal} {Nature Physics}\ }\textbf {\bibinfo {volume} {9}},\ \bibinfo
  {pages} {712} (\bibinfo {year} {2013})}\BibitemShut {NoStop}%
\bibitem [{\citenamefont {Andrews}\ \emph {et~al.}(2014)\citenamefont
  {Andrews}, \citenamefont {Peterson}, \citenamefont {Purdy}, \citenamefont
  {Cicak}, \citenamefont {Simmonds}, \citenamefont {Regal},\ and\ \citenamefont
  {Lehnert}}]{2014andrews}%
  \BibitemOpen
  \bibfield  {author} {\bibinfo {author} {\bibfnamefont {R.W.}\ \bibnamefont
  {Andrews}}, \bibinfo {author} {\bibfnamefont {R.W.}\ \bibnamefont
  {Peterson}}, \bibinfo {author} {\bibfnamefont {T.P.}\ \bibnamefont {Purdy}},
  \bibinfo {author} {\bibfnamefont {K.}~\bibnamefont {Cicak}}, \bibinfo
  {author} {\bibfnamefont {R.W.}\ \bibnamefont {Simmonds}}, \bibinfo {author}
  {\bibfnamefont {C.A.}\ \bibnamefont {Regal}},\ and\ \bibinfo {author}
  {\bibfnamefont {K.W.}\ \bibnamefont {Lehnert}},\ }\bibfield  {title}
  {\bibinfo {title} {Bidirectional and efficient conversion between microwave
  and optical light},\ }\href {https://doi.org/10.1038/nphys2911} {\bibfield
  {journal} {\bibinfo  {journal} {Nature Physics}\ }\textbf {\bibinfo {volume}
  {10}},\ \bibinfo {pages} {321} (\bibinfo {year} {2014})}\BibitemShut
  {NoStop}%
\bibitem [{\citenamefont {O’Connell}\ \emph {et~al.}(2010)\citenamefont
  {O’Connell}, \citenamefont {Hofheinz}, \citenamefont {Ansmann},
  \citenamefont {Bialczak}, \citenamefont {Lenander}, \citenamefont {Lucero},
  \citenamefont {Neeley}, \citenamefont {Sank}, \citenamefont {Wang},
  \citenamefont {Weides}, \citenamefont {Wenner}, \citenamefont {Martinis},\
  and\ \citenamefont {Cleland}}]{2010Connell}%
  \BibitemOpen
  \bibfield  {author} {\bibinfo {author} {\bibfnamefont {A.D.}\ \bibnamefont
  {O’Connell}}, \bibinfo {author} {\bibfnamefont {M.}~\bibnamefont
  {Hofheinz}}, \bibinfo {author} {\bibfnamefont {M.}~\bibnamefont {Ansmann}},
  \bibinfo {author} {\bibfnamefont {R.C.}\ \bibnamefont {Bialczak}}, \bibinfo
  {author} {\bibfnamefont {M.}~\bibnamefont {Lenander}}, \bibinfo {author}
  {\bibfnamefont {E.}~\bibnamefont {Lucero}}, \bibinfo {author} {\bibfnamefont
  {M.}~\bibnamefont {Neeley}}, \bibinfo {author} {\bibfnamefont
  {D.}~\bibnamefont {Sank}}, \bibinfo {author} {\bibfnamefont {H.}~\bibnamefont
  {Wang}}, \bibinfo {author} {\bibfnamefont {M.}~\bibnamefont {Weides}},
  \bibinfo {author} {\bibfnamefont {J.}~\bibnamefont {Wenner}}, \bibinfo
  {author} {\bibfnamefont {J.M.}\ \bibnamefont {Martinis}},\ and\ \bibinfo
  {author} {\bibfnamefont {A.N.}\ \bibnamefont {Cleland}},\ }\bibfield
  {title} {\bibinfo {title} {Quantum ground state and single-phonon control of
  a mechanical resonator},\ }\href {https://doi.org/10.1038/nature08967}
  {\bibfield  {journal} {\bibinfo  {journal} {Nature}\ }\textbf {\bibinfo
  {volume} {464}},\ \bibinfo {pages} {697} (\bibinfo {year}
  {2010})}\BibitemShut {NoStop}%
\bibitem [{\citenamefont {Rouxinol}\ \emph {et~al.}(2016)\citenamefont
  {Rouxinol}, \citenamefont {Hao}, \citenamefont {Brito}, \citenamefont
  {Caldeira}, \citenamefont {Irish},\ and\ \citenamefont
  {LaHaye}}]{2016rouxinol}%
  \BibitemOpen
  \bibfield  {author} {\bibinfo {author} {\bibfnamefont {F.}~\bibnamefont
  {Rouxinol}}, \bibinfo {author} {\bibfnamefont {Y.}~\bibnamefont {Hao}},
  \bibinfo {author} {\bibfnamefont {F.}~\bibnamefont {Brito}}, \bibinfo
  {author} {\bibfnamefont {A.}~\bibnamefont {Caldeira}}, \bibinfo {author}
  {\bibfnamefont {E.}~\bibnamefont {Irish}},\ and\ \bibinfo {author}
  {\bibfnamefont {M.D.}\ \bibnamefont {LaHaye}},\ }\bibfield  {title}
  {\bibinfo {title} {Measurements of nanoresonator-qubit interactions in a
  hybrid quantum electromechanical system},\ }\href
  {https://doi.org/10.1088/0957-4484/27/36/364003} {\bibfield  {journal}
  {\bibinfo  {journal} {Nanotechnology}\ }\textbf {\bibinfo {volume} {27}},\
  \bibinfo {pages} {364003} (\bibinfo {year} {2016})}\BibitemShut {NoStop}%
\bibitem [{\citenamefont {Press}\ \emph {et~al.}(2008)\citenamefont {Press},
  \citenamefont {Ladd}, \citenamefont {Zhang},\ and\ \citenamefont
  {Yamamoto}}]{2008press}%
  \BibitemOpen
  \bibfield  {author} {\bibinfo {author} {\bibfnamefont {D.}~\bibnamefont
  {Press}}, \bibinfo {author} {\bibfnamefont {T.D.}\ \bibnamefont {Ladd}},
  \bibinfo {author} {\bibfnamefont {B.}~\bibnamefont {Zhang}},\ and\ \bibinfo
  {author} {\bibfnamefont {Y.}~\bibnamefont {Yamamoto}},\ }\bibfield  {title}
  {\bibinfo {title} {Complete quantum control of a single quantum dot spin
  using ultrafast optical pulses},\ }\href
  {https://doi.org/10.1038/nature07530} {\bibfield  {journal} {\bibinfo
  {journal} {Nature}\ }\textbf {\bibinfo {volume} {456}},\ \bibinfo {pages}
  {218} (\bibinfo {year} {2008})}\BibitemShut {NoStop}%
\bibitem [{\citenamefont {Sillanp{\"a}{\"a}}\ \emph {et~al.}(2007)\citenamefont
  {Sillanp{\"a}{\"a}}, \citenamefont {Park},\ and\ \citenamefont
  {Simmonds}}]{2007sillanpaa}%
  \BibitemOpen
  \bibfield  {author} {\bibinfo {author} {\bibfnamefont {M.A.}\ \bibnamefont
  {Sillanp{\"a}{\"a}}}, \bibinfo {author} {\bibfnamefont {J.I.}\ \bibnamefont
  {Park}},\ and\ \bibinfo {author} {\bibfnamefont {R.W.}\ \bibnamefont
  {Simmonds}},\ }\bibfield  {title} {\bibinfo {title} {Coherent quantum state
  storage and transfer between two phase qubits via a resonant cavity},\ }\href
  {https://doi.org/10.1038/nature06124} {\bibfield  {journal} {\bibinfo
  {journal} {Nature}\ }\textbf {\bibinfo {volume} {449}},\ \bibinfo {pages}
  {438} (\bibinfo {year} {2007})}\BibitemShut {NoStop}%
\bibitem [{\citenamefont {Hofheinz}\ \emph {et~al.}(2008)\citenamefont
  {Hofheinz}, \citenamefont {Weig}, \citenamefont {Ansmann}, \citenamefont
  {Bialczak}, \citenamefont {Lucero}, \citenamefont {Neeley}, \citenamefont
  {O’Connell}, \citenamefont {Wang}, \citenamefont {Martinis},\ and\
  \citenamefont {Cleland}}]{2008hofheinz}%
  \BibitemOpen
  \bibfield  {author} {\bibinfo {author} {\bibfnamefont {M.}~\bibnamefont
  {Hofheinz}}, \bibinfo {author} {\bibfnamefont {E.}~\bibnamefont {Weig}},
  \bibinfo {author} {\bibfnamefont {M.}~\bibnamefont {Ansmann}}, \bibinfo
  {author} {\bibfnamefont {R.C.}\ \bibnamefont {Bialczak}}, \bibinfo {author}
  {\bibfnamefont {E.}~\bibnamefont {Lucero}}, \bibinfo {author} {\bibfnamefont
  {M.}~\bibnamefont {Neeley}}, \bibinfo {author} {\bibfnamefont {A.D.}\
  \bibnamefont {O’Connell}}, \bibinfo {author} {\bibfnamefont
  {H.}~\bibnamefont {Wang}}, \bibinfo {author} {\bibfnamefont {J.M.}\
  \bibnamefont {Martinis}},\ and\ \bibinfo {author} {\bibfnamefont {A.N.}\
  \bibnamefont {Cleland}},\ }\bibfield  {title} {\bibinfo {title} {Generation
  of {F}ock states in a superconducting quantum circuit},\ }\href
  {https://doi.org/10.1038/nature07136} {\bibfield  {journal} {\bibinfo
  {journal} {Nature}\ }\textbf {\bibinfo {volume} {454}},\ \bibinfo {pages}
  {310} (\bibinfo {year} {2008})}\BibitemShut {NoStop}%
\bibitem [{\citenamefont {Hofheinz}\ \emph {et~al.}(2009)\citenamefont
  {Hofheinz}, \citenamefont {Wang}, \citenamefont {Ansmann}, \citenamefont
  {Bialczak}, \citenamefont {Lucero}, \citenamefont {Neeley}, \citenamefont
  {O’Connell}, \citenamefont {Sank}, \citenamefont {Wenner}, \citenamefont
  {Martinis},\ and\ \citenamefont {Cleland}}]{2009hofheinz}%
  \BibitemOpen
  \bibfield  {author} {\bibinfo {author} {\bibfnamefont {M.}~\bibnamefont
  {Hofheinz}}, \bibinfo {author} {\bibfnamefont {H.}~\bibnamefont {Wang}},
  \bibinfo {author} {\bibfnamefont {M.}~\bibnamefont {Ansmann}}, \bibinfo
  {author} {\bibfnamefont {R.C.}\ \bibnamefont {Bialczak}}, \bibinfo {author}
  {\bibfnamefont {E.}~\bibnamefont {Lucero}}, \bibinfo {author} {\bibfnamefont
  {M.}~\bibnamefont {Neeley}}, \bibinfo {author} {\bibfnamefont {A.D.}\
  \bibnamefont {O’Connell}}, \bibinfo {author} {\bibfnamefont
  {D.}~\bibnamefont {Sank}}, \bibinfo {author} {\bibfnamefont {J.}~\bibnamefont
  {Wenner}}, \bibinfo {author} {\bibfnamefont {J.M.}\ \bibnamefont
  {Martinis}},\ and\ \bibinfo {author} {\bibfnamefont {A.N.}\ \bibnamefont
  {Cleland}},\ }\bibfield  {title} {\bibinfo {title} {Synthesizing arbitrary
  quantum states in a superconducting resonator},\ }\href
  {https://doi.org/10.1038/nature08005} {\bibfield  {journal} {\bibinfo
  {journal} {Nature}\ }\textbf {\bibinfo {volume} {459}},\ \bibinfo {pages}
  {546} (\bibinfo {year} {2009})}\BibitemShut {NoStop}%
\bibitem [{\citenamefont {Wang}\ \emph {et~al.}(2011)\citenamefont {Wang},
  \citenamefont {Mariantoni}, \citenamefont {Bialczak}, \citenamefont
  {Lenander}, \citenamefont {Lucero}, \citenamefont {Neeley}, \citenamefont
  {O’Connell}, \citenamefont {Sank}, \citenamefont {Weides}, \citenamefont
  {Wenner}, \citenamefont {Yamamoto}, \citenamefont {Yin}, \citenamefont
  {Zhao}, \citenamefont {Martinis},\ and\ \citenamefont {Cleland}}]{2011Wang}%
  \BibitemOpen
  \bibfield  {author} {\bibinfo {author} {\bibfnamefont {H.}~\bibnamefont
  {Wang}}, \bibinfo {author} {\bibfnamefont {M.}~\bibnamefont {Mariantoni}},
  \bibinfo {author} {\bibfnamefont {R.C.}\ \bibnamefont {Bialczak}}, \bibinfo
  {author} {\bibfnamefont {M.}~\bibnamefont {Lenander}}, \bibinfo {author}
  {\bibfnamefont {E.}~\bibnamefont {Lucero}}, \bibinfo {author} {\bibfnamefont
  {M.}~\bibnamefont {Neeley}}, \bibinfo {author} {\bibfnamefont {A.D.}\
  \bibnamefont {O’Connell}}, \bibinfo {author} {\bibfnamefont
  {D.}~\bibnamefont {Sank}}, \bibinfo {author} {\bibfnamefont {M.}~\bibnamefont
  {Weides}}, \bibinfo {author} {\bibfnamefont {J.}~\bibnamefont {Wenner}},
  \bibinfo {author} {\bibfnamefont {T.}~\bibnamefont {Yamamoto}}, \bibinfo
  {author} {\bibfnamefont {Y.}~\bibnamefont {Yin}}, \bibinfo {author}
  {\bibfnamefont {J.}~\bibnamefont {Zhao}}, \bibinfo {author} {\bibfnamefont
  {J.M.}\ \bibnamefont {Martinis}},\ and\ \bibinfo {author} {\bibfnamefont
  {A.N.}\ \bibnamefont {Cleland}},\ }\bibfield  {title} {\bibinfo {title}
  {Deterministic entanglement of photons in two superconducting microwave
  resonators},\ }\href {https://doi.org/10.1103/PhysRevLett.106.060401}
  {\bibfield  {journal} {\bibinfo  {journal} {Phys. Rev. Lett.}\ }\textbf
  {\bibinfo {volume} {106}},\ \bibinfo {pages} {060401} (\bibinfo {year}
  {2011})}\BibitemShut {NoStop}%
\bibitem [{\citenamefont {Kockum{,
  }et~al.}(2017{\natexlab{b}})}]{2017kockumConversion}%
  \BibitemOpen
  \bibfield  {author} {\bibinfo {author} {\bibfnamefont {A.F.}\ \bibnamefont
  {Kockum{, }et~al.}},\ }\bibfield  {title} {\bibinfo {title} {Frequency
  conversion in ultrastrong cavity {QED}},\ }\href
  {https://doi.org/10.1038/s41598-017-04225-3} {\bibfield  {journal} {\bibinfo
  {journal} {Scientific reports}\ }\textbf {\bibinfo {volume} {7}},\ \bibinfo
  {pages} {1} (\bibinfo {year} {2017}{\natexlab{b}})}\BibitemShut {NoStop}%
\bibitem [{\citenamefont {Heikkil\"a}\ \emph {et~al.}(2014)\citenamefont
  {Heikkil\"a}, \citenamefont {Massel}, \citenamefont {Tuorila}, \citenamefont
  {Khan},\ and\ \citenamefont {Sillanp\"a\"a}}]{2014Heikkil}%
  \BibitemOpen
  \bibfield  {author} {\bibinfo {author} {\bibfnamefont {T.T.}\ \bibnamefont
  {Heikkil\"a}}, \bibinfo {author} {\bibfnamefont {F.}~\bibnamefont {Massel}},
  \bibinfo {author} {\bibfnamefont {J.}~\bibnamefont {Tuorila}}, \bibinfo
  {author} {\bibfnamefont {R.}~\bibnamefont {Khan}},\ and\ \bibinfo {author}
  {\bibfnamefont {M.A.}\ \bibnamefont {Sillanp\"a\"a}},\ }\bibfield  {title}
  {\bibinfo {title} {Enhancing optomechanical coupling via the {J}osephson
  effect},\ }\href {https://doi.org/10.1103/PhysRevLett.112.203603} {\bibfield
  {journal} {\bibinfo  {journal} {Phys. Rev. Lett.}\ }\textbf {\bibinfo
  {volume} {112}},\ \bibinfo {pages} {203603} (\bibinfo {year}
  {2014})}\BibitemShut {NoStop}%
\bibitem [{\citenamefont {Pirkkalainen}\ \emph {et~al.}(2015)\citenamefont
  {Pirkkalainen}, \citenamefont {Cho}, \citenamefont {Massel}, \citenamefont
  {Tuorila}, \citenamefont {Heikkil{\"a}}, \citenamefont {Hakonen},\ and\
  \citenamefont {Sillanp{\"a}{\"a}}}]{2015pirkkalainen}%
  \BibitemOpen
  \bibfield  {author} {\bibinfo {author} {\bibfnamefont {J.-M.}\ \bibnamefont
  {Pirkkalainen}}, \bibinfo {author} {\bibfnamefont {S.}~\bibnamefont {Cho}},
  \bibinfo {author} {\bibfnamefont {F.}~\bibnamefont {Massel}}, \bibinfo
  {author} {\bibfnamefont {J.}~\bibnamefont {Tuorila}}, \bibinfo {author}
  {\bibfnamefont {T.}~\bibnamefont {Heikkil{\"a}}}, \bibinfo {author}
  {\bibfnamefont {P.}~\bibnamefont {Hakonen}},\ and\ \bibinfo {author}
  {\bibfnamefont {M.}~\bibnamefont {Sillanp{\"a}{\"a}}},\ }\bibfield  {title}
  {\bibinfo {title} {Cavity optomechanics mediated by a quantum two-level
  system},\ }\href {https://doi.org/10.1038/ncomms7981} {\bibfield  {journal}
  {\bibinfo  {journal} {Nature communications}\ }\textbf {\bibinfo {volume}
  {6}},\ \bibinfo {pages} {1} (\bibinfo {year} {2015})}\BibitemShut {NoStop}%
\bibitem [{\citenamefont {Zueco}\ \emph {et~al.}(2009)\citenamefont {Zueco},
  \citenamefont {Reuther}, \citenamefont {Kohler},\ and\ \citenamefont
  {H\"anggi}}]{2009Zueco}%
  \BibitemOpen
  \bibfield  {author} {\bibinfo {author} {\bibfnamefont {D.}~\bibnamefont
  {Zueco}}, \bibinfo {author} {\bibfnamefont {G.M.}\ \bibnamefont {Reuther}},
  \bibinfo {author} {\bibfnamefont {S.}~\bibnamefont {Kohler}},\ and\ \bibinfo
  {author} {\bibfnamefont {P.}~\bibnamefont {H\"anggi}},\ }\bibfield  {title}
  {\bibinfo {title} {Qubit-oscillator dynamics in the dispersive regime:
  Analytical theory beyond the rotating-wave approximation},\ }\href
  {https://doi.org/10.1103/PhysRevA.80.033846} {\bibfield  {journal} {\bibinfo
  {journal} {Phys. Rev. A}\ }\textbf {\bibinfo {volume} {80}},\ \bibinfo
  {pages} {033846} (\bibinfo {year} {2009})}\BibitemShut {NoStop}%
\bibitem [{\citenamefont {Hattermann}\ \emph {et~al.}(2017)\citenamefont
  {Hattermann}, \citenamefont {Bothner}, \citenamefont {Ley}, \citenamefont
  {Ferdinand}, \citenamefont {Wiedmaier}, \citenamefont {S{\'a}rk{\'a}ny},
  \citenamefont {Kleiner}, \citenamefont {Koelle},\ and\ \citenamefont
  {Fort{\'a}gh}}]{2017Hattermanna}%
  \BibitemOpen
  \bibfield  {author} {\bibinfo {author} {\bibfnamefont {H.}~\bibnamefont
  {Hattermann}}, \bibinfo {author} {\bibfnamefont {D.}~\bibnamefont {Bothner}},
  \bibinfo {author} {\bibfnamefont {L.}~\bibnamefont {Ley}}, \bibinfo {author}
  {\bibfnamefont {B.}~\bibnamefont {Ferdinand}}, \bibinfo {author}
  {\bibfnamefont {D.}~\bibnamefont {Wiedmaier}}, \bibinfo {author}
  {\bibfnamefont {L.}~\bibnamefont {S{\'a}rk{\'a}ny}}, \bibinfo {author}
  {\bibfnamefont {R.}~\bibnamefont {Kleiner}}, \bibinfo {author} {\bibfnamefont
  {D.}~\bibnamefont {Koelle}},\ and\ \bibinfo {author} {\bibfnamefont
  {J.}~\bibnamefont {Fort{\'a}gh}},\ }\bibfield  {title} {\bibinfo {title}
  {Coupling ultracold atoms to a superconducting coplanar waveguide
  resonator},\ }\href {https://doi.org/10.1038/s41467-017-02439-7} {\bibfield
  {journal} {\bibinfo  {journal} {Nature communications}\ }\textbf {\bibinfo
  {volume} {8}},\ \bibinfo {pages} {1} (\bibinfo {year} {2017})}\BibitemShut
  {NoStop}%
\bibitem [{\citenamefont {Andr{\'e}}\ \emph {et~al.}(2006)\citenamefont
  {Andr{\'e}}, \citenamefont {DeMille}, \citenamefont {Doyle}, \citenamefont
  {Lukin}, \citenamefont {Maxwell}, \citenamefont {Rabl}, \citenamefont
  {Schoelkopf},\ and\ \citenamefont {Zoller}}]{2006andre}%
  \BibitemOpen
  \bibfield  {author} {\bibinfo {author} {\bibfnamefont {A.}~\bibnamefont
  {Andr{\'e}}}, \bibinfo {author} {\bibfnamefont {D.}~\bibnamefont {DeMille}},
  \bibinfo {author} {\bibfnamefont {J.M.}\ \bibnamefont {Doyle}}, \bibinfo
  {author} {\bibfnamefont {M.D.}\ \bibnamefont {Lukin}}, \bibinfo {author}
  {\bibfnamefont {S.E.}\ \bibnamefont {Maxwell}}, \bibinfo {author}
  {\bibfnamefont {P.}~\bibnamefont {Rabl}}, \bibinfo {author} {\bibfnamefont
  {R.J.}\ \bibnamefont {Schoelkopf}},\ and\ \bibinfo {author} {\bibfnamefont
  {P.}~\bibnamefont {Zoller}},\ }\bibfield  {title} {\bibinfo {title} {A
  coherent all-electrical interface between polar molecules and mesoscopic
  superconducting resonators},\ }\href {https://doi.org/10.1038/nphys386}
  {\bibfield  {journal} {\bibinfo  {journal} {Nature Physics}\ }\textbf
  {\bibinfo {volume} {2}},\ \bibinfo {pages} {636} (\bibinfo {year}
  {2006})}\BibitemShut {NoStop}%
\bibitem [{\citenamefont {Fink}\ \emph {et~al.}(2009)\citenamefont {Fink},
  \citenamefont {Bianchetti}, \citenamefont {Baur}, \citenamefont {G\"oppl},
  \citenamefont {Steffen}, \citenamefont {Filipp}, \citenamefont {Leek},
  \citenamefont {Blais},\ and\ \citenamefont {Wallraff}}]{2009Finka}%
  \BibitemOpen
  \bibfield  {author} {\bibinfo {author} {\bibfnamefont {J.M.}\ \bibnamefont
  {Fink}}, \bibinfo {author} {\bibfnamefont {R.}~\bibnamefont {Bianchetti}},
  \bibinfo {author} {\bibfnamefont {M.}~\bibnamefont {Baur}}, \bibinfo {author}
  {\bibfnamefont {M.}~\bibnamefont {G\"oppl}}, \bibinfo {author} {\bibfnamefont
  {L.}~\bibnamefont {Steffen}}, \bibinfo {author} {\bibfnamefont
  {S.}~\bibnamefont {Filipp}}, \bibinfo {author} {\bibfnamefont {P.J.}\
  \bibnamefont {Leek}}, \bibinfo {author} {\bibfnamefont {A.}~\bibnamefont
  {Blais}},\ and\ \bibinfo {author} {\bibfnamefont {A.}~\bibnamefont
  {Wallraff}},\ }\bibfield  {title} {\bibinfo {title} {Dressed collective qubit
  states and the {T}avis-{C}ummings model in circuit {QED}},\ }\href
  {https://doi.org/10.1103/PhysRevLett.103.083601} {\bibfield  {journal}
  {\bibinfo  {journal} {Phys. Rev. Lett.}\ }\textbf {\bibinfo {volume} {103}},\
  \bibinfo {pages} {083601} (\bibinfo {year} {2009})}\BibitemShut {NoStop}%
\bibitem [{\citenamefont {Verd\'u}\ \emph {et~al.}(2009)\citenamefont
  {Verd\'u}, \citenamefont {Zoubi}, \citenamefont {Koller}, \citenamefont
  {Majer}, \citenamefont {Ritsch},\ and\ \citenamefont
  {Schmiedmayer}}]{2009Verda}%
  \BibitemOpen
  \bibfield  {author} {\bibinfo {author} {\bibfnamefont {J.}~\bibnamefont
  {Verd\'u}}, \bibinfo {author} {\bibfnamefont {H.}~\bibnamefont {Zoubi}},
  \bibinfo {author} {\bibfnamefont {C.}~\bibnamefont {Koller}}, \bibinfo
  {author} {\bibfnamefont {J.}~\bibnamefont {Majer}}, \bibinfo {author}
  {\bibfnamefont {H.}~\bibnamefont {Ritsch}},\ and\ \bibinfo {author}
  {\bibfnamefont {J.}~\bibnamefont {Schmiedmayer}},\ }\bibfield  {title}
  {\bibinfo {title} {Strong magnetic coupling of an ultracold gas to a
  superconducting waveguide cavity},\ }\href
  {https://doi.org/10.1103/PhysRevLett.103.043603} {\bibfield  {journal}
  {\bibinfo  {journal} {Phys. Rev. Lett.}\ }\textbf {\bibinfo {volume} {103}},\
  \bibinfo {pages} {043603} (\bibinfo {year} {2009})}\BibitemShut {NoStop}%
\bibitem [{\citenamefont {Schuster}\ \emph {et~al.}(2010)\citenamefont
  {Schuster}, \citenamefont {Sears}, \citenamefont {Ginossar}, \citenamefont
  {DiCarlo}, \citenamefont {Frunzio}, \citenamefont {Morton}, \citenamefont
  {Wu}, \citenamefont {Briggs}, \citenamefont {Buckley}, \citenamefont
  {Awschalom},\ and\ \citenamefont {Schoelkopf}}]{2010Schustera}%
  \BibitemOpen
  \bibfield  {author} {\bibinfo {author} {\bibfnamefont {D.I.}\ \bibnamefont
  {Schuster}}, \bibinfo {author} {\bibfnamefont {A.P.}\ \bibnamefont {Sears}},
  \bibinfo {author} {\bibfnamefont {E.}~\bibnamefont {Ginossar}}, \bibinfo
  {author} {\bibfnamefont {L.}~\bibnamefont {DiCarlo}}, \bibinfo {author}
  {\bibfnamefont {L.}~\bibnamefont {Frunzio}}, \bibinfo {author} {\bibfnamefont
  {J.J.L.}\ \bibnamefont {Morton}}, \bibinfo {author} {\bibfnamefont
  {H.}~\bibnamefont {Wu}}, \bibinfo {author} {\bibfnamefont {G.A.D.}\
  \bibnamefont {Briggs}}, \bibinfo {author} {\bibfnamefont {B.B.}\
  \bibnamefont {Buckley}}, \bibinfo {author} {\bibfnamefont {D.D.}\
  \bibnamefont {Awschalom}},\ and\ \bibinfo {author} {\bibfnamefont {R.J.}\
  \bibnamefont {Schoelkopf}},\ }\bibfield  {title} {\bibinfo {title}
  {High-cooperativity coupling of electron-spin ensembles to superconducting
  cavities},\ }\href {https://doi.org/10.1103/PhysRevLett.105.140501}
  {\bibfield  {journal} {\bibinfo  {journal} {Phys. Rev. Lett.}\ }\textbf
  {\bibinfo {volume} {105}},\ \bibinfo {pages} {140501} (\bibinfo {year}
  {2010})}\BibitemShut {NoStop}%
\bibitem [{\citenamefont {Kubo}\ \emph {et~al.}(2010)\citenamefont {Kubo},
  \citenamefont {Ong}, \citenamefont {Bertet}, \citenamefont {Vion},
  \citenamefont {Jacques}, \citenamefont {Zheng}, \citenamefont {Dr\'eau},
  \citenamefont {Roch}, \citenamefont {Auffeves}, \citenamefont {Jelezko},
  \citenamefont {Wrachtrup}, \citenamefont {Barthe}, \citenamefont {Bergonzo},\
  and\ \citenamefont {Esteve}}]{2010Kuboa}%
  \BibitemOpen
  \bibfield  {author} {\bibinfo {author} {\bibfnamefont {Y.}~\bibnamefont
  {Kubo}}, \bibinfo {author} {\bibfnamefont {F.R.}\ \bibnamefont {Ong}},
  \bibinfo {author} {\bibfnamefont {P.}~\bibnamefont {Bertet}}, \bibinfo
  {author} {\bibfnamefont {D.}~\bibnamefont {Vion}}, \bibinfo {author}
  {\bibfnamefont {V.}~\bibnamefont {Jacques}}, \bibinfo {author} {\bibfnamefont
  {D.}~\bibnamefont {Zheng}}, \bibinfo {author} {\bibfnamefont
  {A.}~\bibnamefont {Dr\'eau}}, \bibinfo {author} {\bibfnamefont {J.-F.}\
  \bibnamefont {Roch}}, \bibinfo {author} {\bibfnamefont {A.}~\bibnamefont
  {Auffeves}}, \bibinfo {author} {\bibfnamefont {F.}~\bibnamefont {Jelezko}},
  \bibinfo {author} {\bibfnamefont {J.}~\bibnamefont {Wrachtrup}}, \bibinfo
  {author} {\bibfnamefont {M.F.}\ \bibnamefont {Barthe}}, \bibinfo {author}
  {\bibfnamefont {P.}~\bibnamefont {Bergonzo}},\ and\ \bibinfo {author}
  {\bibfnamefont {D.}~\bibnamefont {Esteve}},\ }\bibfield  {title} {\bibinfo
  {title} {Strong coupling of a spin ensemble to a superconducting resonator},\
  }\href {https://doi.org/10.1103/PhysRevLett.105.140502} {\bibfield  {journal}
  {\bibinfo  {journal} {Phys. Rev. Lett.}\ }\textbf {\bibinfo {volume} {105}},\
  \bibinfo {pages} {140502} (\bibinfo {year} {2010})}\BibitemShut {NoStop}%
\bibitem [{\citenamefont {Zhu}\ \emph {et~al.}(2011)\citenamefont {Zhu},
  \citenamefont {Saito}, \citenamefont {Kemp}, \citenamefont {Kakuyanagi},
  \citenamefont {Karimoto}, \citenamefont {Nakano}, \citenamefont {Munro},
  \citenamefont {Tokura}, \citenamefont {Everitt}, \citenamefont {Nemoto},
  \citenamefont {Kasu}, \citenamefont {Mizuochi},\ and\ \citenamefont
  {Semba}}]{2011zhu}%
  \BibitemOpen
  \bibfield  {author} {\bibinfo {author} {\bibfnamefont {X.}~\bibnamefont
  {Zhu}}, \bibinfo {author} {\bibfnamefont {S.}~\bibnamefont {Saito}}, \bibinfo
  {author} {\bibfnamefont {A.}~\bibnamefont {Kemp}}, \bibinfo {author}
  {\bibfnamefont {K.}~\bibnamefont {Kakuyanagi}}, \bibinfo {author}
  {\bibfnamefont {S.-i.}\ \bibnamefont {Karimoto}}, \bibinfo {author}
  {\bibfnamefont {H.}~\bibnamefont {Nakano}}, \bibinfo {author} {\bibfnamefont
  {W.J.}\ \bibnamefont {Munro}}, \bibinfo {author} {\bibfnamefont
  {Y.}~\bibnamefont {Tokura}}, \bibinfo {author} {\bibfnamefont {M.S.}\
  \bibnamefont {Everitt}}, \bibinfo {author} {\bibfnamefont {K.}~\bibnamefont
  {Nemoto}}, \bibinfo {author} {\bibfnamefont {M.}~\bibnamefont {Kasu}},
  \bibinfo {author} {\bibfnamefont {N.}~\bibnamefont {Mizuochi}},\ and\
  \bibinfo {author} {\bibfnamefont {K.}~\bibnamefont {Semba}},\ }\bibfield
  {title} {\bibinfo {title} {Coherent coupling of a superconducting flux qubit
  to an electron spin ensemble in diamond},\ }\href
  {https://doi.org/10.1038/nature10462} {\bibfield  {journal} {\bibinfo
  {journal} {Nature}\ }\textbf {\bibinfo {volume} {478}},\ \bibinfo {pages}
  {221} (\bibinfo {year} {2011})}\BibitemShut {NoStop}%
\bibitem [{\citenamefont {Chu}\ \emph {et~al.}(2017)\citenamefont {Chu},
  \citenamefont {Kharel}, \citenamefont {Renninger}, \citenamefont {Burkhart},
  \citenamefont {Frunzio}, \citenamefont {Rakich},\ and\ \citenamefont
  {Schoelkopf}}]{2017chu}%
  \BibitemOpen
  \bibfield  {author} {\bibinfo {author} {\bibfnamefont {Y.}~\bibnamefont
  {Chu}}, \bibinfo {author} {\bibfnamefont {P.}~\bibnamefont {Kharel}},
  \bibinfo {author} {\bibfnamefont {W.H.}\ \bibnamefont {Renninger}}, \bibinfo
  {author} {\bibfnamefont {L.D.}\ \bibnamefont {Burkhart}}, \bibinfo {author}
  {\bibfnamefont {L.}~\bibnamefont {Frunzio}}, \bibinfo {author} {\bibfnamefont
  {P.T.}\ \bibnamefont {Rakich}},\ and\ \bibinfo {author} {\bibfnamefont
  {R.J.}\ \bibnamefont {Schoelkopf}},\ }\bibfield  {title} {\bibinfo {title}
  {Quantum acoustics with superconducting qubits},\ }\href
  {https://doi.org/10.1126/science.aao1511} {\bibfield  {journal} {\bibinfo
  {journal} {Science}\ }\textbf {\bibinfo {volume} {358}},\ \bibinfo {pages}
  {199} (\bibinfo {year} {2017})}\BibitemShut {NoStop}%
\bibitem [{\citenamefont {Scarlino}\ \emph {et~al.}(2019)\citenamefont
  {Scarlino}, \citenamefont {van Woerkom}, \citenamefont {Stockklauser},
  \citenamefont {Koski}, \citenamefont {Collodo}, \citenamefont {Gasparinetti},
  \citenamefont {Reichl}, \citenamefont {Wegscheider}, \citenamefont {Ihn},
  \citenamefont {Ensslin},\ and\ \citenamefont {Wallraff}}]{2019Scarlino}%
  \BibitemOpen
  \bibfield  {author} {\bibinfo {author} {\bibfnamefont {P.}~\bibnamefont
  {Scarlino}}, \bibinfo {author} {\bibfnamefont {D.J.}\ \bibnamefont {van
  Woerkom}}, \bibinfo {author} {\bibfnamefont {A.}~\bibnamefont
  {Stockklauser}}, \bibinfo {author} {\bibfnamefont {J.V.}\ \bibnamefont
  {Koski}}, \bibinfo {author} {\bibfnamefont {M.C.}\ \bibnamefont {Collodo}},
  \bibinfo {author} {\bibfnamefont {S.}~\bibnamefont {Gasparinetti}}, \bibinfo
  {author} {\bibfnamefont {C.}~\bibnamefont {Reichl}}, \bibinfo {author}
  {\bibfnamefont {W.}~\bibnamefont {Wegscheider}}, \bibinfo {author}
  {\bibfnamefont {T.}~\bibnamefont {Ihn}}, \bibinfo {author} {\bibfnamefont
  {K.}~\bibnamefont {Ensslin}},\ and\ \bibinfo {author} {\bibfnamefont
  {A.}~\bibnamefont {Wallraff}},\ }\bibfield  {title} {\bibinfo {title}
  {All-microwave control and dispersive readout of gate-defined quantum dot
  qubits in circuit quantum electrodynamics},\ }\href
  {https://doi.org/10.1103/PhysRevLett.122.206802} {\bibfield  {journal}
  {\bibinfo  {journal} {Phys. Rev. Lett.}\ }\textbf {\bibinfo {volume} {122}},\
  \bibinfo {pages} {206802} (\bibinfo {year} {2019})}\BibitemShut {NoStop}%
\bibitem [{\citenamefont {Wollman}\ \emph {et~al.}(2015)\citenamefont
  {Wollman}, \citenamefont {Lei}, \citenamefont {Weinstein}, \citenamefont
  {Suh}, \citenamefont {Kronwald}, \citenamefont {Marquardt}, \citenamefont
  {Clerk},\ and\ \citenamefont {Schwab}}]{2015wollman}%
  \BibitemOpen
  \bibfield  {author} {\bibinfo {author} {\bibfnamefont {E.E.}\ \bibnamefont
  {Wollman}}, \bibinfo {author} {\bibfnamefont {C.}~\bibnamefont {Lei}},
  \bibinfo {author} {\bibfnamefont {A.}~\bibnamefont {Weinstein}}, \bibinfo
  {author} {\bibfnamefont {J.}~\bibnamefont {Suh}}, \bibinfo {author}
  {\bibfnamefont {A.}~\bibnamefont {Kronwald}}, \bibinfo {author}
  {\bibfnamefont {F.}~\bibnamefont {Marquardt}}, \bibinfo {author}
  {\bibfnamefont {A.A.}\ \bibnamefont {Clerk}},\ and\ \bibinfo {author}
  {\bibfnamefont {K.}~\bibnamefont {Schwab}},\ }\bibfield  {title} {\bibinfo
  {title} {Quantum squeezing of motion in a mechanical resonator},\ }\href
  {https://doi.org/10.1126/science.aac5138} {\bibfield  {journal} {\bibinfo
  {journal} {Science}\ }\textbf {\bibinfo {volume} {349}},\ \bibinfo {pages}
  {952} (\bibinfo {year} {2015})}\BibitemShut {NoStop}%
\bibitem [{\citenamefont {Teufel}\ \emph {et~al.}(2011)\citenamefont {Teufel},
  \citenamefont {Donner}, \citenamefont {Li}, \citenamefont {Harlow},
  \citenamefont {Allman}, \citenamefont {Cicak}, \citenamefont {Sirois},
  \citenamefont {Whittaker}, \citenamefont {Lehnert},\ and\ \citenamefont
  {Simmonds}}]{2011teufel}%
  \BibitemOpen
  \bibfield  {author} {\bibinfo {author} {\bibfnamefont {J.D.}\ \bibnamefont
  {Teufel}}, \bibinfo {author} {\bibfnamefont {T.}~\bibnamefont {Donner}},
  \bibinfo {author} {\bibfnamefont {D.}~\bibnamefont {Li}}, \bibinfo {author}
  {\bibfnamefont {J.W.}\ \bibnamefont {Harlow}}, \bibinfo {author}
  {\bibfnamefont {M.}~\bibnamefont {Allman}}, \bibinfo {author} {\bibfnamefont
  {K.}~\bibnamefont {Cicak}}, \bibinfo {author} {\bibfnamefont {A.J.}\
  \bibnamefont {Sirois}}, \bibinfo {author} {\bibfnamefont {J.D.}\
  \bibnamefont {Whittaker}}, \bibinfo {author} {\bibfnamefont {K.W.}\
  \bibnamefont {Lehnert}},\ and\ \bibinfo {author} {\bibfnamefont {R.W.}\
  \bibnamefont {Simmonds}},\ }\bibfield  {title} {\bibinfo {title} {Sideband
  cooling of micromechanical motion to the quantum ground state},\ }\href
  {https://doi.org/10.1038/nature10261} {\bibfield  {journal} {\bibinfo
  {journal} {Nature}\ }\textbf {\bibinfo {volume} {475}},\ \bibinfo {pages}
  {359} (\bibinfo {year} {2011})}\BibitemShut {NoStop}%
\bibitem [{\citenamefont {Breuer}\ and\ \citenamefont
  {Petruccione}(2002)}]{2002theory}%
  \BibitemOpen
  \bibfield  {author} {\bibinfo {author} {\bibfnamefont {H.-P.}\ \bibnamefont
  {Breuer}}\ and\ \bibinfo {author} {\bibfnamefont {F.}~\bibnamefont
  {Petruccione}},\ }\href@noop {} {\emph {\bibinfo {title} {The theory of open
  quantum systems}}}\ (\bibinfo  {publisher} {Oxford University Press on
  Demand},\ \bibinfo {year} {2002})\BibitemShut {NoStop}%
\bibitem [{\citenamefont {Yin}(2021)}]{2021yin}%
  \BibitemOpen
  \bibfield  {author} {\bibinfo {author} {\bibfnamefont {M.}~\bibnamefont
  {Yin}},\ }\href@noop {} {\bibinfo {title} {Mechanical oscillator can excite
  an atom through the quantum vacuum}} (\bibinfo {year} {2021}),\ \Eprint
  {https://arxiv.org/abs/2106.14206} {arXiv:2106.14206 [quant-ph]} \BibitemShut
  {NoStop}%
\end{thebibliography}%

\end{document}